\documentclass[preprint,authoryear,3p,times,12pt]{elsarticle}

\usepackage{graphicx}
\usepackage{array,multirow}
\usepackage{amsmath}
\usepackage{booktabs}
\usepackage{mdwmath}
\usepackage{epstopdf}
\usepackage[breaklinks]{hyperref}

\usepackage[table]{xcolor}
\definecolor{gray}{rgb}{0.5,0.5,0.5}
\definecolor{dgreen}{rgb}{0,0.5,0}
\definecolor{dred}{rgb}{0.6,0,0}

\usepackage{subcaption}

\newcommand{\V}{{\mathbb V}}

\newcommand{\X}{{\mathbb X}}

\newcommand{\EE}{{\mathcal{E}}}

\newcommand{\XX}{{\mathcal{X}}}
\newcommand{\YY}{{\mathcal{Y}}}

\newcommand{\bsf}{\boldsymbol f}

\newcommand{\bsP}{\boldsymbol P}

\newcommand{\bsY}{\boldsymbol Y}

\newcommand{\bsone}{\boldsymbol 1}

\newcommand{\bsnull}{\boldsymbol 0}

\newcommand{\bsbeta}{\boldsymbol \beta}

\newcommand{\bseps}{\boldsymbol \varepsilon}

\newcommand{\bsphi}{\boldsymbol \phi}

\newcommand{\bsPhi}{\boldsymbol \Phi}

\newcommand{\eps}{{\varepsilon}}

\DeclareMathOperator*{\argmin}{arg\,min}

\DeclareMathOperator{\modulo}{mod}
\DeclareMathOperator{\sign}{sign}

\DeclareMathOperator{\var}{\V ar}

\newcommand{\ov}\overline
\newcommand{\what}{\widehat}
\newcommand{\wtilde}{\widetilde}
\newcommand{\ow}{\text{otherwise}}
\newcommand{\rig}\right
\newcommand{\lef}\left
\newcommand{\nf}\normalfont

\journal{Energy Economics}

\begin{document}
	
\begin{frontmatter}

\title{Day-ahead electricity price forecasting with high-dimensional structures: Univariate vs.\ multivariate modeling frameworks}

\author[UDE]{Florian Ziel}
\ead{florian.ziel@uni-due.de}
\author[WUT]{Rafa{\l} Weron}
\ead{rafal.weron@pwr.edu.pl}

\address[UDE]{Faculty of Business Administration and Economics, University Duisburg-Essen, Essen, Germany}
\address[WUT]{Department of Operations Research, Wroc{\l}aw University of Technology, Wroc{\l}aw, Poland}

\date{This version: \today}

\begin{abstract}
{We conduct an extensive empirical study on short-term electricity price forecasting (EPF) to address the long-standing question if the optimal model structure for EPF is univariate or multivariate.} We provide evidence that despite a minor edge in predictive performance overall, the multivariate modeling framework does not uniformly outperform the univariate one across all {12 considered} datasets, seasons of the year or hours of the day, and at times is outperformed by the latter. This {is} an indication that combining advanced structures or the corresponding forecasts from both modeling approaches can bring a further improvement in forecasting accuracy. {We show that this indeed can be the case, even for a simple averaging scheme involving only two models.}
Finally, we also analyze variable selection for the best performing high-dimensional lasso-type models, thus provide guidelines to structuring better performing forecasting model designs.
%We conduct an extensive empirical study on short-term electricity price forecasting (EPF), involving datasets from 12 power markets, state-of-the-art parsimonious expert models, univariate and multivariate autoregressive benchmarks, and multi-parameter regression models estimated via the lasso. We show that using the latter shrinkage approach can bring statistically significant accuracy gains compared to commonly used EPF models. Additionally, we address the long-standing question on the optimal model structure for EPF. We provide evidence that despite a  minor edge in predictive performance overall, the multivariate modeling framework does not uniformly outperform the univariate one across all datasets, seasons of the year or hours of the day, and at times is outperformed by the latter. This may be an indication that combining advanced structures or the corresponding forecasts from both modeling approaches can bring a further improvement in forecasting accuracy. Finally, we also analyze variable selection for the best performing high-dimensional lasso-type models, thus provide guidelines to structuring better performing forecasting model designs.
\end{abstract}


\end{frontmatter}

\section{Introduction} 
\label{Introduction}

There is no consensus in the existing literature on short-term \emph{electricity price forecasting} (EPF) as to the representation of the price series \cite[see][for a recent review]{wer:14}. Should the modeling be implemented in a \emph{multivariate} fashion, i.e., with separate but possibly interdependent models for each of the 24 (48 or more) load periods, or within a \emph{univariate} framework, where one large model is constructed and the same set of parameters is used to produce one- to 24-step ahead predictions for all load periods of the next day?

Surprisingly, though, there are very few and very limited studies in the EPF literature where the univariate and multivariate frameworks are compared. \cite{cua:hlo:kos:obe:04} apply variants of AR(1) and general ARMA processes (including ARMA with jumps) to short-term EPF in the German EEX market. They conclude that specifications in which each hour of the day is modeled separately (i.e., a \emph{multivariate} framework) present uniformly better forecasting properties than \emph{univariate} time series models. More recently, \cite{zie:16:TPWRS} notes that, when we compare the forecasting performance of relatively simple time series models implemented either {in a \emph{multivariate}} or a \emph{univariate} framework, the latter generally perform better for the first half of the day, whereas the former are better in the second half of the day. However, there has been no through, empirical study to date, involving many fine-tuned specifications from both groups. With this paper we want to fill the gap and provide much needed evidence. In particular we want to address three pertinent questions:
\begin{enumerate}
	\itemsep0em
	\item Which modeling framework -- \emph{multivariate} or \emph{univariate} --  is better for EPF?
	\item If one of them is better, is it better across all hours, seasons of the year and markets?
	\item How many and which past values of the spot price process should be used in EPF models?
\end{enumerate}

The remainder of the paper is structured as follows. In Section \ref{sec:Frameworks} we thoroughly discuss the univariate and multivariate modeling frameworks, which are driven by different data-format perspectives. This is a crucial, conceptual part of the paper, which sets ground for the empirical analysis in the following Sections. 
In Section \ref{sec:Data} we briefly describe the 12 price series used and present the \emph{area hyperbolic sine} transform for stabilizing the variance of spot price data. In Section \ref{sec:Models} we define 10 forecasting models representing eight model classes:  (C1) the mean values of the past prices, (C2) similar-day techniques, (C3) sets of 24 parsimonious, interrelated autoregressive (AR) structures (so-called \emph{expert} models), (C4) sets of 24 univariate AR models, (C5) vector autoregressive (VAR) models, (C6) sets of 24 parameter-rich, interrelated AR models estimated using the \emph{least absolute shrinkage and selection operator} (i.e., lasso or LASSO; which shrinks to zero the coefficients of redundant explanatory variables), (C7) univariate AR models and (C8) univariate, parameter-rich AR models estimated using the lasso. In Section \ref{sec:Empirical} we evaluate their performance on the basis of the Mean Absolute Error (MAE), the mean percentage deviation from the best (m.p.d.f.b.) model and using two variants of the \cite{die:mar:95} test for significant differences in the forecasting performance. We also discuss variable selection for the best performing lasso-type models. In Section \ref{sec:Conclusions} we wrap up the results and provide guidelines for energy modelers and forecasters. Finally, in the Appendixes we define the full set of 58 forecasting models considered in our empirical study (for clarity of exposition in Section \ref{sec:Empirical} we report detailed results only for 10 representative models), provide formulas for alternative representations of some of the models, and summarize the predictive performance of all 58 models.

\section{The univariate and multivariate modeling frameworks}	
\label{sec:Frameworks}
	
{
Recall, that the day-ahead price series is a result of conducted once per day (usually around noon) auctions for the 24 hours of the next day \citep{bur:gra:sch:07,hui:huu:mah:07,wer:zie:18}. Consequently, the electricity prices $P_{d,1}, \ldots, P_{d, 24}$ for day $d$ and hours $1,\ldots,24$ are disclosed at once, and can be regarded as a \emph{multivariate} time series of the 24-dimensional random vector $\bsP_d = [P_{d,1}, \ldots, P_{d, 24}]'$. Next to the daily auction argument there are {two other} practical reasons for the multivariate modeling framework:
(i) the demand forecasting literature, which has generally favored the multivariate framework for short-term predictions, and (ii) {the fact} that each load period (hour, half-hour) displays a rather distinct price profile, reflecting the daily variation of demand, costs, operational constraints and bidding strategies \citep{gia:par:pel:16,kar:bun:08,sha:yam:li:02}. On the other hand, the electricity prices can be rewritten as one `high-frequency' (hourly, half-hourly) \emph{univariate} time series: $P_t=P_{24d+h}=P_{d,h}$, hence are prone to modeling within a univariate framework. 
The univariate approach is more popular in the engineering EPF literature, dominated by neural network models \cite[see][for a review]{agg:sai:kum:09a}, but has its roots also in the traditional time series analysis of financial and commodity markets. 

Both approaches have their proponents. For instance, \cite{cua:hlo:kos:obe:04}, \cite{mis:tru:wer:06}, \cite{zho:yan:ni:li:nie:06}, \cite{g-m:rod:san:07}, \cite{kar:bun:08}, \cite{lis:nan:14}, \cite{alo:bas:gar:16}, \cite{gai:gou:ned:16}, \cite{hag:etal:16}, \cite{mac:now:wer:16}, \cite{now:wer:16}, \cite{uni:now:wer:16}, and \cite{zie:16:TPWRS}, among others, advocate the use of sets of 24 (48 or more) models estimated independently for each load period, typically using Ordinary Least Squares (OLS). 
In the neural network literature, \cite{amj:key:09:ECM}, {\cite{mar:uni:wer:18}} and \cite{pan:dag:16}, among others, use a separate network (i.e., a different parameter set) for each hour of the next day.

Studies where univariate statistical time series models are used include \cite{nog:con:con:esp:02}, \cite{con:esp:nog:con:03}, \cite{con:con:esp:pla:05}, \cite{zar:can:bha:tho:06}, \cite{par:fle:sch:15} and \cite{zie:ste:hus:15}, while papers where neural networks are put to work include \cite{rod:and:04}, \cite{amj:06}, \cite{pao:07}, \cite{amj:dar:key:10}, \cite{abe:amj:s-h:cat:15}, \cite{kim:15}, \cite{dud:16}, \cite{kel:etal:16} and \cite{raf:nik:kho:17}, among others.
}
	
\subsection{The multivariate modeling framework}	
\label{ssec:MultivariateFramework}
	
{The simplest, yet surprisingly often used structure for the 24-dimensional price time series} is a \emph{set of 24 univariate models}:
\begin{equation}\label{eqn:multivariate:model1}
\begin{cases}
\begin{array}{lcl}
P_{d,1} = f_1(P_{d-1,1},P_{d-2,1},...) + \varepsilon_{d,1} & \longrightarrow & \hat{P}_{d,1},\\
\multicolumn{1}{c}{\vdots}  & & \multicolumn{1}{c}{\vdots} \\
P_{d,24} = f_{24}(P_{d-1,24},P_{d-2,24},...) + \varepsilon_{d,24} & \longrightarrow & \hat{P}_{d,24},
\end{array}
\end{cases}
\end{equation}
where $\varepsilon_{d,h}$ is the innovation (noise) term for day $d$ and hour $h$, and $f_h(\cdot)$ are some functions of the explanatory variables of the past prices in the same load period.
A commonly raised argument in favor of this approach is that it is simple to implement, involves only a small number of parameters for each load period and hence is computationally non-demanding. The downside, however, is that the estimated set of models does not take into account the potentially important dependencies between the variables across the load periods.
Still, by increasing the set of dependent explanatory variables such interrelationships can be added. For instance, \cite{gai:gou:ned:16}, \cite{uni:now:wer:16} and \cite{zie:16:TPWRS} consider the previous day's price for midnight, i.e., $P_{d-1,24}$, as an explanatory variable in each of the 24 single models. 
Formally such a \emph{set of 24 interrelated models} can be written as:
\begin{equation}\label{eqn:multivariate:model1b}
\begin{cases}
\begin{array}{lcl}
P_{d,1} = f_1(P_{d-1,1},P_{d-2,1},\ldots,P_{d-1,24},P_{d-2,24},\ldots) + \varepsilon_{d,1} & \longrightarrow & \hat{P}_{d,1},\\
\multicolumn{1}{c}{\vdots}  & & \multicolumn{1}{c}{\vdots} \\
P_{d,24} = f_{24}(P_{d-1,1},P_{d-2,1},\ldots,P_{d-1,24},P_{d-2,24},\ldots) + \varepsilon_{d,24} & \longrightarrow & \hat{P}_{d,24},
\end{array}
\end{cases}
\end{equation}
which according to \cite{cha:00} and \cite{die:04} can be regraded as a \textit{multivariate} model, since the dependency structure is interrelated.

It should be emphasized that both frameworks, defined by Eqns.\ \eqref{eqn:multivariate:model1} and \eqref{eqn:multivariate:model1b}, make explicitly (or implicitly) assumptions on the innovations for individual load periods. For instance, that for each hour $\eps_{d,h}$ follows a normal distribution with zero mean, i.e., $\eps_{d,h}\sim N(0,\sigma_h^2)$. However, they do not assume anything about the joint distribution of the innovations for different hours. 
To mitigate this unwanted feature, a \emph{fully multivariate} modeling framework may be implemented which treats the price series as panel data:
\begin{equation}\label{eqn:multivariate:model2}
\begin{bmatrix}
P_{d,1}\\ 
\vdots \\
P_{d,24}\\ 
\end{bmatrix}
= \bsf \left(
\begin{bmatrix}
P_{d-1,1}\\ 
\vdots \\
P_{d-1,24}\\ 
\end{bmatrix},
\begin{bmatrix}
P_{d-2,1}\\  
\vdots \\
P_{d-2,24}\\ 
\end{bmatrix}, ... \right) + 
\begin{bmatrix}
\varepsilon_{d,1}\\  
\vdots \\
\varepsilon_{d,24}\\ 
\end{bmatrix}
\quad \longrightarrow \quad
\begin{bmatrix}
\hat{P}_{d,1}\\ 
\vdots \\
\hat{P}_{d,24}\\ 
\end{bmatrix}.
\end{equation}
The model structure may be identical to that in Eqns.\ \eqref{eqn:multivariate:model1}-\eqref{eqn:multivariate:model1b}, but it allows for a joint estimation for all load periods, e.g., via multivariate Least Squares, multivariate Yule-Walker equations (as in this paper) or maximum likelihood \citep{lue:05}. 
Hence, there is an explicit (or implicit) joint distribution assumption on the error vector $\bseps_d = [\eps_{d,1},\ldots, \eps_{d,24}]'$, e.g., that 
$\bseps_d \sim N_{24}(\bsnull, \boldsymbol{\Sigma})$, where $\boldsymbol{\Sigma}$ is a 24-dimensional covariance matrix of a multivariate normal distribution.
Comparing Eqn.\ \eqref{eqn:multivariate:model2} with \eqref{eqn:multivariate:model1b} it is clear that the former is more general and the \emph{set of 24 interrelated models} can be nested in the \emph{fully multivariate} model.
From the statistical point of view, Eqn.\ \eqref{eqn:multivariate:model1b} describes only the marginal distribution of $\bsP_d$, but in an interrelated way.
	
Note, that it is also possible to estimate a \emph{fully multivariate} model using a two-step procedure. First estimating separately the 24 rows of Eqn.\ \eqref{eqn:multivariate:model2}, then estimating the noise terms, e.g., $\bseps_d \sim N_{24}(\bsnull, \boldsymbol{\Sigma})$, using the residuals of the 24 models. However, in such a case the dependencies between the 24 single model specifications are ignored and the estimation procedure may lead to a suboptimal in-sample fit. But not necessarily worse forecasts. In general, as \cite{cha:00} emphasizes, while \emph{fully multivariate} models are usually found to yield a better in-sample fit, their forecasts need not necessarily be better.

Vector AutoRegression (VAR) is the basic modeling structure in the \emph{fully multivariate} context, for sample EPF applications see \cite{hui:huu:mah:07}, \cite{pan:smi:08}, \cite{hal:nie:nie:10} and \cite{he:yu:tan:15}. However, if the number of parameters is very large it may be a good idea to reduce dimensionality of the problem first and consider factor models, as in \cite{g-m:rod:san:12}, \cite{wu:cha:tsu:hou:13}, \cite{mac:wer:15,mac:wer:16} and \cite{rav:bou:dij:15}. In one of the few applications in the computational intelligence EPF literature, \cite{yam:sha:li:04} use a neural network with 24 nodes in the output layer, hence consider a \emph{fully multivariate} approach.

\subsection{The univariate modeling framework}	
\label{ssec:UnivariateFramework}

In the second stream of EPF literature, the day-ahead electricity prices are modeled by a \emph{univariate} time series model for $P_t$:
\begin{equation}\label{eqn:univariate:model}
P_t = f(P_{t-1},P_{t-2},...) + \varepsilon_t,
\end{equation}
where $\varepsilon_t$ is the innovation term at time $t$, and $f(\cdot)$ is some function of the explanatory variables.
Similarly as in Eqns.\ \eqref{eqn:multivariate:model1}-\eqref{eqn:multivariate:model2}, the univariate time series model for the hourly prices $P_t$ makes either an explicit or an implicit assumption on the innovations, e.g., that $\eps_t \sim N(0,\sigma^2)$. However, in contrast to the multivariate modeling framework, predicting the next day's 24 hourly prices now involves computing one- to 24-step ahead forecasts:
\begin{itemize}
	\itemsep0em
	\item either in a \emph{recursive} or \emph{iterative} scheme (as in this paper), where the price forecast for hour $t$ is used as an explanatory variable when forecasting the price for hour $(t+1)$, i.e., $\hat{P}_{t} = f(P_{t-1},P_{t-2},...) \longrightarrow \hat{P}_{t+1} = f(\hat{P}_{t},P_{t-1},...)$,
	
	\item or \emph{directly} by computing 24-step ahead forecasts for each load period: $\hat{P}_{t} = f(P_{t-24},P_{t-25},...)$, $\hat{P}_{t+1} = f(P_{t-23},P_{t-24},...)$, etc., as in \cite{kel:etal:16}; note, that this approach is a special case of the recursive scheme with $\hat{P}_{t}$ not depending on $P_{t-1},\ldots,P_{t-23}$.
\end{itemize}
The drawback of the recursive scheme is that it is sensitive to the accumulation of errors because they propagate forward, while the latter method does not use the most recent information (which may decrease performance for the late night and early morning hours).
%Studies where univariate statistical time series models are used include \cite{nog:con:con:esp:02}, \cite{con:esp:nog:con:03}, \cite{con:con:esp:pla:05}, \cite{zar:can:bha:tho:06}, \cite{par:fle:sch:15} and \cite{zie:ste:hus:15}, while papers where neural networks are put to work include \cite{rod:and:04}, \cite{amj:06}, \cite{pao:07}, \cite{amj:dar:key:10}, \cite{abe:amj:s-h:cat:15}, \cite{kim:15}, \cite{dud:16}, \cite{kel:etal:16} and \cite{raf:nik:kho:17}, among others. 

\subsection{Frameworks vs.\ models}	
\label{ssec:Classification}

We consider two major modeling frameworks -- multivariate and univariate. The \emph{multivariate framework} uses an explicit `day $\times$ hour', matrix-like structure for the 24-dimensional electricity price vector $\bsP_d$, as introduced in Eqns.\ \eqref{eqn:multivariate:model1}-\eqref{eqn:multivariate:model2}, and either explicitly or implicitly assumes that the residual variance is different for each of the 24 (48 or more) load periods, i.e., $\var(\varepsilon_{d,h1})\ne \var(\varepsilon_{d,h2})$. The 24 individual models can be either estimated independently or jointly in a \emph{fully multivariate} framework, see Eqn.\ \eqref{eqn:multivariate:model2}, while prices for all load periods of the next day are predicted at once as one-step (\emph{de facto} one-day) ahead forecasts. 
	
Note, that since the estimation in Eqn.\ \eqref{eqn:multivariate:model1} is conducted independently for each load period and not jointly for all, and there are no interdependencies between the models for individual load periods, many authors would not call such models multivariate. However, sets of estimated independently, but interdependent models, as defined by Eqn.\ \eqref{eqn:multivariate:model1b}, are often regarded as multivariate \citep{cha:00,die:04}. Finally, the framework defined in Eqn.\ \eqref{eqn:multivariate:model2} clearly leads to multivariate models \citep{lue:05}.
	
The \emph{univariate framework} treats prices as one `high-frequency' hourly time series, as in Eqn.\ \eqref{eqn:univariate:model}. The univariate models are estimated jointly for all load periods and either explicitly or implicitly assume that the residual variance, i.e., $\var(\varepsilon_{t})$, is the same across all hours (unless an additional variance model is specified). Forecasting with these models requires either using a recursive scheme and computing a series of 24 one-step ahead forecasts or calculating 24-step ahead predictions for each load period.  

\subsection{Converting univariate to multivariate frameworks and vice versa}	
\label{ssec:Converting}

We should also note, that in general a univariate framework can be converted into a multivariate one and vice versa, similarly as $P_{d,h}$ can be rewritten into $P_{t}=P_{24d+h}$.
However, when doing so the implicit (and partially the explicit) error structure changes. 
For instance, when changing a multivariate framework into a univariate one, an implicit assumption on innovations $\eps_{d,h}$ turns into an implicit assumption on univariate innovations $\eps_t$. Here, the innovation specification gets simplified, since $\eps_{d,h}$ may have a different variance for each hour $h$, but $\eps_t$ has a constant variance for all load periods.
If we have an explicit innovation specification within a multivariate framework, e.g., $\bseps_d \sim N(\bsnull, \boldsymbol{\Sigma})$, then it is preserved on the marginal distribution level. Hence, the resulting innovation specification is $\eps_t \sim N(0, \sigma_t^2)$ with $\sigma^2_t$ being the diagonal elements of $\boldsymbol{\Sigma}$.

On the other hand, when changing a univariate framework to a multivariate one, the implicit assumption that innovations $\eps_t$ follow the same distribution, e.g., $\var(\eps_t) = \sigma^2$, turns to an implicit assumption that the innovations have a different distribution for every hour of the day, e.g., $\var(\eps_{d,h}) =\sigma_h^2$. If we have an explicit distribution assumption, e.g., $\eps_t\sim N(0,\sigma^2)$, it turns into a multivariate distribution assumption, e.g., $\bseps_d \sim N(\bsnull, \boldsymbol{\Sigma})$ 
with a model specific restriction on the diagonal of $\boldsymbol{\Sigma}$. However, this representation is usually not unique as a multivariate distribution cannot be represented uniquely by its marginal distributions. 

In practice, the forecasting impact of changing the model representation for implicit innovation specifications is marginal, since many estimation methods (e.g., OLS) are asymptotically consistent for homoscedastic but also for heteroscedastic innovation specifications. For illustration purposes, in \ref{sec:App:AlternativeRepresentations} we show alternative representations of selected models.

\section{Datasets and data preprocessing}
\label{sec:Data}

\subsection{Datasets}

To conduct a thorough empirical study we consider a total of 12 electricity spot price datasets, see Table \ref{tab_prices_summary}. 
Note, that like \cite{uni:now:wer:16}, we use the terms \emph{spot} and \emph{day-ahead} interchangeably, which is line with the majority of literature on European electricity markets. However, in the U.S., the spot market is another name for the real-time market, while the day-ahead market is usually called the forward market \citep{bur:gra:sch:07,wer:zie:18}.  

Eleven datasets come from six major European power markets, including the European Power Exchange (EPEX SPOT) for power spot trading in Germany, France, Austria, Switzerland and Luxembourg, the Nordic power exchange Nord Pool and OMIE, which manages the Iberian markets (Spain and Portugal).
All eleven concern day-ahead markets with 24 hourly load periods and cover a six year period from 30 July 2010 to 28 July 2016. The last dataset comes from the price track of the Global Energy Forecasting Competition 2014 (GEFCom2014), the largest energy forecasting competition to date \citep{hon:pin:fan:etal:16}, and includes locational marginal prices (LMPs, i.e. zonal prices) at an hourly resolution from 1 January 2011 to 17 December 2013. The exact origin of the data has never been revealed by the organizers but --  given its features -- comes from one of the U.S. markets. 

\begin{table}[tbp]
\caption{Summary table of the considered electricity spot price series, i.e.,  day-ahead prices at hourly resolution. The GEFCom2014 dataset covers a three year period from 1 January 2011 to 17 December 2013, the remaining datasets -- a six year period from 30 July 2010 to 28 July 2016.}
\label{tab_prices_summary}
\centering
\footnotesize
\begin{tabular}{llll}
\toprule
Electricity market and region & Acronym & Unit & Source\\ 
\midrule
BELPEX price for Belgium & BELPEX.BE & EUR/MWh & belpex.be \\ 
EPEX price for Switzerland & EPEX.CH & EUR/MWh & epexspot.com \\ 
EPEX price for Germany and Austria & EPEX.DE+AT & EUR/MWh & epexspot.com \\ 
EPEX price for France & EPEX.FR & EUR/MWh & epexspot.com \\ 
EXAA price for Germany and Austria & EXAA.DE+AT & EUR/MWh & exaa.at\\ 
GEFCom2014 competition data & GEFCOM2014 & USD/MWh & \cite{hon:pin:fan:etal:16} \\ 
Nord Pool price for West Denmark & NP.DK1 & EUR/MWh & nordpoolspot.com \\ 
Nord Pool price for East Denmark & NP.DK2 & EUR/MWh & nordpoolspot.com \\ 
Nord Pool System price & NP.SYS & EUR/MWh & nordpoolspot.com \\ 
OMIE price for Spain  & OMIE.ES & EUR/MWh & omie.es \\ 
OMIE price for Portugal & OMIE.PT & EUR/MWh & omie.es  \\ 
OTE price for the Czech Republic & OTE.CZ & EUR/MWh & ote-cr.cz \\ 
\bottomrule
\end{tabular}
\end{table}

In the empirical analysis we use a 730-day (ca. two-year) rolling calibration window.  
First, all considered models are estimated using data from the initial calibration period (i.e., from 31 July 2010 to 30 July 2012 for the European datasets and from 1 January 2011 to 30 December 2012 for GEFCom2014) and forecasts for all 24 hours of the next day (respectively 31 July 2012 and 31 December 2012) are determined. Then the window is rolled forward by one day, the models are reestimated and forecasts for all 24 hours of the next day are computed. This procedure is repeated until the predictions for the 24 hours of the last day in the out-of-sample test period (respectively 28 July 2016 and 17 December 2013) are made. Note, that for the European datasets we are left with roughly four years (1459 days) of data for out-of-sample testing and for the GEFCom2014 dataset with only 352 days.
Note also, that because of the clock-change issue, we have to do minor adjustments to the data to obtain well defined price processes. We interpolate the missing hour in March and average the doubled hour in October for the European data. The GEFCom2014 data was released clock-change adjusted, however, the used adjustment methodology is not reported in \cite{hon:pin:fan:etal:16}.

\subsection{Variance stabilizing transformation}

It is widely known that many electricity price series exhibit price spikes, mostly positive but in some markets also negative \citep{fan:gam:pro:13,now:rav:tru:wer:14}. For electricity prices with only positive values the logarithmic transform is very popular to reduce spike severity and consequently stabilize the variance. However, for datasets with very close to zero or negative prices the log-transform is not feasible. In such cases, typically no transformation is used. This is reasonable for moderately spiky data like the German/Austrian EPEX prices. But for datasets with extreme spikes, like the French EPEX prices, such a `raw data approach' requires robust estimation algorithms \cite[see e.g.][]{hub:ron:09} or models with embedded spike components \cite[like in][]{wer:09}. However, such robust techniques are not popular in EPF and most studies utilize standard least-squares methods.

\begin{figure}[tbp]
	\centering
	\includegraphics[width=1\textwidth]{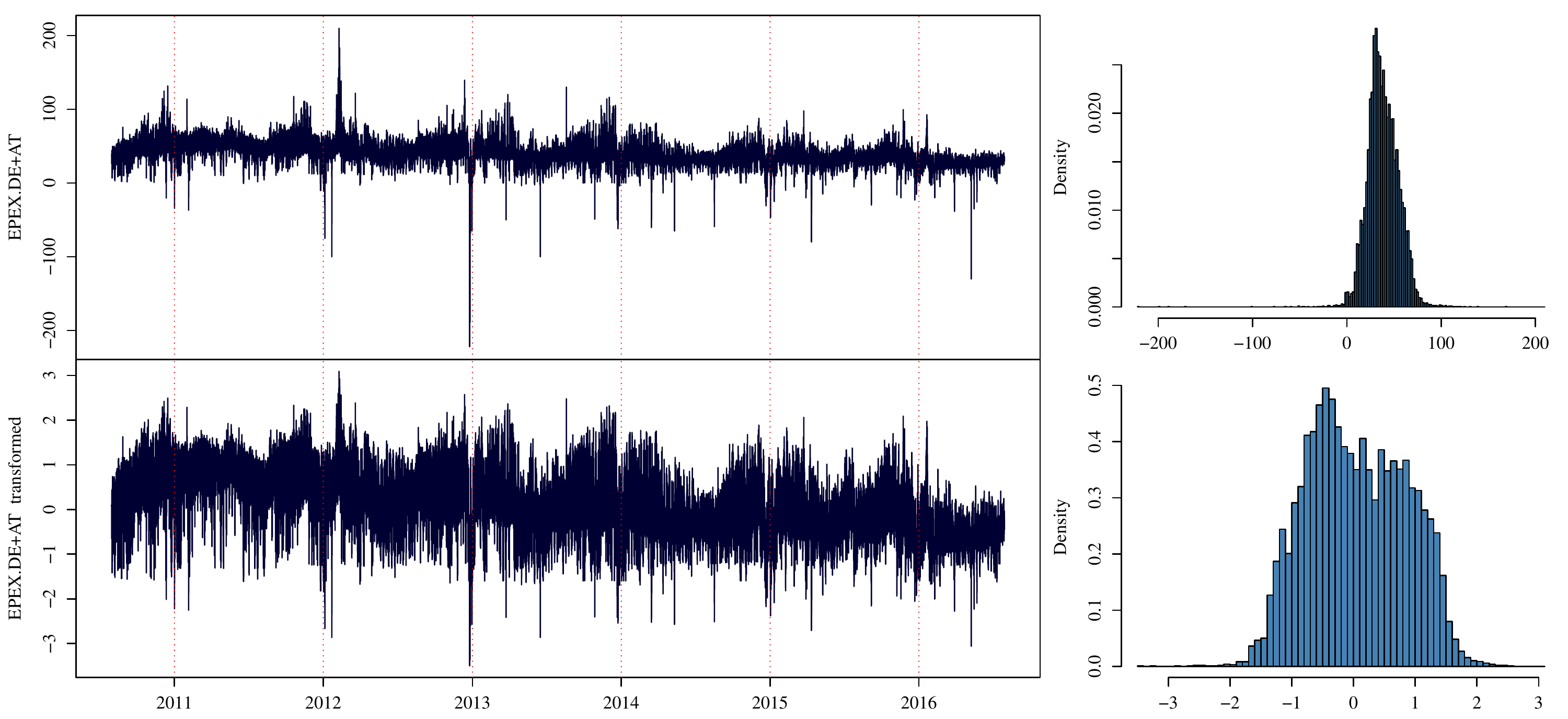} 
	\includegraphics[width=1\textwidth]{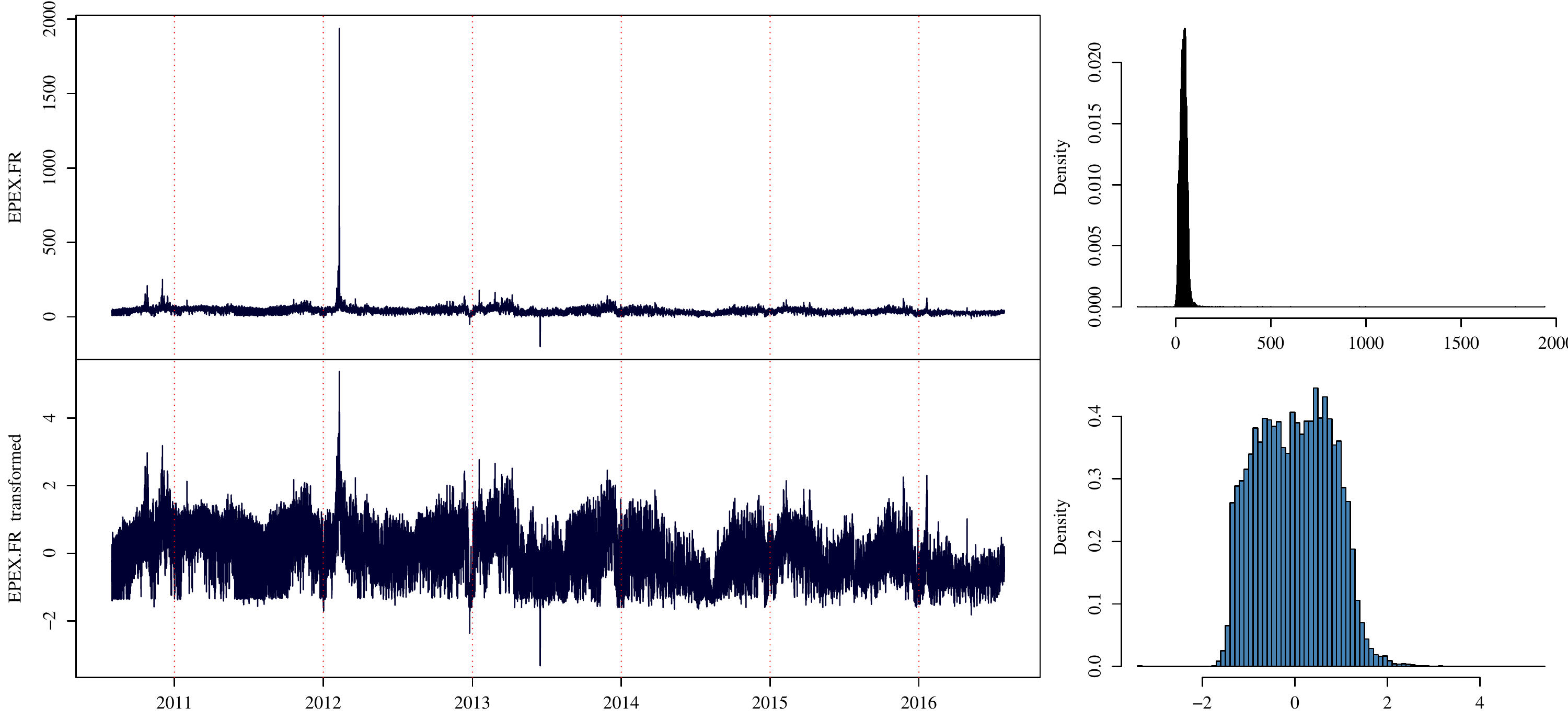} 
	\caption{EPEX spot price $P_{d,h}$ in EUR/MWh for Germany and Austria (EPEX.DE+AT; \emph{top row}) and France (EPEX.FR; \emph{third row}), and the asinh-transformed prices, i.e., $Y_{d,h}$, for both markets (\emph{second} and \emph{bottow rows}). The marginal densities are depicted in the right panels.}
	\label{fig_price_example}
\end{figure}

As we want to conduct a comprehensive forecasting study involving many diverse datasets, we have to deal with this problem in an automated way. The time series  forecasting literature usually suggests the Box-Cox transform \citep{hyn:ath:13}. However, it has the disadvantage of returning a bi-modal marginal distribution of the transformed prices. As {a viable alternative}, we propose the \emph{area} (or \emph{inverse}) \emph{hyperbolic sine} transformation {\cite[see][for a recent review of variance stabilizing transformations]{uni:wer:zie:17}}:
\begin{equation} 
\text{asinh}(x) = \log\left(x + \sqrt{x^2 + 1} \right),
\end{equation}
for standardized spot prices $x=\frac{1}{b}\{P_{d,h}-a\}$. In Figure \ref{fig_price_example} the original and transformed electricity prices are visualized for two series that exhibit positive and negative price spikes. We can see that the spikes become less severe, but do not vanish completely. The area hyperbolic sine transformation {has been originally used in the EPF context} by \cite{sch:11}, but the article went unnoticed. 

In our empirical study in Section \ref{sec:Empirical} {we use the (median, MAD) normalization, i.e.,} we set the shift parameter, $a$, equal to the median of the 730-day calibration sample and the scale parameter, $b$, equal to the sample \emph{median absolute deviation} (MAD) around the sample median adjusted by a factor for asymptotically normal consistency to the standard deviation {\cite[see][]{uni:wer:zie:17}}. This factor is $\frac{1}{z_{0.75}} \approx 1.4826$ where $z_{0.75}$ is the 75\% quantile of the normal distribution; in R this is the default option if one runs \texttt{mad(x)}, in Matlab this corresponds to \texttt{1.4826*mad(x,1)}.
The transform acts so that close to $a$ the transformation is almost linear, whereas positive and negative price spikes are pulled towards the center in a logarithmic way; asymptotically $\text{asinh}(x) \approx \sign(x) \log(2|x|)$ as $|x|\rightarrow\infty$.

In what follows, we denote by $Y_{d,h}$ (or $Y_t$ in the univariate context) the transformed data, i.e. 
\begin{equation}
Y_{d,h} = \text{asinh} \left( \frac{P_{d,h}-a}{b} \right) 
\end{equation}
and calibrate all models to the asinh-transformed prices (except for the naive model defined in Section \ref{ssec:Benchmarks}). Once the forecasts $\hat{Y}_{d,h}$ are computed we apply the inverse transform, i.e., the \emph{hyperbolic sine}, to obtain the day-ahead electricity price forecasts:
\begin{equation}
\hat{P}_{d,h} = b \cdot \text{sinh}\left(\hat{Y}_{d,h}\right) + a
\end{equation}
and use the latter to evaluate and compare the models in Section \ref{sec:Empirical}.

\section{Models}
\label{sec:Models}

Our choice of the forecasting models is guided by the existing literature on short-term EPF and the desire to perform a comprehensive study that addresses the three pertinent questions put forward in the Introduction. As we want to focus on the explanatory power of the past spot prices, we consider `pure price' or `price only' models, i.e., models without exogenous (stochastic) variables, like weather, load or renewable energy generation forecasts.\footnote{Strictly speaking, however, our models include also other non-price variables -- dummies representing calendar effects. Yet, as is common in the EPF literature \citep{wer:14}, we do not treat them as exogenous variables since their nature is deterministic and they can be removed prior to fitting a stochastic model to prices, like in the \textbf{24AR} and \textbf{AR}-type models considered here.}
Overall we consider 58 models from eight classes, see \ref{sec:App:AllModels}. However, for clarity of exposition, in the main body of the text we focus only on 10 models -- one best performing model from each of the eight classes and the second best performer from the two best classes (i.e., C6 and C8): 
\begin{enumerate}
	\itemsep0em
	\item[C1.] the weekly mean of hourly frequency benchmark, which is a simple periodic function $\rightarrow$ denoted by \textbf{mean$_{\text{HoW}}$},
		
	\item[C2.] the so-called \emph{naive} benchmark of \cite{nog:con:con:esp:02}, which belongs to the class of similar-day techniques $\rightarrow$ denoted by \textbf{naive},
	
	\item[C3.] 16 parsimonious autoregressive (AR) models within a multivariate framework, built on some prior knowledge of experts and following \cite{uni:now:wer:16} and \cite{zie:16:TPWRS} called \emph{expert models} or \emph{experts}, estimated using Ordinary Least Squares (OLS) $\rightarrow$ represented by \textbf{expert$_{\text{DoW,nl}}$},
	
	\item[C4.] two AR specifications composed of sets of 24 independent models (for each hour of the day) and estimated using Yule-Walker equations $\rightarrow$ represented by \textbf{24AR$_{\text{HoW}}$},
	
	\item[C5.] two vector autoregressive (VAR) models estimated using multivariate Yule-Walker equations, i.e., the only fully multivariate models in this study $\rightarrow$ represented by \textbf{VAR$_{\text{HoW}}$},
	
	\item[C6.] 16 parameter-rich AR structures within a multivariate framework, estimated using the least absolute shrinkage and selection operator (i.e., lasso or LASSO), which shrinks to zero the coefficients of redundant explanatory variables $\rightarrow$ represented by  \textbf{24lasso$_{\text{DoW,p,nl}}^{\text{HQC}}$} and \textbf{24lasso$_{\text{DoW,nl}}^{\text{HQC}}$},
	
	\item[C7.] four univariate AR models estimated using Yule-Walker equations $\rightarrow$ represented by \textbf{AR$_{\text{HoW}}$},
	
	\item[C8.] 16 univariate parameter-rich AR specifications estimated using the LASSO $\rightarrow$ represented by \textbf{lasso$_{\text{HoW,p}}^{\text{HQC}}$} and \textbf{lasso$_{\text{HoW}}^{\text{HQC}}$}.
\end{enumerate} 
%where the star ($*$) is a wildcard character that represents any set of model attributes.
In Section \ref{ssec:Dummies} we define the seasonal dummies and means used later in the text. Next in Sections \ref{ssec:Benchmarks}-\ref{ssec:UnivariateModels} we describe the 10 representative models, starting with the two benchmarks, then moving on to autoregressive structures considered within a multivariate framework and concluding with univariate specifications. Model definitions for the remaining models can be found in \ref{sec:App:AllModels}, while a summary of results in \ref{sec:App:ModelSelection}.

\subsection{Seasonal dummies and means}
\label{ssec:Dummies}

Before we introduce the forecasting  models let us briefly define three types of dummy variables and the corresponding (time-varying) means. Since the daily and weekly seasonalities are the most pronounced for electricity prices we define:
{
\begin{itemize}
\itemsep0em

\item the \emph{hour-of-the-day dummy} for $k = 1, \ldots, 24$ and arbitrary $d$: 
$$
\text{HoD}^k_{d,h}=\begin{cases}
1 & \mbox{if $h$ is the $k$-th hour of the day,} \\
0 & \mbox{otherwise,} 
\end{cases}
$$

\item the \emph{day-of-the-week dummy} for $k = 1~(\mbox{Monday}), \ldots, 7~(\mbox{Sunday})$ and arbitrary $h$: 
$$
\text{DoW}^k_{d,h}=\begin{cases}
1 & \mbox{if $d$ is the $k$-th day of the week,} \\
0 & \mbox{otherwise,} 
\end{cases}
$$ 

\item the \emph{hour-of-the-week dummy} for $k = 1~(\mbox{Monday, hour 1}), \ldots, 168~(\mbox{Sunday, hour 24})$: 
$$
\text{HoW}^k_{d,h}=\begin{cases}
1 & \mbox{if $24(d-1)+h$ is the $k$-th hour of the week,} \\
0 & \mbox{otherwise.} 
\end{cases}
$$ 

%\item the \emph{hour-of-the-day dummy}: $\text{HoD}^k_{d,h}$ for $k = 1, \ldots, 24$, where $\text{HoD}^k_{d,h}=1$ if $h$ is the $k$-th hour of the day ($d$ is arbitrary) and zero otherwise,
%
%\item the \emph{day-of-the-week dummy}: $\text{DoW}^k_{d,h}$ for $k = 1, \ldots, 7$, where $\text{DoW}^k_{d,h}=1$ if $d$ is the $k$-th day of the week ($k=1$ refers to Monday, $k=2$ to Tuesday, etc.; $h$ is arbitrary) and zero otherwise,
%
%\item the \emph{hour-of-the-week dummy}: $\text{HoW}^k_{d,h}$ for $k = 1, \ldots, 168$, where $\text{HoW}^k_{d,h}=1$ if $(d-1)*24+h$ is the $k$-th hour of the week (we start counting on Monday at hour 1am) and zero otherwise.
	
\end{itemize}
% If the time index, $(d,h)$, matches the corresponding time unit, $k$, then the dummy takes value $1$; for all other $k$'s it takes value $0$. For example, $\text{HoD}^1_{d,h}=1$ if $h=1$ and $0$ otherwise. Without loss of generality, we assume that the week begins on Monday, i.e., $\text{DoW}^1_{d,h}=1$ only if $d$ is a Monday. 
In the univariate context we use the single time index notation ($\text{HoD}^k_t$, $\text{DoW}^k_t$ and $\text{HoW}^k_t$) and sometimes we omit the time index at all. 
}
Note, that it always holds: $\text{HoW}^{j+24(k-1)} = \text{DoW}^{k}\text{HoD}^{j}$.
Note also, that we implicitly assume that the data resolution is hourly, as for the 12 datasets described in Section \ref{sec:Data}. However, all methods considered here can be easily modified to work with half-hourly, 15 minute or even shorter load periods.

\begin{figure}[tb]
	\centering
	\includegraphics[width=1\textwidth]{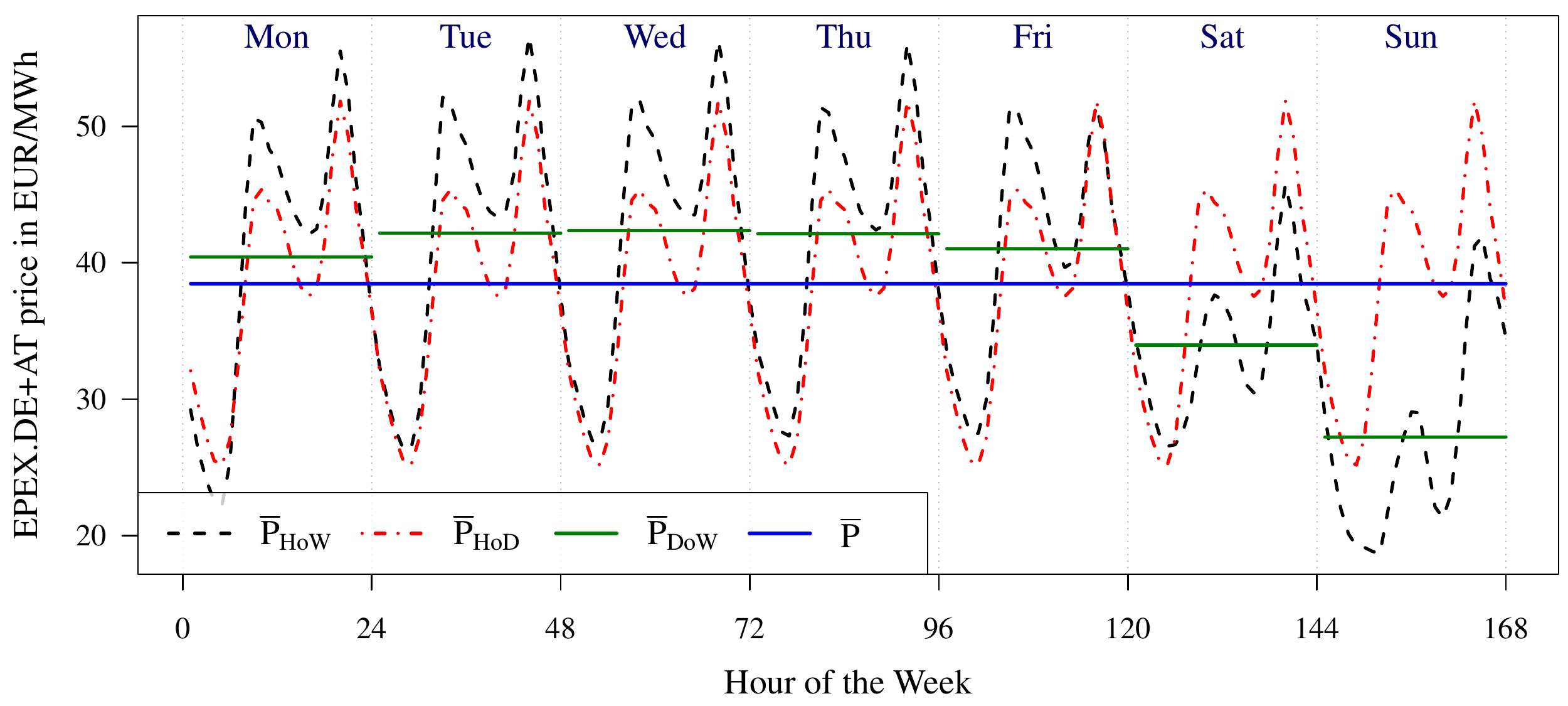} 
	\caption{Illustration of the four means defined in Section \ref{ssec:Dummies}  for the EPEX spot price in Germany and Austria (EPEX.DE+AT), and the whole 6-year sample. Note, that back-transformed mean prices are plotted here, e.g., $\ov{P} = b\cdot\text{sinh}(\ov{Y})+a$. 
	}
	\label{fig_mean_example}
\end{figure}

Given the three dummies -- $\text{HoW}^k$, $\text{HoD}^k$ and $\text{DoW}^k$ -- we  introduce three time-varying means estimated with OLS within a standard linear regression framework:
\begin{itemize}
	\itemsep0em
	{
	\item $\ov{Y}_{\text{HoW},d,h}$ -- the weekly mean of hourly frequency,
	\item $\ov{Y}_{\text{HoD},d,h}$ -- the daily mean of hourly frequency,
	\item $\ov{Y}_{\text{DoW},d,h}$ -- the weekly mean of daily frequency.}
\end{itemize}
In the same manner, we introduce the sample mean of the full sample, $\ov{Y}$. In the empirical study of Section \ref{sec:Empirical} all four means are computed iteratively, for each 730-day calibration window. However, in Figure \ref{fig_mean_example} we illustrate the four means using the whole 6-year sample of the EPEX spot price in Germany and Austria (EPEX.DE+AT).

\subsection{Two simple benchmarks}
\label{ssec:Benchmarks}

%\subsubsection{The `hour of the week' benchmark}
%\label{ssec:MeanModels}

The first benchmark, denoted by \textbf{mean$_{\text{HoW}}$}, is given by the time-varying, i.e., estimated for each calibration window, weekly mean of hourly frequency of 
the asinh-transformed price $Y_{d,h}$:
\begin{equation}\label{eqn:mean:HoW}
Y_{d,h} = \ov{Y}_{\text{HoW},d,h} + \eps_{d,h} = \sum_{k=1}^{168} \beta_k \text{HoW}^k_{d,h}+ \eps_{d,h}.
\end{equation}
Unless stated otherwise, the error terms ($\eps_{d,h}$ or $\eps_t$) in all models considered in this study are assumed to have zero mean, finite variance and are uncorrelated. Note, that we have also considered two other time-varying means defined in Section \ref{ssec:Dummies}, $\ov{Y}_{\text{HoD},d,h}$ and $\ov{Y}_{\text{DoW},d,h}$, but their predictive performance was worse and is not reported in this study.
%\begin{eqnarray}
%Y_{d,h} & = & \ov{Y}_{\text{HoD},d,h} + \eps_{d,h} = \sum_{k=1}^{24} \beta_k \text{HoD}^k_{d,h} + \eps_{d,h} \label{eqn:mean:HoD}\\
%Y_{d,h} & = & \ov{Y}_{\text{HoW},d,h} + \eps_{d,h} = \sum_{k=1}^{168} \beta_k \text{HoW}^k_{d,h}+ \eps_{d,h} \label{eqn:mean:HoW} \\
%Y_{d,h} & = & \ov{Y}_{\text{DoW},d,h} + \eps_{d,h} = \sum_{k=1}^{7} \beta_k \text{DoW}^k_{d,h} + \eps_{d,h},  \label{eqn:mean:DoW}
%\end{eqnarray}

%\subsubsection{The naive benchmark}
%\label{ssec:NaiveModels}

The second benchmark belongs to the class of similar-day techniques \cite[for a taxonomy of EPF approaches see e.g.][]{wer:14}:
\begin{eqnarray}
 P_{d,h} & = & \begin{cases} 
      P_{d-7,h} + \eps_{d,h}, & \text{DoW}^k_{d,h} = 1 \text{ for } k=1,6,7,  \\   
      P_{d-1,h} + \eps_{d,h}, & \ow,  \\   
           \end{cases}  \label{eqn:naive} 
\end{eqnarray}
and denote it by \textbf{naive}. Most likely it has been introduced to the EPF literature by \cite{nog:con:con:esp:02}. It proceeds as follows: the forecast for hour $h$ on Monday, Saturday and Sunday is set equal to the price for the same hour a week ago; the forecast for hour $h$ on the remaining days is set equal to the price for the same hour yesterday. As was argued by \cite{nog:con:con:esp:02} and \cite{con:con:esp:pla:05}, forecasting procedures that are not calibrated carefully fail to outperform the naive method surprisingly often. Note, that we have also considered two simpler similar-day techniques: (i) $P_{d,h} = P_{d-1,h} + \eps_{d,h}$ and (ii) $P_{d,h} = P_{d-7,h} + \eps_{d,h}$, but their predictive performance was worse and is not reported in this study.

\subsection{The multivariate modeling framework}
\label{ssec:MultivariateModels}

Having discussed two simple benchmarks, we are now ready to define models considered in this study 
within the multivariate framework. They all use an explicit `day $\times$ hour', matrix-like structure, 
implicitly assume that the residual variance is different for each load period, and are estimated:
\begin{itemize}
	\item independently for each load period and do not admit interdependencies between the hours of the day, see Eqn.\ \eqref{eqn:multivariate:model1},
	
	\item independently for each load period but admit interdependencies between the hours of the day, see Eqn.\ \eqref{eqn:multivariate:model1b},
	
	\item or jointly in a fully multivariate framework, see Eqn.\ \eqref{eqn:multivariate:model2}.
\end{itemize}

\noindent
In contrast to the univariate models discussed in Section \ref{ssec:UnivariateModels}, here the prices for all load periods of the next day are predicted at once as one-step (i.e, one-day) ahead forecasts. 

We should note, that recently reported empirical evidence provides a fundamental justification of the multivariate framework, as opposed to the univariate. Namely, analyzing the bidding behavior in the Italian power market, \cite{gia:par:pel:16} find that since the solar production suddenly decreases in the evening and the merit order curve rapidly shifts, the thermal and hydro producers are able to exert market power at this time of the day. They further speculate that these generators apply different bidding strategies for different hours, hence significantly change the respective price formation mechanisms.

\subsubsection{An expert model}
\label{sssec:ExpertModels}

This class of models is based on a parsimonious autoregressive structure originally proposed by \cite{mis:tru:wer:06} and later used in a number of EPF studies \citep{wer:mis:08,ser:11,kri:12,now:rav:tru:wer:14,gai:gou:ned:16,mac:now:wer:16,mar:uni:wer:18,now:wer:18,uni:now:wer:16,zie:16:TPWRS}. Since these models are built on some prior knowledge of experts, following \cite{uni:now:wer:16} and \cite{zie:16:TPWRS}, we refer to them as \emph{expert} models. In the empirical comparison in Section \ref{sec:Empirical}, this class is represented by the \textbf{expert$_{\text{DoW,nl}}$} model. Within this autoregressive structure the asinh-transformed price on day $d$ and hour $h$ is given by the following formula:
\begin{align}
Y_{d,h} = & ~~\beta_{h,1} + \underbrace{\beta_{h,2} Y_{d-1,h} + \beta_{h,3} Y_{d-2,h} + \beta_{h,4} Y_{d-7,h}}_{\text{autoregressive effects}} + \underbrace{\beta_{h,5} Y_{d-1,\min} + \beta_{h,6} Y_{d-1,\max}}_{\text{non-linear effects}} +  \beta_{h,7} Y_{d-1,24} \nonumber \\ 
&+ \underbrace{\sum_{j=1}^7 \beta_{h,7+j} \text{DoW}^{j}_{d,h}}_{\text{weekday dummies}}
+ ~\eps_{d,h},
\label{eq_expert-wd:SEL}  
\end{align}
where 
\begin{itemize}
	\itemsep0em
	\item the lagged prices $Y_{d-1,h}$, $Y_{d-2,h}$ and $Y_{d-7,h}$ account for the autoregressive effects of the previous days (the same hour yesterday, two days ago and one week ago);
	\item $Y_{d-1,\min} = \min_{h=1,...,24} \{Y_{d-1,h}\}$ and $Y_{d-1,\max} = \max_{h=1,...,24} \{Y_{d-1,h}\}$ are respectively the minimum and the maximum of the previous day's 24 hourly prices and create a link with all yesterday's prices, not just the prices for the same hour; note, that the minimum and maximum are non-linear due to $\min(x,y) = 0.5(x+y-|x-y|)$ and $\max(x,y) = 0.5(x+y+ |x-y|)$, hence subscript \textbf{nl} in the model name;
	\item $Y_{d-1,24}$ is the price for the last load period of the previous day and is included in \eqref{eq_expert-wd:SEL} to take advantage of the fact that prices for early morning hours depend more on the previous day's price at midnight than on the price for the same hour, as recently emphasized by \cite{mac:now:16} and \cite{zie:16:TPWRS};
	\item and $\text{DoW}^{i}_{d,h}$, $i=1,...,7$ are the daily dummies, hence subscript \textbf{DoW} in the model name.
\end{itemize}
Note, that if $h=24$ then the term which includes $\beta_{h,7}$ is collinear with $\beta_{h,2}$. Hence, in this case, the model has fewer parameters. We estimate the parameters using OLS.

The \textbf{expert$_{\text{DoW,nl}}$} model is inspired by the \textbf{mAR1hm} and \textbf{AR2hm} models of \cite{uni:now:wer:16}, but does not include a dummy for public holidays (for the sake of parsimony) and a term that depends on the average price of the previous day. The impact of the latter, however, has been shown by \cite{uni:now:wer:16} to be negligible, hence the change.

\subsubsection{A set of 24 AR models}
\label{sssec:24AR}

This class of benchmarks is very popular in the EPF literature. In our setup, the demeaned with respect to $\ov{Y}_{\text{HoW},d,h}$ asinh-transformed price is modeled as a standard autoregressive process of order $p_h$, i.e., AR($p_h$), independently for each hour $h$. Formally, the model -- denoted later in the text by \textbf{24AR$_{\text{HoW}}$} -- is given by:
\begin{equation}
Y_{d,h} = \ov{Y}_{\text{HoW},d,h} + \phi_{0,h} + \sum_{k=1}^{p_h} \phi_{k,h}( Y_{d-k,h} - \ov{Y}_{\text{HoW},d,h}) + \eps_{d,h},
\label{eq_24ARwd}
\end{equation}
where $\phi_{k,h}$ are the autoregressive parameters. We estimate the model by solving the Yule-Walker equations with a maximum order $p_{h,\max}=8$, to cover a potential dependency of up to 8 days. Each $p_h$ is chosen based on the Akaike Information Criterion (AIC), see e.g. \cite{hyn:ath:13} or \cite{zie:ste:16}.

In contrast to the expert model defined in Section \ref{sssec:ExpertModels}, the \textbf{24AR$_{\text{HoW}}$} model does not admit any interdependencies between the prices for different load periods. Although it is written in a multivariate framework, it actually is a set 24 independent univariate models at daily frequency, one for each hour, like in Eqn.\ \eqref{eqn:multivariate:model1}. Note also, that this model is a special case of a 24-dimensional Vector AutoRegressive (VAR) model with diagonal parameter matrices, see Section \ref{sssec:VAR} below.

\subsubsection{A VAR model}
\label{sssec:VAR}

Using matrix notation we introduce a Vector AutoRegressive (VAR) structure, denoted later in the text by \textbf{VAR$_{\text{HoW}}$}:
\begin{equation}
\bsY_d = \ov{\bsY}_{\text{HoW},d} + \bsphi_0 + \sum_{k=1}^{p} \bsPhi_k (\bsY_{d-k} - \ov{\bsY}_{\text{HoW},d}) + \bseps_d,
\label{eq_VAR_HoW}
\end{equation}
where $\bsY_d = [Y_{d,1}, \ldots, Y_{d,24}]$ with its mean vector $\ov{\bsY}_{\text{HoW},d}$ across all available days in the calibration sample (which corresponds to the 168 possible values of $\ov{Y}_{\text{HoW},d,h}$, 24 for each of the seven days of the week), $\bsPhi_k$ is a parameter matrix, $\bsphi_0$ is the intercept vector and $\bseps_d = [\eps_{d,1}, \ldots, \eps_{d,24}]$. 
We calibrate the model by solving the multivariate Yule-Walker equations, see e.g. \cite{lue:05}, with $p_{\max}=8$ to cover the same memory as for the model in Section \ref{sssec:24AR}. 
%Here the univariate representation of the latter is given by:
%\begin{equation}
%Y_t = \phi_{0,t} + \sum_{k=1}^{24p} \phi_{k,t} Y_{t-k} + \eps_{t} \ \text{ with } \ \phi_{k,t} = \sum_{j=1}^{24} \phi_{k,j} \text{HoD}^j_t \text{ for } k\geq 1 \text{ and } \phi_{0,t} =  \sum_{j=1}^{168} \phi_{0,j} \text{HoW}^j_t.
%\end{equation} 

\subsubsection{Multivariate lasso models}
\label{sssec:24lasso}

As has been noted in a number of studies, both statistical and computational intelligence, a key point in EPF is the appropriate choice of explanatory variables \citep{amj:key:09,gia:gro:12,gon:m-m:jua:15,kar:bun:08,kel:etal:16,mac:14,vor:par:13,wer:14}. The typical approach has been to select predictors in an ad hoc fashion, sometimes using expert knowledge, seldom based on formal selection or shrinkage procedures for high-dimensional model specifications \cite[like in][]{gai:gou:ned:16,lud:feu:neu:15,uni:now:wer:16,zie:16:TPWRS,zie:ste:hus:15}. 

Recall, that \emph{shrinkage} (also known as \emph{regularization}) fits the full model with all predictors using an algorithm that shrinks coefficients of the less important explanatory variables towards zero \citep{jam:wit:has:tib:13}. Some shrinkage methods, like the \emph{least absolute shrinkage and selection operator} (i.e., \emph{lasso} or \emph{LASSO}) introduced by \cite{tib:96}, may actually shrink some of the coefficients to zero itself, thus \emph{de facto} performing variable selection.
It should be noted, however, that while variable selection is beneficial for interpretability, for reducing the forecasting errors only the shrinkage property is crucial \citep{uni:now:wer:16}.

Let us first introduce a general regression model, somewhat inspired by the \textbf{full ARX} or \textbf{fARX} model of \cite{uni:now:wer:16}, and then define two special cases considered later in Section \ref{sec:Empirical} (the remaining models from this class are defined in \ref{ssec:App:24lasso}). For each hour of the day, $h=1,\ldots,24$, the model is given by:
\begin{align}
Y_{d,h} =
& \underbrace{\sum_{l = 1}^{24} \sum_{k=1}^8 \phi_{h,k,l, 0} Y_{d-k,l}}_{\text{autoregressive effects}} 
+ \underbrace{\sum_{k=1}^8 \phi_{h,k,\min, 0} Y_{d-k,\min}  
+ \sum_{k=1}^8 \phi_{h,k,\max, 0} Y_{d-k,\max}}_{\text{non-linear effects}}  \nonumber \\
&  + \underbrace{\sum_{j=1}^7 \phi_{h,0,0, j} \text{DoW}^{j}_{d,h}}_{\text{weekday dummies}} 
+ \underbrace{\sum_{j=1}^7 \phi_{h,1,h, j} \text{DoW}^{j}_{d,h} Y_{d-1,h} 
+ \sum_{j=1}^7 \phi_{h,1,24,j} \text{DoW}^{j}_{d,h} Y_{d-1,24}}_{\text{periodic effects}}  + \eps_{d,h}.
 \label{eq_model1} 
\end{align}
The first term describes the autoregressive effects up to eight days ago, the second and third terms specify the lagged non-linear effects using the minimum and maximum price of the day (again up to eight days ago), the fourth term describes the day-of-the-week effect and the fifth and sixth terms are the periodic effects (as seen in the periodic expert models, see Section \ref{sssec:ExpertModels}). Note, that the $\phi$'s are indexed by four variables, with the first one indicating the target hour (i.e., $h$). The remaining three refer to the time lag in days ($k=1,\ldots,8$ or 0), the hour of the day ($l=1,\ldots,24$) or an aggregate value for all 24 hours of the day ($l=\min, \max$), and the day of the week ($j=1,\ldots,7$ or 0).

We denote this model by \textbf{24lasso$_{\text{DoW,p,nl}}^{\text{HQC}}$}, because it is embedded in a multivariate framework (one equation for each of the 24 hours), estimated via the lasso (independently for each load period, see below), with day-of-the-week, periodic and non-linear effects. The superscript \textbf{HQC} denotes the Hannan-Quinn Information Criterion used to select the tuning (or regularization) parameter $\lambda$ within the calibration window, see \ref{ssec:App:24lasso} for details.
Based on Eqn.\ \eqref{eq_model1}, we introduce a variant without the periodic parameter terms (i.e., with $\phi_{h,1,h, j}=\phi_{h,1,24,j}=0$) and denote it by \textbf{24lasso$_{\text{DoW,nl}}^{\text{HQC}}$}.

Now, let us comment on the calibration procedure. For this purpose, let us write Eqn.\ \eqref{eq_model1} in a more compact form:
\begin{equation}
 Y_{d,h} = \X_{d,h}' \bsbeta_h + \eps_{d,h},
\end{equation}
where $\X_{d,h}$ is a vector of the regressors and $\bsbeta_h$ is a vector of their coefficients. Given a $D=730$ observation sample $\YY_{h} = [Y_{1,h}, \ldots, Y_{D,h} ]'$, the sample representation is as follows:
\begin{equation}
 \YY_{h} = \XX_{h}' \bsbeta_h + \EE_{h}, 
\end{equation}
where $\XX_{h}' = [\X_{1,h}', \ldots, \X_{D,h}']' $ and $\EE_{h} = [\eps_{1,h}, \ldots, \eps_{D,h} ]'$.
To efficiently estimate the model using the lasso, we require a scaled version of the OLS equation, i.e.:
\begin{equation}
 \wtilde{\YY}_{h} = \wtilde{\XX}_{h}' \wtilde{\bsbeta}_h + \wtilde{\EE}_{h},
\end{equation}
where $\wtilde{\YY}_{h}$ and $\wtilde{\XX}_{h}$ are the scaled versions of $\YY_{h}$ and $\XX_{h}$, respectively, i.e. the $\|\cdot\|_2$-norm of each column is 1.
Then the lasso estimator is given by \citep{has:tib:wai:15}:
\begin{equation}
 \what{\wtilde{\bsbeta}}_{h,\lambda} = \argmin_{\bsbeta} \| \wtilde{\YY}_{h} -  \wtilde{\XX}_{h}' \bsbeta  \|^2_2 + \lambda \|\bsbeta\|_1, 
 \label{eqn:lasso}
\end{equation}
with the tuning (or regularization) parameter $\lambda\geq 0$. For $\lambda=0$ we receive the OLS estimator of $\wtilde{\bsbeta}_h$. 
Given $\what{\wtilde{\bsbeta}}_{h,\lambda}$, by rescaling we can easily compute  $\what{\bsbeta}_h$.

\subsection{The univariate modeling framework}
\label{ssec:UnivariateModels}

Now, let us turn to the \emph{univariate} framework and consider models for which the forecasts are computed recursively -- the price forecast for hour 1 is used as input (i.e., an explanatory variable) when making the prediction for hour 2, etc. We will start with relatively standard AR model, then continue with parameter rich structures estimated via the lasso.

\subsubsection{A univariate AR model}
\label{sssec:AR}

This univariate model, denoted later in the text by \textbf{AR$_{\text{HoW}}$}, is a counterpart of the \textbf{24AR$_{\text{HoW}}$} model defined in Section \ref{sssec:24AR}:
\begin{equation}
Y_t = \ov{Y}_{\text{HoW},t} + \phi_0 + \sum_{k=1}^p \phi_k( Y_{t-k} - \ov{Y}_{\text{HoW},t}) + \eps_t,  \label{eq_ARHoW}
\end{equation}
where $\ov{Y}_{\text{HoW},t}$ is the hour-of-the-week mean in the calibration period, $\phi_k$ are the autoregressive parameters and $p$ is the order of the AR process. Again, we estimate the model by solving the Yule-Walker equations and minimizing the AIC with respect to the maximum considered AR order $p_{\max}=196$. Note, that $p_{\max}^{\text{AR}}=196$ hours corresponds to $p_{\max}^{\text{24AR}}=8$ days, i.e., a potential memory of eight days. 
Interestingly, in \cite{zie:ste:hus:15}, \textbf{AR$_{\text{HoW}}$} served as a very strong benchmark for the EPEX.DE+AT market and outperformed a number of sophisticated model structures.
% We have seen that a full VAR process can be rewritten as $S$-periodic univariate AR. Here, we know that the univariate representation represents a stationary process the same holds for the special VAR model given through equation \eqref{eq_uni_AR_as_mult}.

\subsubsection{Univariate lasso models}
\label{sssec:lasso}

Like \textbf{AR$_{\text{HoW}}$} is a counterpart of \textbf{24AR$_{\text{HoW}}$}, the two  univariate models considered here are similar to the multivariate lasso models discussed in Section \ref{sssec:24lasso}. The \textbf{lasso$_{\text{HoW, p}}^{\text{HQC}}$} model is given by: 
\begin{align}
Y_{t} =
& \sum_{k =1 }^{168} \phi_{0,k} \text{HoW}^k_t  
+ \sum_{k =1 }^{196} \phi_{1,k} Y_{t-k} 
+ \underbrace{ \sum_{k =1 }^{168} \phi_{2,k} \text{HoW}^k_t Y_{t-1} 
+ \sum_{k =1 }^{168} \phi_{3,k} \text{HoW}^k_t Y_{t-24}  }_{\text{periodic effects}} + \eps_{t},
\label{eq_model_uni}
\end{align}
where superscript \textbf{HQC} denotes the information criterion used for the lasso estimation method. A variant without periodic effects  (i.e., with $\phi_{2,k}=\phi_{3,k}=0$) is denoted by \textbf{lasso$_{\text{HoW}}^{\text{HQC}}$}. 
Note, that the \textbf{AR$_{\text{HoW}}$} model defined in Section \ref{sssec:AR} has the same structure as \textbf{lasso$_{\text{HoW}}^{\text{HQC}}$}, but is estimated in a different way.

\section{Empirical results}
\label{sec:Empirical}

\subsection{Performance evaluation in terms of MAE}
\label{ssec:MAE}

For each of the eleven European datasets we have $D=1459$ days (or approximately 4 years) in the out-of-sample test period. For the GEFCom2014 only about 1 year (exactly $D=352$ days) of out-of-sample data is available, however, given its popularity (gained during the competition), data origin (the U.S.) and availability \cite[see the online Appendix of][]{hon:pin:fan:etal:16} we include it for comparative purposes.

As the main evaluation criterion we consider the Mean Absolute Error (MAE) for the full out-of-sample period. It is computed for each model and dataset as:
\begin{equation}
\label{eqn:MAE}
\text{MAE} = \frac{1}{24D} \sum_{d=1}^D \sum_{h=1}^{24} |\what{\eps}_{d,h}|,
\end{equation}
where $\what{\eps}_{d,h}$ denotes the estimated forecasting error for day $d$ and hour $h$. The MAE errors for the 10 models defined in Section \ref{sec:Models} are reported for all 12 datasets in Table \ref{tab:MAE}.\footnote{The MAE errors for all 58 models and four major markets are visualized in Figure \ref{fig:MAE:4} in \ref{sec:App:ModelSelection}.}
We have also analyzed Root Mean Square Errors (RMSE), but the results were qualitatively the same and, hence, are not reported nor analyzed here due to space limitations (but are available from the authors upon request). Only in Table \ref{tab:MAE} we provide for comparison an aggregate measure of fit based on the RMSE -- the m.p.d.f.b. -- as defined in \eqref{eqn:mpdfb} below. Clearly, the relative forecasting performance of the models and their ranking is nearly identical, irrespective of whether MAE or RMSE is used.
 
\begin{table}[tb]
\caption{Mean Absolute Errors (MAE) in EUR/MWh (or USD/MWh for GEFCom2014) for the full out-of-sample period, as defined by Eqn.\ \eqref{eqn:MAE}, for the 10 models defined in Section \ref{sec:Models}} and all 12 datasets. A heat map	is used to indicate better ($\rightarrow$ green) and worse ($\rightarrow$ red) performing models. In the bottom rows we report the mean percentage deviation from the best (m.p.d.f.b.) model, as defined by Eqn.\ \eqref{eqn:mpdfb}, for the MAE and the Root Mean Square Error (RMSE; the individual RMSE's are not reported due to space limitations, but are available from the authors upon request).
\label{tab:MAE}
\centering
\setlength{\tabcolsep}{5pt}
\footnotesize
\begin{tabular}{lrrrrrrrrrr}
	\toprule
	Market 
	& \rotatebox{90}{mean$_{\text{HoW}}$} 
	& \rotatebox{90}{naive} 
	& \rotatebox{90}{ { expert$_{\text{DoW,nl}}$ } } 
	& \rotatebox{90}{24AR$_{\text{HoW}}$} 
	& \rotatebox{90}{VAR$_{\text{HoW}}$}
	& \rotatebox{90}{{ 24lasso$_{\text{DoW,nl}}^{\text{HQC}}$ } } 
	& \rotatebox{90}{24lasso$_{\text{DoW,p,nl}}^{\text{HQC}}$} 
	& \rotatebox{90}{AR$_{\text{HoW}}$} 
	& \rotatebox{90}{lasso$_{\text{HoW}}^{\text{HQC}}$} 
	& \rotatebox{90}{lasso$_{\text{HoW}, \text{p}}^{\text{HQC}}$} \\ 
	\midrule
BELPEX.BE & \cellcolor[rgb]{1,0.2,0.3} 9.693 & \cellcolor[rgb]{1,0.385,0.3} 8.362 & \cellcolor[rgb]{0.821,1,0.3} 6.063 & \cellcolor[rgb]{1,0.966,0.3} 6.293 & \cellcolor[rgb]{1,0.955,0.3} 6.312 & \cellcolor[rgb]{0.537,0.809,0.3} 5.973 & \cellcolor[rgb]{0.2,0.7,0.3} 5.948 & \cellcolor[rgb]{0.747,0.988,0.3} 6.032 & \cellcolor[rgb]{0.377,0.73,0.3} 5.953 & \cellcolor[rgb]{0.434,0.752,0.3} 5.958 \\ 
  EPEX.CH & \cellcolor[rgb]{1,0.2,0.3} 9.909 & \cellcolor[rgb]{1,0.656,0.3} 5.716 & \cellcolor[rgb]{0.766,1,0.3} 4.072 & \cellcolor[rgb]{1,1,0.3} 4.300 & \cellcolor[rgb]{0.858,1,0.3} 4.139 & \cellcolor[rgb]{0.45,0.76,0.3} 3.945 & \cellcolor[rgb]{0.2,0.7,0.3} 3.926 & \cellcolor[rgb]{0.704,0.944,0.3} 4.035 & \cellcolor[rgb]{0.615,0.865,0.3} 3.993 & \cellcolor[rgb]{0.66,0.903,0.3} 4.013 \\ 
  EPEX.DE+AT & \cellcolor[rgb]{1,0.2,0.3} 8.110 & \cellcolor[rgb]{1,0.255,0.3} 7.751 & \cellcolor[rgb]{0.999,1,0.3} 5.225 & \cellcolor[rgb]{1,0.786,0.3} 5.636 & \cellcolor[rgb]{0.952,1,0.3} 5.200 & \cellcolor[rgb]{0.708,0.948,0.3} 5.105 & \cellcolor[rgb]{0.752,0.992,0.3} 5.118 & \cellcolor[rgb]{0.593,0.848,0.3} 5.078 & \cellcolor[rgb]{0.453,0.762,0.3} 5.058 & \cellcolor[rgb]{0.2,0.7,0.3} 5.048 \\ 
  EPEX.FR & \cellcolor[rgb]{1,0.2,0.3} 9.508 & \cellcolor[rgb]{1,0.49,0.3} 7.111 & \cellcolor[rgb]{0.879,1,0.3} 4.998 & \cellcolor[rgb]{1,0.983,0.3} 5.218 & \cellcolor[rgb]{0.926,1,0.3} 5.031 & \cellcolor[rgb]{0.586,0.843,0.3} 4.862 & \cellcolor[rgb]{0.2,0.7,0.3} 4.818 & \cellcolor[rgb]{0.747,0.987,0.3} 4.923 & \cellcolor[rgb]{0.654,0.898,0.3} 4.884 & \cellcolor[rgb]{0.643,0.888,0.3} 4.880 \\ 
  EXAA.DE+AT & \cellcolor[rgb]{1,0.2,0.3} 7.490 & \cellcolor[rgb]{1,0.357,0.3} 6.450 & \cellcolor[rgb]{0.955,1,0.3} 4.289 & \cellcolor[rgb]{1,0.835,0.3} 4.657 & \cellcolor[rgb]{0.845,1,0.3} 4.234 & \cellcolor[rgb]{0.557,0.822,0.3} 4.146 & \cellcolor[rgb]{0.588,0.845,0.3} 4.152 & \cellcolor[rgb]{0.602,0.855,0.3} 4.155 & \cellcolor[rgb]{0.487,0.779,0.3} 4.135 & \cellcolor[rgb]{0.2,0.7,0.3} 4.120 \\ 
  GEFCom2014 & \cellcolor[rgb]{1,0.2,0.3} 15.119 & \cellcolor[rgb]{1,0.55,0.3} 10.170 & \cellcolor[rgb]{0.885,1,0.3}  7.022 & \cellcolor[rgb]{1,0.964,0.3}  7.475 & \cellcolor[rgb]{0.913,1,0.3}  7.057 & \cellcolor[rgb]{0.2,0.7,0.3}  6.691 & \cellcolor[rgb]{0.472,0.771,0.3}  6.724 & \cellcolor[rgb]{0.626,0.874,0.3}  6.792 & \cellcolor[rgb]{0.725,0.964,0.3}  6.861 & \cellcolor[rgb]{0.642,0.888,0.3}  6.802 \\ 
  NP.DK1 & \cellcolor[rgb]{1,0.2,0.3} 8.059 & \cellcolor[rgb]{1,0.319,0.3} 7.340 & \cellcolor[rgb]{0.873,1,0.3} 5.191 & \cellcolor[rgb]{1,0.778,0.3} 5.668 & \cellcolor[rgb]{0.993,1,0.3} 5.248 & \cellcolor[rgb]{0.581,0.839,0.3} 5.106 & \cellcolor[rgb]{0.683,0.924,0.3} 5.128 & \cellcolor[rgb]{0.764,1,0.3} 5.151 & \cellcolor[rgb]{0.422,0.747,0.3} 5.086 & \cellcolor[rgb]{0.2,0.7,0.3} 5.079 \\ 
  NP.DK2 & \cellcolor[rgb]{1,0.2,0.3} 7.960 & \cellcolor[rgb]{1,0.457,0.3} 6.465 & \cellcolor[rgb]{0.893,1,0.3} 4.881 & \cellcolor[rgb]{1,0.878,0.3} 5.184 & \cellcolor[rgb]{1,1,0.3} 4.947 & \cellcolor[rgb]{0.61,0.862,0.3} 4.786 & \cellcolor[rgb]{0.684,0.925,0.3} 4.804 & \cellcolor[rgb]{0.812,1,0.3} 4.846 & \cellcolor[rgb]{0.534,0.807,0.3} 4.772 & \cellcolor[rgb]{0.2,0.7,0.3} 4.751 \\ 
  NP.SYS & \cellcolor[rgb]{1,0.2,0.3} 6.441 & \cellcolor[rgb]{1,0.762,0.3} 2.680 & \cellcolor[rgb]{0.774,1,0.3} 1.806 & \cellcolor[rgb]{1,1,0.3} 1.975 & \cellcolor[rgb]{1,0.999,0.3} 2.062 & \cellcolor[rgb]{0.2,0.7,0.3} 1.686 & \cellcolor[rgb]{0.373,0.729,0.3} 1.692 & \cellcolor[rgb]{0.727,0.967,0.3} 1.783 & \cellcolor[rgb]{0.68,0.922,0.3} 1.763 & \cellcolor[rgb]{0.652,0.896,0.3} 1.752 \\ 
  OMIE.ES & \cellcolor[rgb]{1,0.2,0.3} 12.243 & \cellcolor[rgb]{1,0.557,0.3}  8.529 & \cellcolor[rgb]{0.712,0.951,0.3}  6.134 & \cellcolor[rgb]{1,0.817,0.3}  7.080 & \cellcolor[rgb]{0.886,1,0.3}  6.262 & \cellcolor[rgb]{0.566,0.828,0.3}  6.067 & \cellcolor[rgb]{0.2,0.7,0.3}  6.016 & \cellcolor[rgb]{0.746,0.987,0.3}  6.155 & \cellcolor[rgb]{0.717,0.956,0.3}  6.137 & \cellcolor[rgb]{0.692,0.932,0.3}  6.123 \\ 
  OMIE.PT & \cellcolor[rgb]{1,0.2,0.3} 12.079 & \cellcolor[rgb]{1,0.536,0.3}  8.744 & \cellcolor[rgb]{0.578,0.837,0.3}  6.289 & \cellcolor[rgb]{1,0.83,0.3}  7.189 & \cellcolor[rgb]{0.751,0.991,0.3}  6.370 & \cellcolor[rgb]{0.403,0.74,0.3}  6.248 & \cellcolor[rgb]{0.2,0.7,0.3}  6.237 & \cellcolor[rgb]{0.645,0.89,0.3}  6.315 & \cellcolor[rgb]{0.64,0.886,0.3}  6.313 & \cellcolor[rgb]{0.656,0.9,0.3}  6.320 \\ 
  OTE.CZ & \cellcolor[rgb]{1,0.2,0.3} 7.687 & \cellcolor[rgb]{1,0.331,0.3} 6.835 & \cellcolor[rgb]{0.928,1,0.3} 4.600 & \cellcolor[rgb]{1,0.848,0.3} 4.942 & \cellcolor[rgb]{0.957,1,0.3} 4.615 & \cellcolor[rgb]{0.707,0.947,0.3} 4.512 & \cellcolor[rgb]{0.746,0.986,0.3} 4.524 & \cellcolor[rgb]{0.533,0.807,0.3} 4.473 & \cellcolor[rgb]{0.427,0.749,0.3} 4.460 & \cellcolor[rgb]{0.2,0.7,0.3} 4.452 \\ 
\midrule
  m.p.d.f.b.$_\text{MAE}$ (\%) & \cellcolor[rgb]{1,0.2,0.3} 104.96 & \cellcolor[rgb]{1,0.519,0.3}  47.60 & \cellcolor[rgb]{0.787,1,0.3}   3.36 & \cellcolor[rgb]{1,0.931,0.3}  11.84 & \cellcolor[rgb]{0.942,1,0.3}   5.59 & \cellcolor[rgb]{0.311,0.712,0.3}   0.62 & \cellcolor[rgb]{0.2,0.7,0.3}   0.58 & \cellcolor[rgb]{0.635,0.882,0.3}   1.90 & \cellcolor[rgb]{0.532,0.806,0.3}   1.25 & \cellcolor[rgb]{0.487,0.779,0.3}   1.04 \\ 
 m.p.d.f.b.$_\text{RMSE}$ (\%) & \cellcolor[rgb]{1,0.2,0.3} 77.24 & \cellcolor[rgb]{1,0.402,0.3} 47.93 & \cellcolor[rgb]{0.78,1,0.3}  2.61 & \cellcolor[rgb]{1,0.912,0.3}  9.57 & \cellcolor[rgb]{0.961,1,0.3}  4.53 & \cellcolor[rgb]{0.2,0.7,0.3}  0.62 & \cellcolor[rgb]{0.419,0.746,0.3}  0.79 & \cellcolor[rgb]{0.702,0.942,0.3}  2.00 & \cellcolor[rgb]{0.693,0.934,0.3}  1.95 & \cellcolor[rgb]{0.678,0.919,0.3}  1.84 \\ 
 \bottomrule
\end{tabular}
\end{table}

In Table \ref{tab:MAE} we can clearly see the dominance of the lasso models over the competitors. However, there is no single lasso model that is the best for all datasets. The multivariate  \textbf{24lasso$_{\text{DoW,nl}}^{\text{HQC}}$} and \textbf{24lasso$_{\text{DoW,p,nl}}^{\text{HQC}}$} models are the best for seven datasets: BELPEX.BE, EPEX.CH, EPEX.FR, GEFCom2014, NP.SYS, OMIE.ES and OMIE.PT, while the univariate \textbf{lasso$_{\text{DoW,p}}^{\text{HQC}}$} model is the best for the remaining five datasets: EPEX.DE+AT, EXAA.DE+AT, NP.DK1, NP.DK2 and OTE.CZ. 

Given the full set of results for all 58 models (see \ref{sec:App:AllModels}) and all 12 datasets it is hard to rank the models. In particular, it is not possible to make conclusive statements about the outperformance of the univariate modeling framework by the multivariate one or vice versa. To tackle this issue we introduce the \emph{mean percentage deviation from the best} (m.p.d.f.b.) model, which is inspired by the m.d.f.b.\ measure used by \cite{wer:mis:08} and \cite{now:rav:tru:wer:14}, among others. The m.p.d.f.b.\ measure for model $i$ indicates how similar is this model's performance to the `optimal model' composed of the best performing model for each of the 12 datasets:
\begin{equation}
\label{eqn:mpdfb}
 \text{m.p.d.f.b.}_{\text{ERR}} = \frac{1}{12} \sum_{j=1}^{12}  \frac{\left|\text{ERR}_{i,j} - \text{ERR}_{\text{best model},j} \right|}{\text{ERR}_{\text{best model},j}} \times 100\%,
\end{equation}
where $\text{ERR}_{\text{best model},j} = \min_{1\le i \le 58} \text{ERR}_{i,j}$ and ERR can be the MAE, as defined in Eqn.\ \eqref{eqn:MAE}, the Root Mean Square Error (RMSE) or any other error measure for point forecasts. 

The m.p.d.f.b.\ measures are reported in the bottom rows of Table \ref{tab:MAE}.\footnote{The m.p.d.f.b.\ measure for all 58 models is plotted in Fig.\ \ref{fig:mpdfb} in \ref{sec:App:ModelSelection}.}
Somewhat surprisingly, we find that the relatively simple, univariate \textbf{AR$_{\text{HoW}}$} model with a m.p.d.f.b.$_\text{MAE}$ of 1.90\% is only slightly worse than the more sophisticated lasso structures. For two markets (EPEX.DE+AT and OTE.CZ) it even beats the overall best performing \textbf{24lasso$_{\text{DoW,p,nl}}^{\text{HQC}}$} model. This confirms the findings of \cite{zie:ste:hus:15} who found this model to be a very strong benchmark for the EPEX.DE+AT market, that outperformed a number of sophisticated model structures.

The next in terms of m.p.d.f.b.\ are the { \textbf{expert}$_{\text{DoW,nl}}$ and \textbf{expert}$_{\text{DoW,p,nl}}$} models with nearly identical performance (on average, but not across all datasets). They are extremely competitive for the Iberian markets. In particular, \textbf{expert}$_{\text{DoW,p,nl}}$ is second best in terms of MAE (only after \textbf{24lasso$_{\text{DoW,p,nl}}^{\text{HQC}}$} and just ahead of \textbf{24lasso$_{\text{DoW,nl}}^{\text{HQC}}$}) for the OMIE.PT dataset. 

The performance of the AR and VAR-type structures considered within a multivariate framework is rather disappointing. The \textbf{24AR}$_{\text{HoW}}$ model is worse than all expert specifications and \textbf{VAR}$_{\text{HoW}}$ is not much better (see \ref{sec:App:ModelSelection} for details). %The univariate AR structures do not fare much better, except for the already mentioned \textbf{AR$_{\text{HoW}}$} model. 
At the very end, which is not surprising, are the simple benchmarks --  \textbf{mean$_{\text{HoW}}$} and \textbf{naive} are significantly (see also the discussion in Section \ref{ssec:DM}) outperformed by the more sophisticated models across all hours, seasons of the year and markets. %Generally, the \textbf{naive} model defined by Eqn.\  \eqref{eqn:naive} is the best out of the six benchmarks, but for six datasets (GEFCom2014 and the Nordic and Iberian markets) it is beaten by the \textbf{naive$_{\text{D}}$} model.

\begin{table}[tbp]
	\caption{Mean percentage deviation from the best model for the MAE errors, i.e., m.p.d.f.b.$_\text{MAE}$ (in \%), in the four seasons of the year: Spring (March, April, May), Summer (June, July, August), Fall (September, October, November) and Winter (December, January, February), for the same 10 representative models as in Table \ref{tab:MAE}. The values correspond to all 12 datasets and the full out-of-sample period. A heat map is used to indicate better ($\rightarrow$ green) and worse ($\rightarrow$ red) performing models.
	}
	\label{tab:mpdfb:season}
	\centering
	\setlength{\tabcolsep}{5pt}
	\footnotesize
	\begin{tabular}{lrrrrrrrrrr}
		\toprule
		Season 
		& \rotatebox{90}{mean$_{\text{HoW}}$} 
		& \rotatebox{90}{naive} 
		& \rotatebox{90}{ { expert$_{\text{DoW,nl}}$ } } 
		& \rotatebox{90}{24AR$_{\text{HoW}}$} 
		& \rotatebox{90}{VAR$_{\text{HoW}}$}
		& \rotatebox{90}{ { 24lasso$_{\text{DoW,nl}}^{\text{HQC}}$ }} 
		& \rotatebox{90}{24lasso$_{\text{DoW,p,nl}}^{\text{HQC}}$} 
		& \rotatebox{90}{AR$_{\text{HoW}}$} 
		& \rotatebox{90}{lasso$_{\text{HoW}}^{\text{HQC}}$} 
		& \rotatebox{90}{lasso$_{\text{HoW}, \text{p}}^{\text{HQC}}$} \\ 
		\midrule
		Spring & \cellcolor[rgb]{1,0.2,0.3} 121.21 & \cellcolor[rgb]{1,0.597,0.3} 44.35 & \cellcolor[rgb]{0.837,1,0.3} 5.34 & \cellcolor[rgb]{1,0.974,0.3} 12.09 & \cellcolor[rgb]{0.989,1,0.3} 8.12 & \cellcolor[rgb]{0.473,0.772,0.3} 1.89 & \cellcolor[rgb]{0.2,0.7,0.3} 1.42 & \cellcolor[rgb]{0.733,0.973,0.3} 3.93 & \cellcolor[rgb]{0.565,0.828,0.3} 2.39 & \cellcolor[rgb]{0.475,0.773,0.3} 1.90 \\
		Summer & \cellcolor[rgb]{1,0.2,0.3} 144.52 & \cellcolor[rgb]{1,0.621,0.3} 48.77 & \cellcolor[rgb]{0.722,0.961,0.3} 3.67 & \cellcolor[rgb]{1,1,0.3} 11.09 & \cellcolor[rgb]{0.952,1,0.3} 7.94 & \cellcolor[rgb]{0.2,0.7,0.3} 0.82 & \cellcolor[rgb]{0.449,0.759,0.3} 1.26 & \cellcolor[rgb]{0.65,0.895,0.3} 2.80 & \cellcolor[rgb]{0.506,0.79,0.3} 1.57 & \cellcolor[rgb]{0.58,0.839,0.3} 2.11 \\
		Fall & \cellcolor[rgb]{1,0.2,0.3} 74.78 & \cellcolor[rgb]{1,0.378,0.3} 49.40 & \cellcolor[rgb]{0.855,1,0.3} 3.38 & \cellcolor[rgb]{1,0.886,0.3} 10.41 & \cellcolor[rgb]{1,1,0.3} 5.73 & \cellcolor[rgb]{0.59,0.846,0.3} 1.49 & \cellcolor[rgb]{0.565,0.828,0.3} 1.38 & \cellcolor[rgb]{0.2,0.7,0.3} 0.78 & \cellcolor[rgb]{0.464,0.767,0.3} 1.04 & \cellcolor[rgb]{0.308,0.711,0.3} 0.80 \\
		Winter & \cellcolor[rgb]{1,0.2,0.3} 92.08 & \cellcolor[rgb]{1,0.45,0.3} 50.65 & \cellcolor[rgb]{0.776,1,0.3} 3.52 & \cellcolor[rgb]{1,0.845,0.3} 15.19 & \cellcolor[rgb]{0.841,1,0.3} 4.23 & \cellcolor[rgb]{0.2,0.7,0.3} 1.21 & \cellcolor[rgb]{0.24,0.702,0.3} 1.21 & \cellcolor[rgb]{0.681,0.923,0.3} 2.69 & \cellcolor[rgb]{0.63,0.877,0.3} 2.32 & \cellcolor[rgb]{0.547,0.816,0.3} 1.86 \\
		\bottomrule
	\end{tabular}
\end{table}

Finally, let us look at model performance in the four seasons of the year: Spring (March, April, May), Summer (June, July, August), Fall (September, October, November) and Winter (December, January, February). Of course, there is some variability in forecasting accuracy across the seasons. For instance, in the Summer, {\textbf{expert$_\text{DoW,nl}$}} is the best performing model for both Iberian datasets, but {\textbf{lasso$_{\text{HoW}}^{\text{HQC}}$}} and \textbf{24lasso$_{\text{DoW,nl}}^{\text{HQC}}$} follow closely by. In the Fall, \textbf{AR$_\text{HoW}$} is the best for EPEX.DE+AT and -- somewhat surprisingly -- the best overall in terms of m.p.d.f.b.$_\text{MAE}$ (see Table \ref{tab:mpdfb:season}), but the univariate {\textbf{lasso$_{\text{HoW,p}}^{\text{HQC}}$}} is right next to it (being the best for EPEX.CH and NP.DK2). In general, looking at the aggregate  m.p.d.f.b.$_\text{MAE}$ values in Table \ref{tab:mpdfb:season}, we can observe that the annual ranking of the models (see Table \ref{tab:MAE}) is preserved in the 
Spring, Summer and Winter. However, in the Fall the univariate models have an edge over the multivariate lasso models. A plausible explanation may be that in the majority of analyzed markets the electricity prices and their volatility tend to increase towards the end of the calendar year. The univariate models, by taking into account all hourly prices in the past week, seem to be able to adapt quicker to the increasing prices.

\subsection{Testing for significant differences in the forecasting performance}
\label{ssec:DM}

The MAE values analyzed in Section \ref{ssec:MAE} can be used to provide a ranking of models. However, the power to draw statistically significant conclusions on the outperformance of the forecasts of one model by those of another is limited -- even given their standard errors -- since the dependency structure between the errors (or the MAE's) is neglected. Therefore, we also computed the \cite{die:mar:95} test (abbreviated DM) which takes the correlation structure into account. It tests forecasts of each pair of models against each other.

\subsubsection{The `multivariate' DM test}

In the EPF literature, the DM test is usually performed separately for each of the 24 hours of the day \cite[see][]{wer:14}. We also do this here. But first let us introduce a different approach, where only one statistic for each pair of models is computed based on the 24-dimensional vector of errors for each day; we call the resulting DM test \emph{multivariate} or \emph{vectorized}. 
Therefore, denote by $\what{\bseps}_{X,d} = [\what{\eps}_{X,d,1}, \ldots, \what{\eps}_{X,d,24}]'$ and $\what{\bseps}_{Y,d} = [\what{\eps}_{Y,d,1}, \ldots, \what{\eps}_{Y,d,24}]'$ the vectors of out-of-sample errors for day $d$ of models $X$ and $Y$, respectively. Then the multivariate loss differential series:
\begin{equation}
\label{eqn:Delta:DM}
\Delta_{X,Y,d} = \|\what{\bseps}_{X,d}\|_1 - \|\what{\bseps}_{Y,d}\|_1,
\end{equation}
defines the differences of errors in the $\|\cdot\|_1$-norm, i.e., $\|\what{\bseps}_{X,d}\|_1 = \sum_{i=1}^{24} |\what{\eps}_{X,d,h}|$.
For each model pair and each dataset we compute the $p$-value of two one-sided DM tests: (i) a test with the null hypothesis $H_0: E(\Delta_{X,Y,d}) \leq 0$, i.e., the outperformance of the forecasts of model $Y$ by those of model $X$, and (ii) the complementary test with the reverse null $H^R_0: E(\Delta_{X,Y,d}) \geq 0$, i.e., the outperformance of the forecasts of model $X$ by those of model $Y$. As in the standard DM test, we assume that the loss differential series is covariance stationary.

\begin{figure*}[p]
\centering
 \includegraphics[width=.32\textwidth, height=.22\textheight]{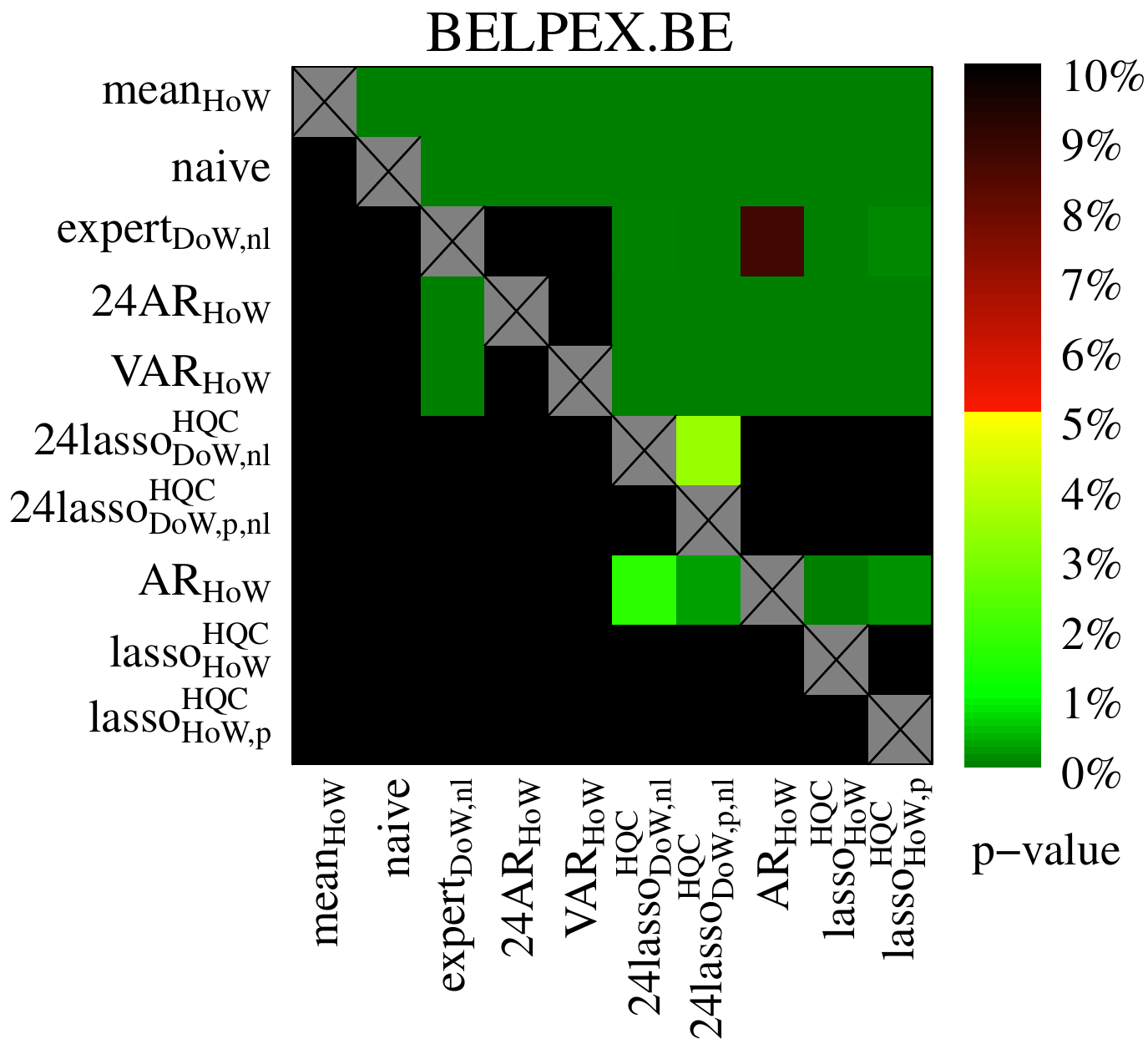} 
 \includegraphics[width=.32\textwidth, height=.22\textheight]{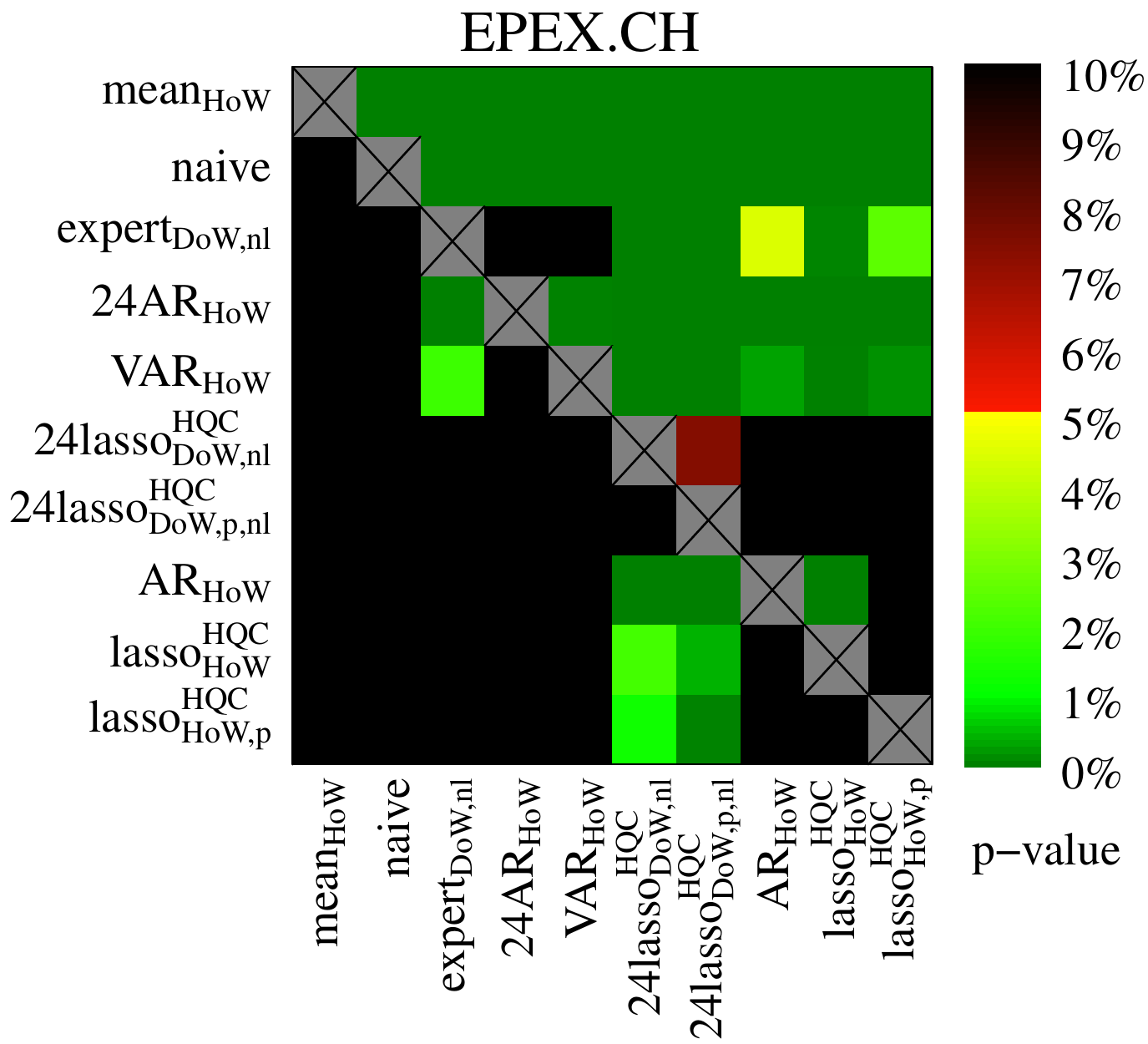} 
  \includegraphics[width=.32\textwidth, height=.22\textheight]{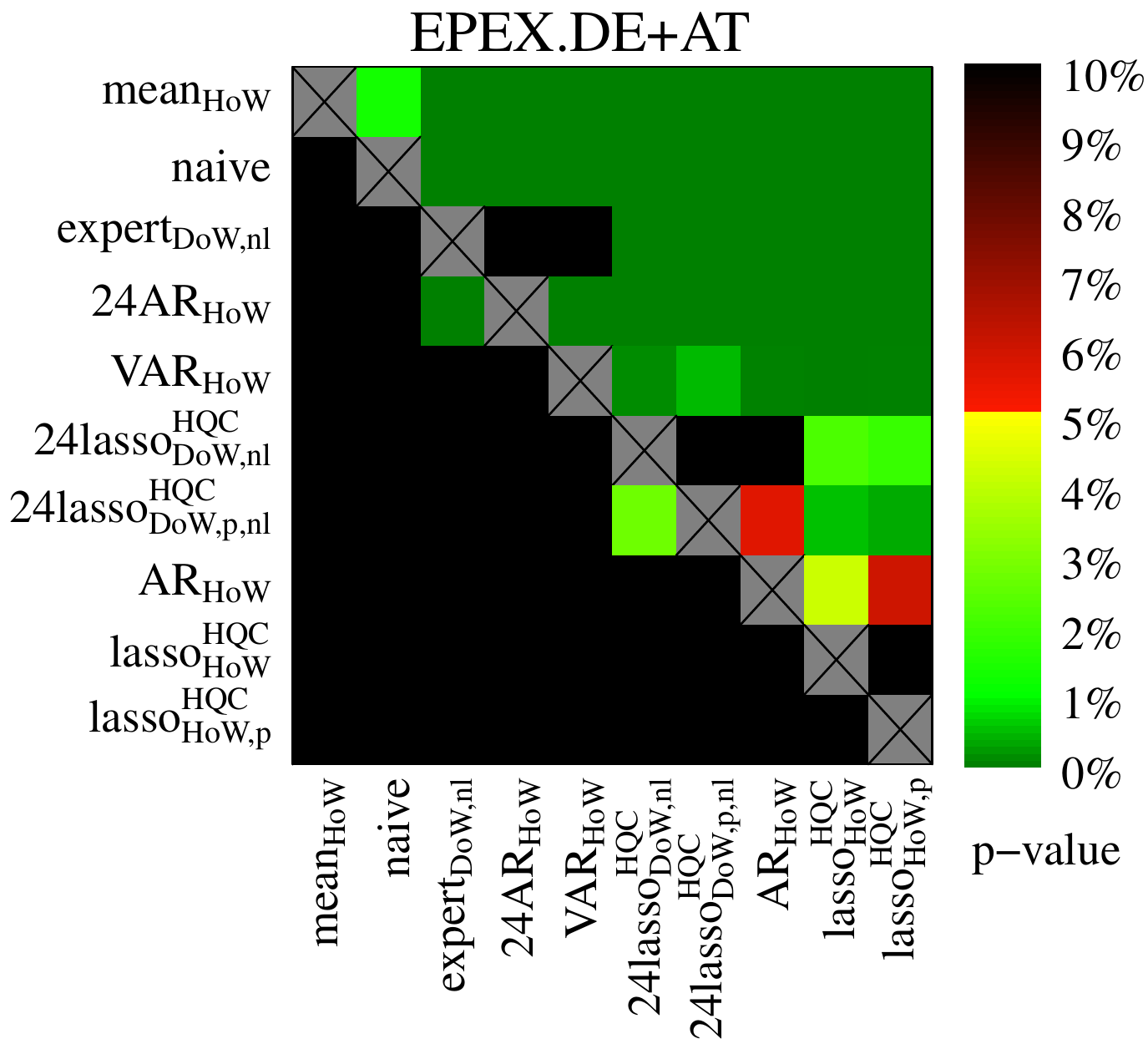} 
  \includegraphics[width=.32\textwidth, height=.22\textheight]{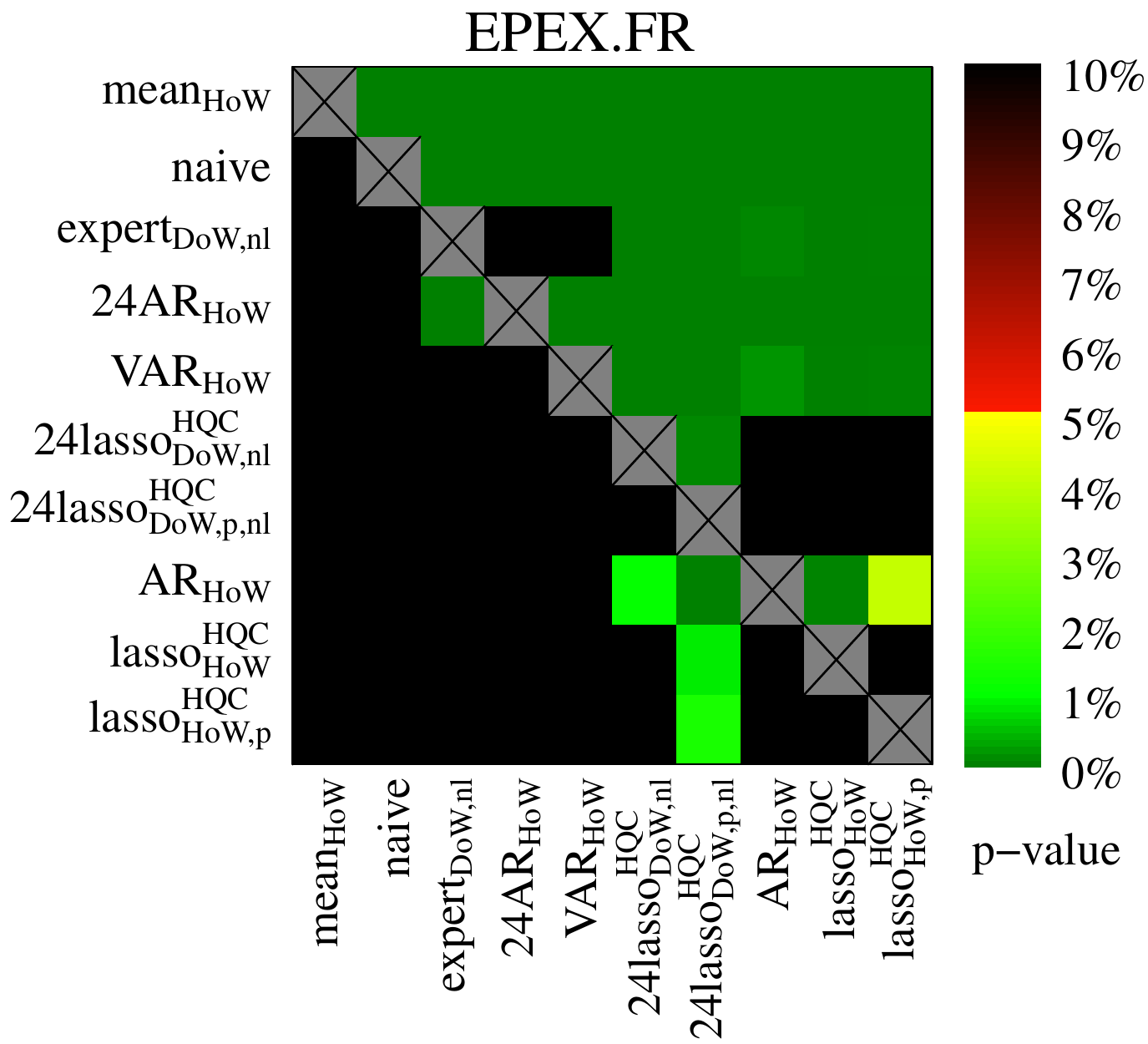} 
  \includegraphics[width=.32\textwidth, height=.22\textheight]{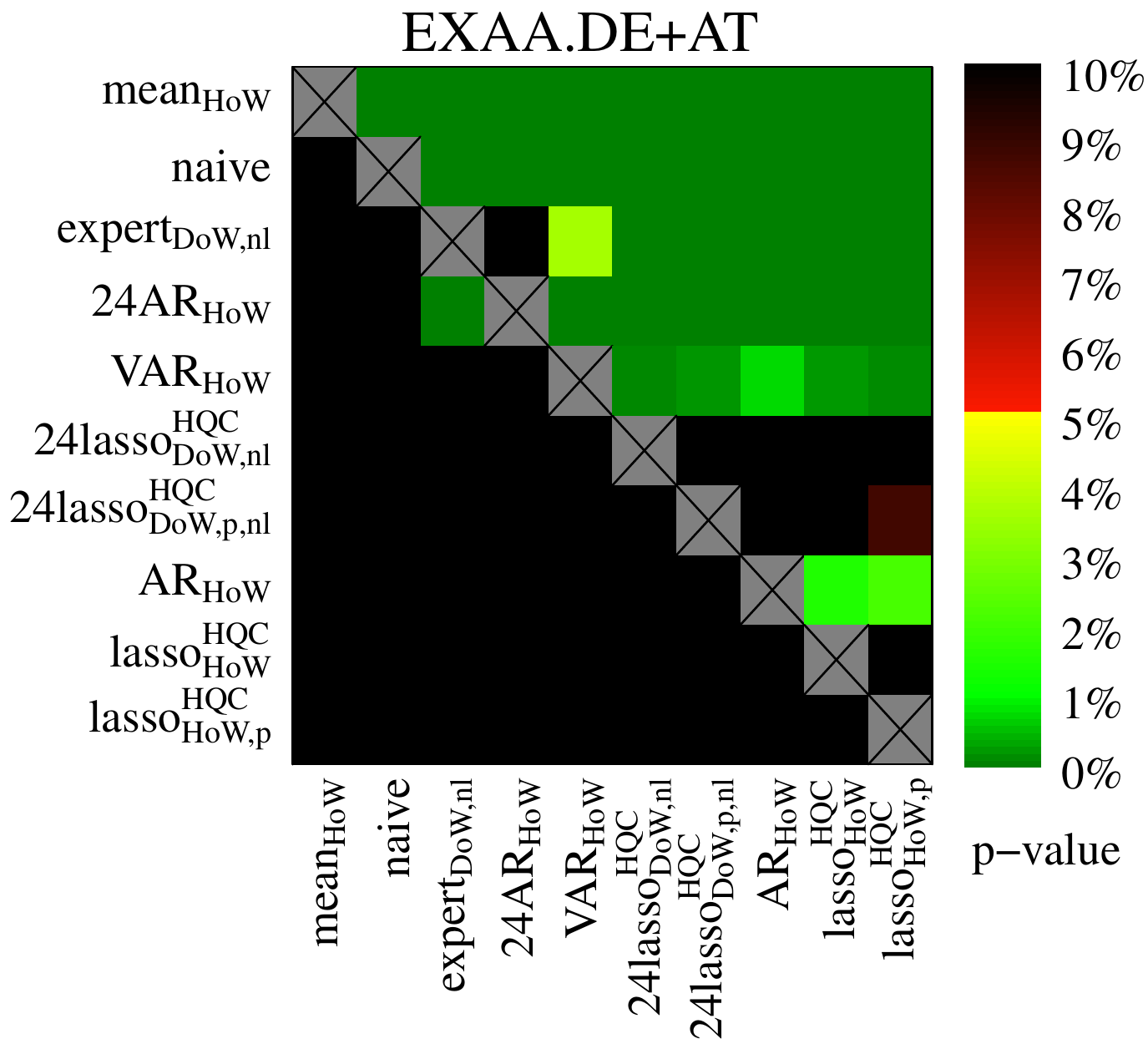} 
  \includegraphics[width=.32\textwidth, height=.22\textheight]{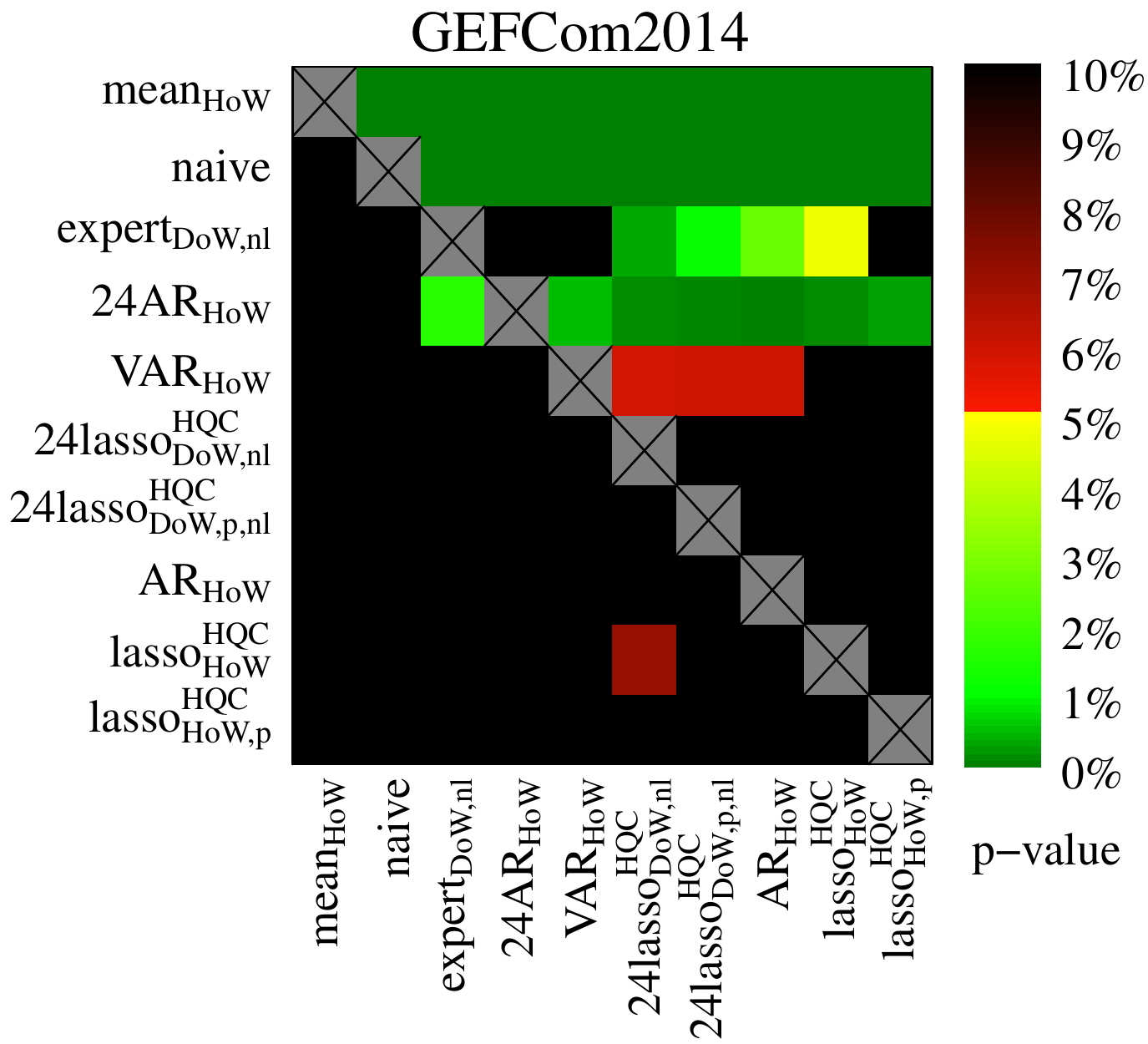} 
  \includegraphics[width=.32\textwidth, height=.22\textheight]{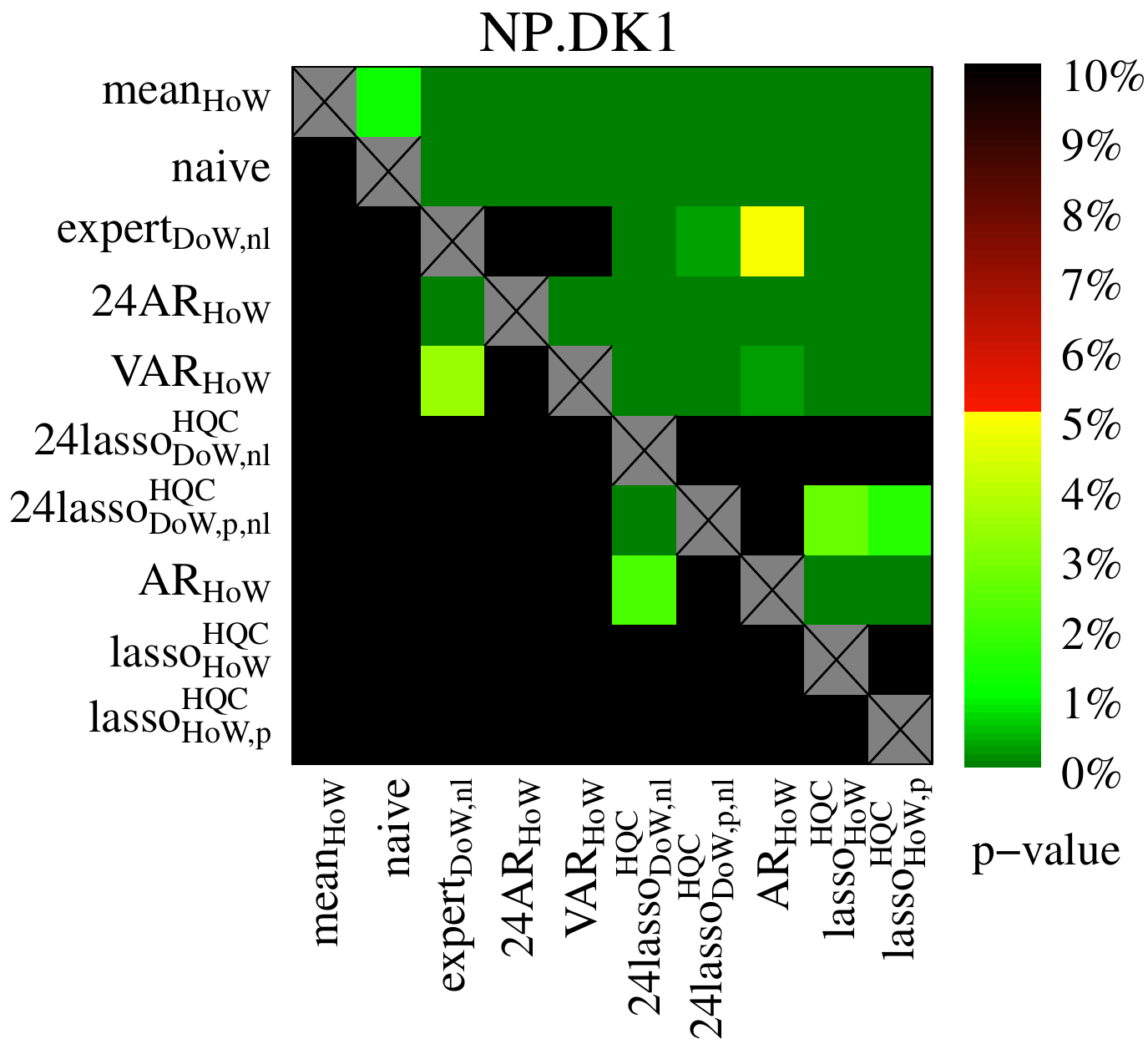} 
  \includegraphics[width=.32\textwidth, height=.22\textheight]{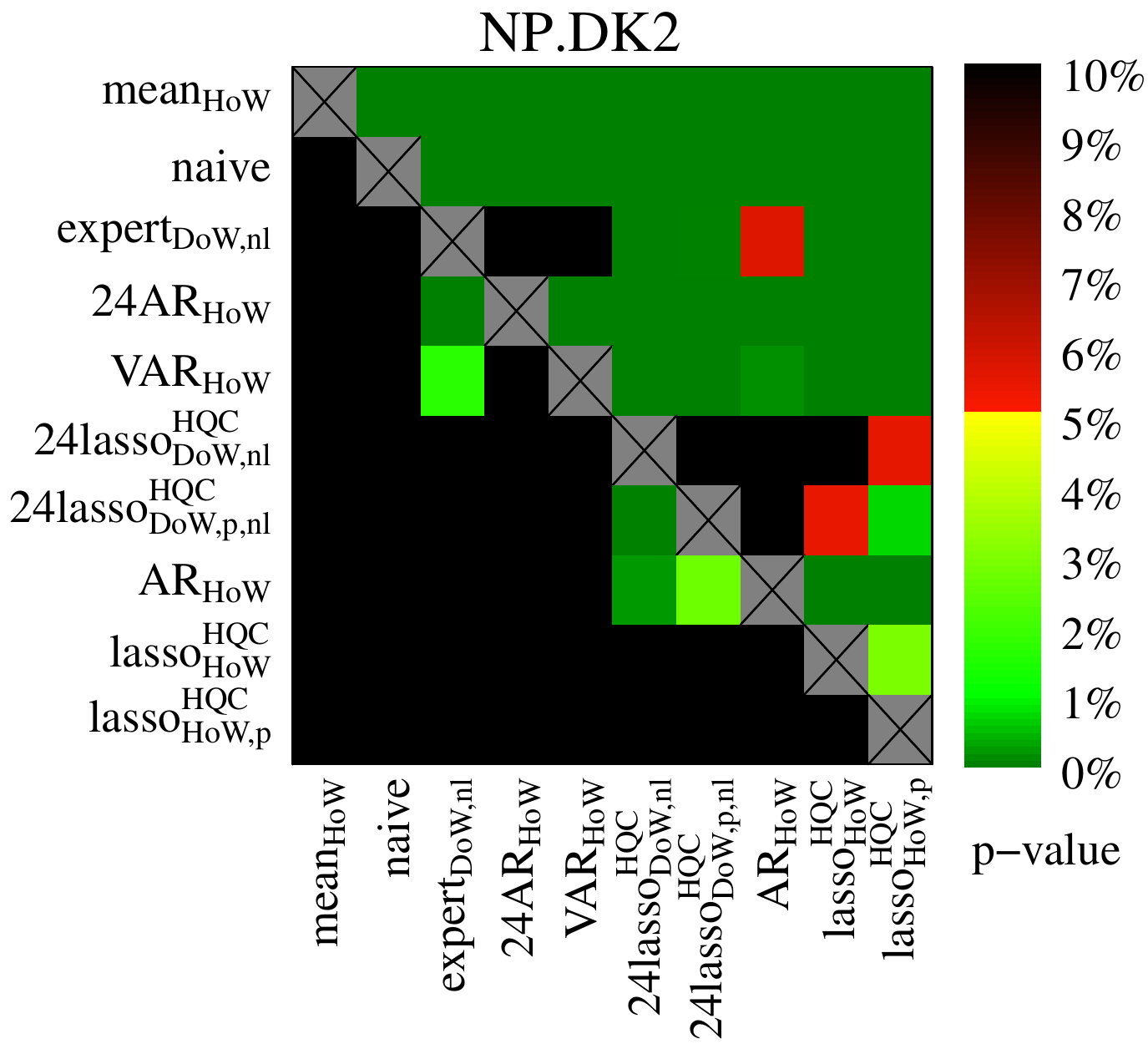} 
  \includegraphics[width=.32\textwidth, height=.22\textheight]{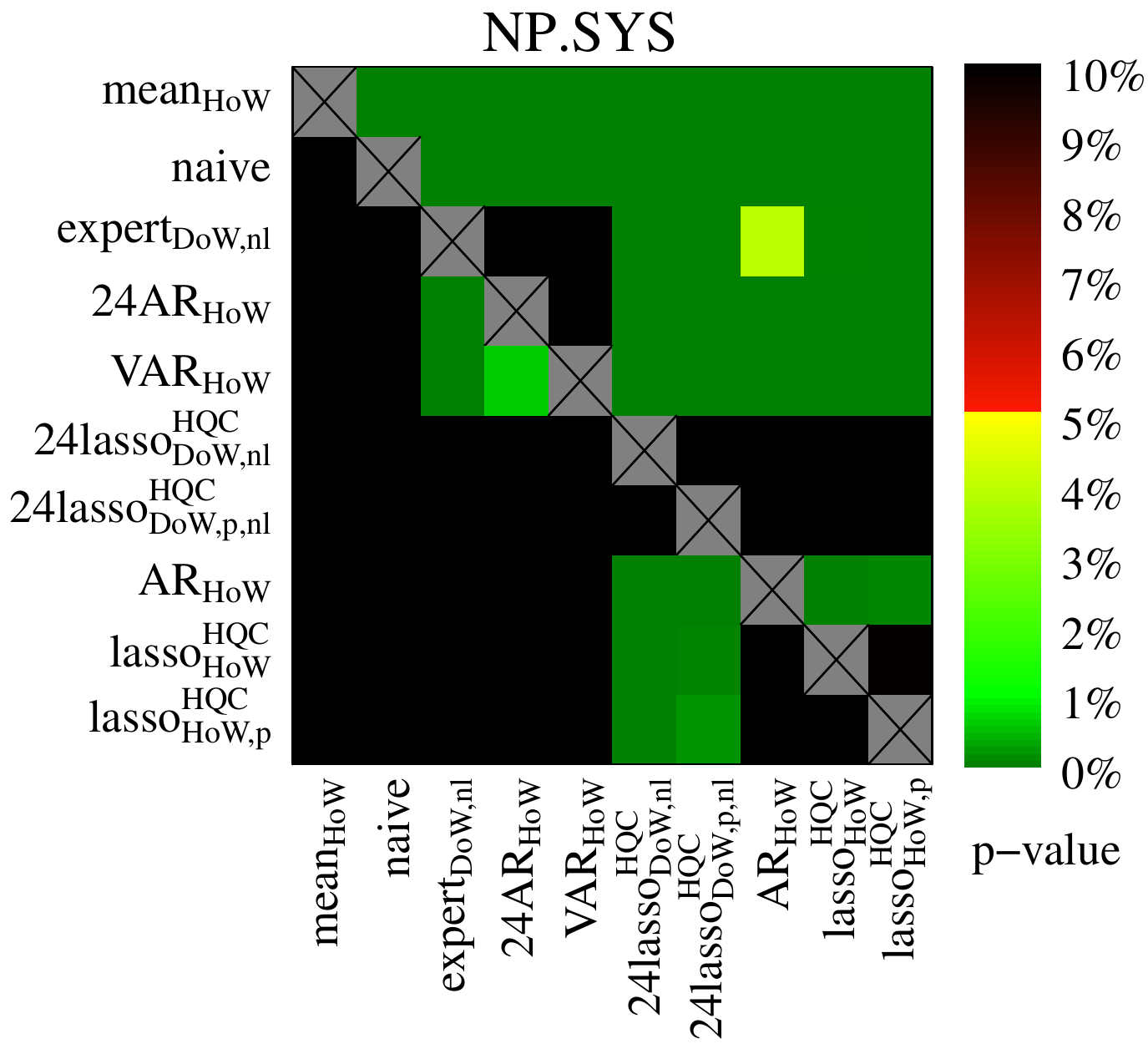} 
  \includegraphics[width=.32\textwidth, height=.22\textheight]{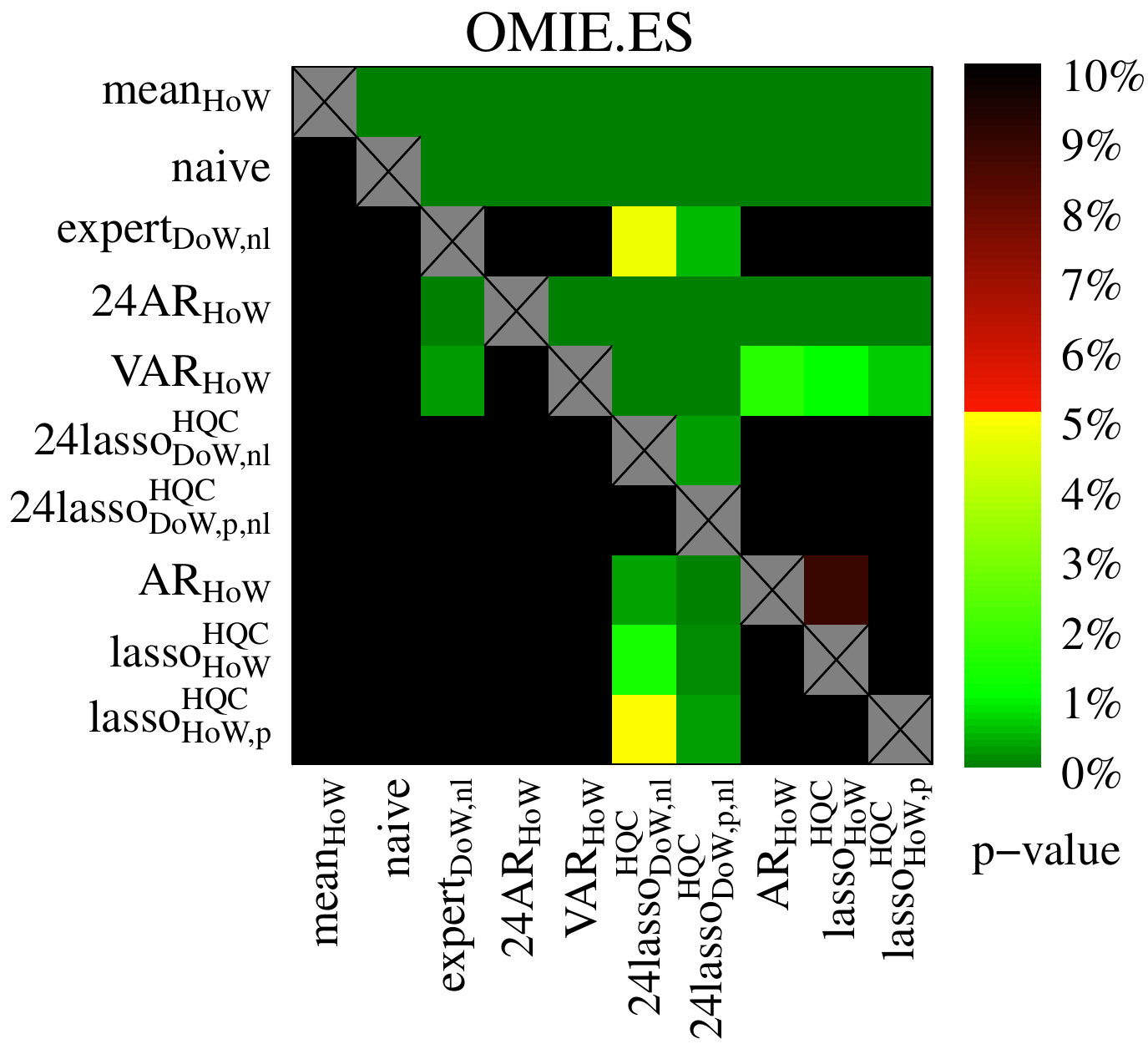} 
  \includegraphics[width=.32\textwidth, height=.22\textheight]{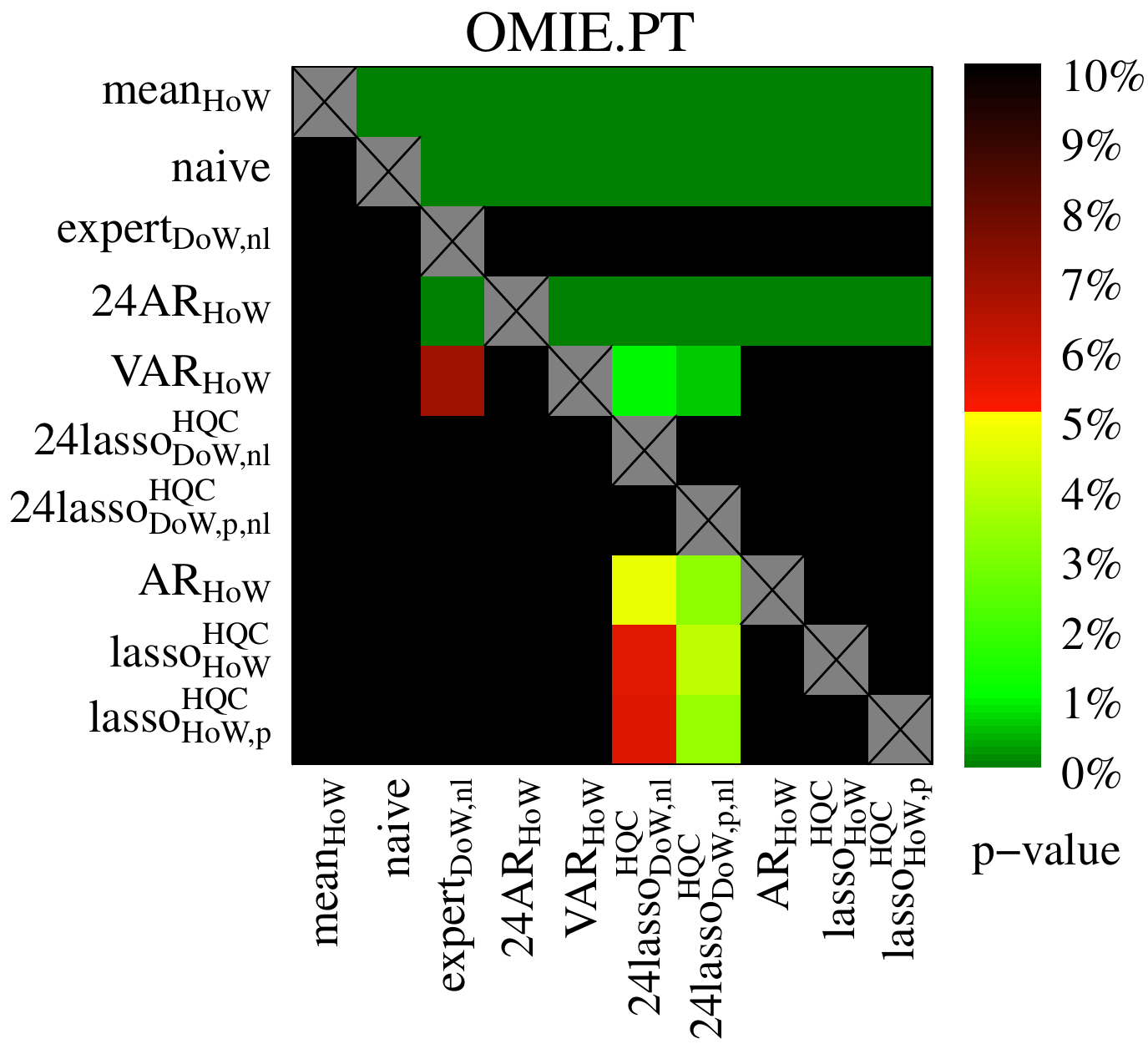} 
  \includegraphics[width=.32\textwidth, height=.22\textheight]{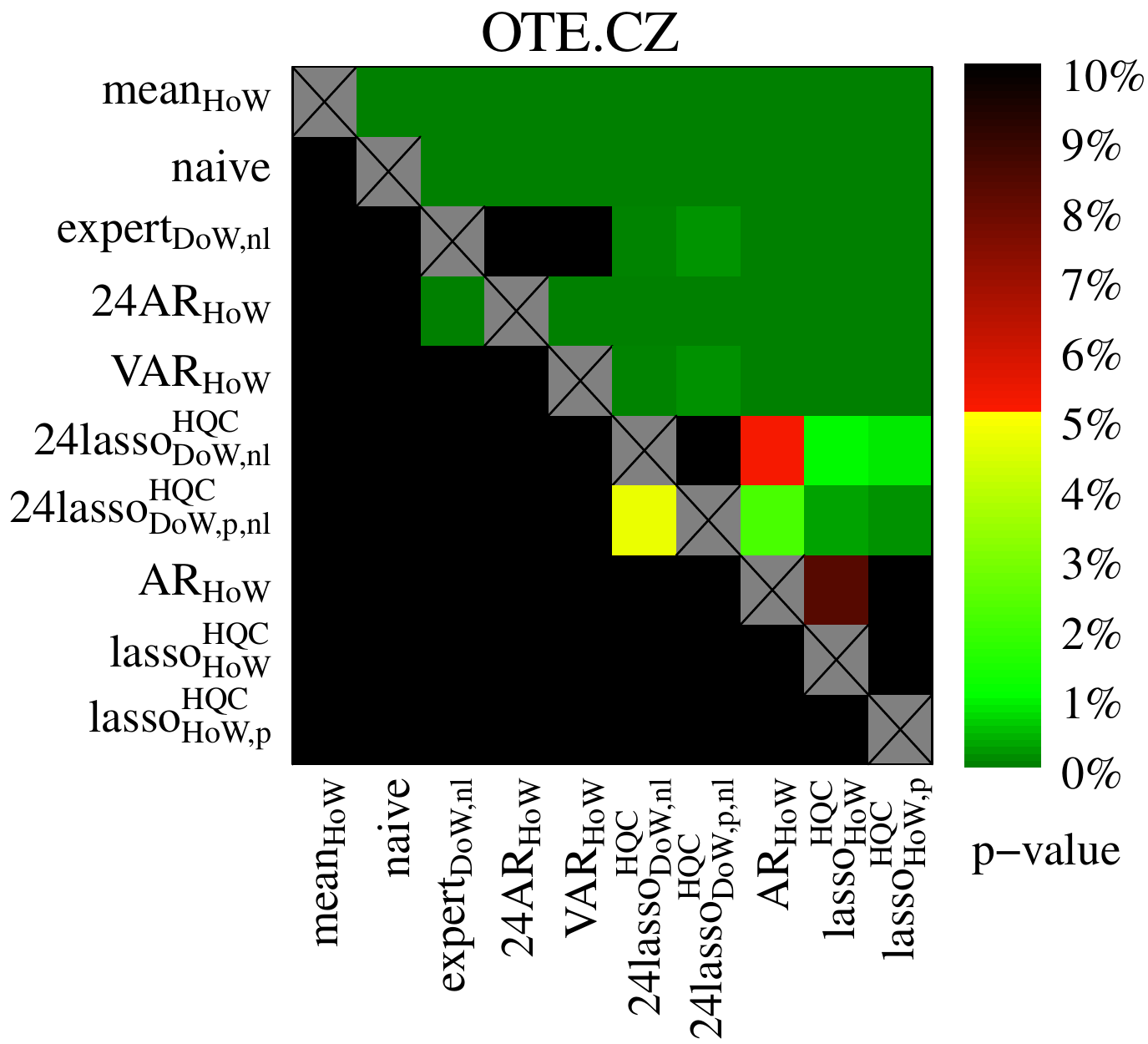} 

\caption{Results of the `multivariate' (or `vectorized') DM test defined by the multivariate loss differential series in Eqn. \eqref{eqn:Delta:DM}. We use a heat map to indicate the range of the $p$-values -- the closer they are to zero ($\rightarrow$ dark green) the more significant is the difference between the forecasts of a model on the X-axis (better) and the forecasts of a model on the Y-axis (worse).
}
 \label{fig_dmtest}
\end{figure*}

In Figure \ref{fig_dmtest} we summarize the results for the 10 considered models, for each market. We use a heat map to indicate the range of the $p$-values -- the closer they are to zero ($\rightarrow$ dark green) the more significant is the difference between the forecasts of a model on the X-axis (better) and the forecasts of a model on the Y-axis (worse). In other words, if in a given row each square is green then the forecasts of models on the X-axis are significantly better then those of the model on the Y-axis.
For instance, for all markets the first row is always dark green indicating that the forecasts of \textbf{mean$_{\text{HoW}}$} are significantly outperformed by those of all other models. Likewise, for all markets the second row is dark green except for the first column indicating that the forecasts of the \textbf{naive} model are always significantly outperformed by those of all other models but \textbf{mean$_{\text{HoW}}$}.

Note, that the $p$-values of the DM test are symmetric around 0.5, so if the standard test with null $H_0$ yields a $p$-value of 0.05, then the complementary test with the reverse null, i.e., $H^R_0$, returns a $p$-value of 0.95. However, the charts in Figure \ref{fig_dmtest} break the symmetry around the diagonal since the scale is capped at 0.1 (i.e., 10\%; for better exposition of the relevant results). Thus if the standard test returns a $p$-value of, say, 0.85, then the complementary test yields a $p$-value of 0.15, and both are indicated by black squares. So neither the forecasts of the model on the X-axis are significantly better than those of the model on the Y-axis nor vice versa, as is the case, e.g., for BELPEX.BE and models \textbf{24AR$_{\text{HoW}}$} and  \textbf{VAR$_{\text{HoW}}$}.

Looking closely at Figure \ref{fig_dmtest} we can observe that for eight datasets (BELPEX.BE, EPEX.CH, EPEX.FR, EXAA.DE+AT, GEFCom2014, NP.SYS, OMIE.ES and OMIE.PT) the forecasts of the best according to MAE multivariate lasso model, i.e.,  \textbf{24lasso$_{\text{DoW,p,nl}}^{\text{HQC}}$}, are not significantly worse than those of the best univariate lasso model, i.e., \textbf{lasso$_{\text{HoW,p}}^{\text{HQC}}$}. On the other hand, the forecasts of the latter are for eight markets (BELPEX.BE, EPEX.DE+AT, EPEX.FR, EXAA.DE+AT, GEFCom2014, NP.DK1, NP.DK2 and OTE.CZ) not significantly worse than the forecasts of \textbf{24lasso$_{\text{DoW,p,nl}}^{\text{HQC}}$}. Thus, there are four markets (BELPEX.BE, EPEX.FR, EXAA.DE+AT and GEFCom2014) where we cannot decide which model is better based on the considered DM test.

\subsubsection{The standard DM test and the performance across the hours}

To get a better understanding of how the significance changes across the hours we now consider the standard DM test, for which the loss differential series is given by \cite[see also][]{bor:bun:lis:nan:13,now:rav:tru:wer:14,zie:ste:hus:15:EXAA,now:wer:16,uni:now:wer:16}:
\begin{equation}
\label{eqn:Delta:DM:std}
\Delta_{X,Y,d,h} = |\what{\eps}_{X,d,h}| - |\what{\eps}_{Y,d,h}|.
\end{equation}
We perform two one-sided DM tests at the 5\% significance level: (i) a test with the null hypothesis $H_0: E(\Delta_{X,Y,d,h}) \leq 0$, i.e. the outperformance of the forecasts of model $Y$ by those of model $X$, and (ii) the complementary test with the reverse null $H^R_0: E(\Delta_{X,Y,d,h}) \geq 0$, i.e. the outperformance of the forecasts of model $X$ by those of model $Y$. Note, that like in the above cited studies, we assume that forecasts for consecutive days, hence loss differentials are covariance stationary. Note also, that compared to the multivariate DM test approach discussed above, we now perform the test at one significance level (5\%) and focus on binary variables representing the passing or not of the test for a particular hour, not the $p$-values themselves.

\begin{figure*}[p]
\centering
%DMtesth-chessboard
\includegraphics[width=.32\textwidth, height=.22\textheight]{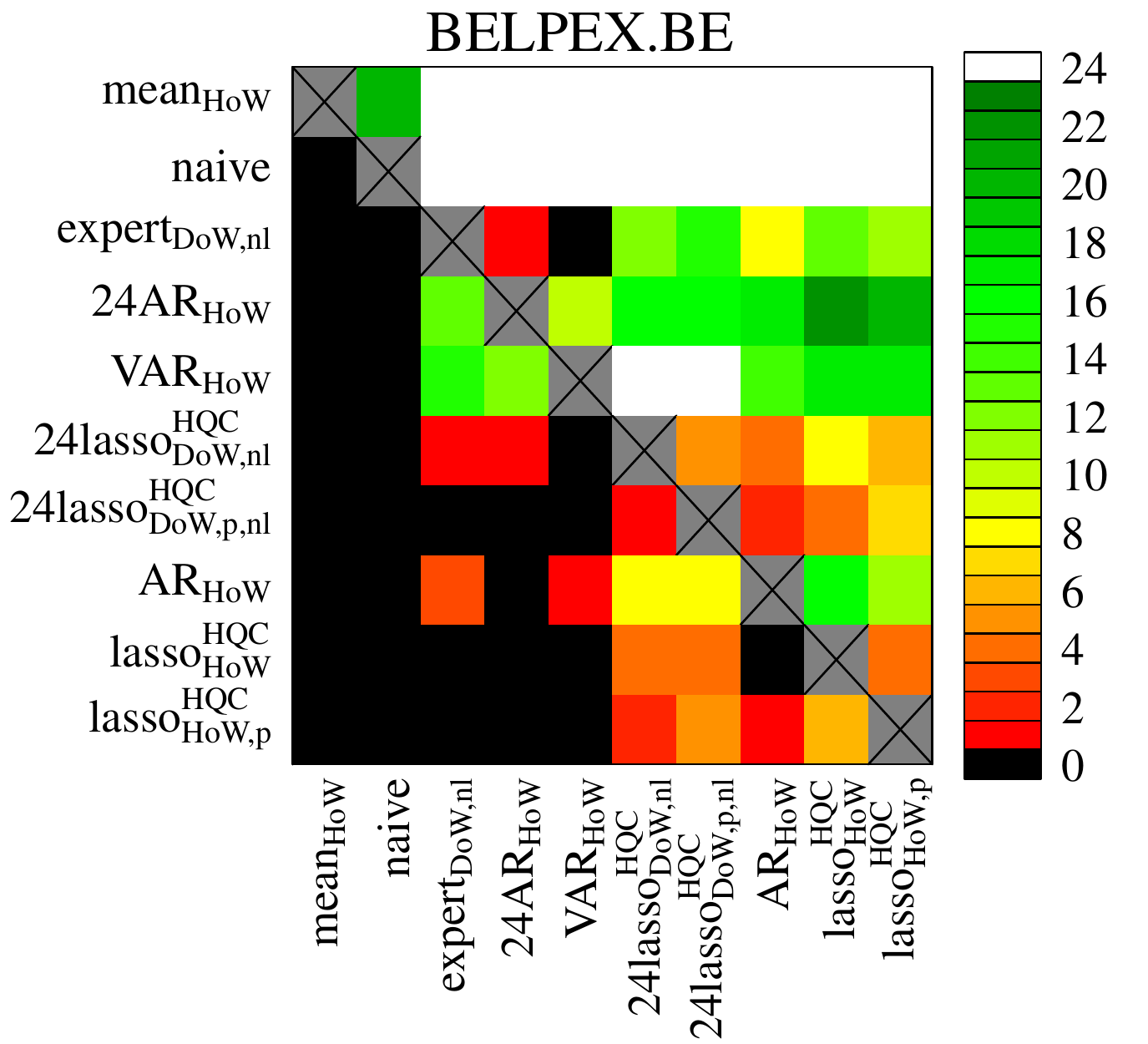} 
 \includegraphics[width=.32\textwidth, height=.22\textheight]{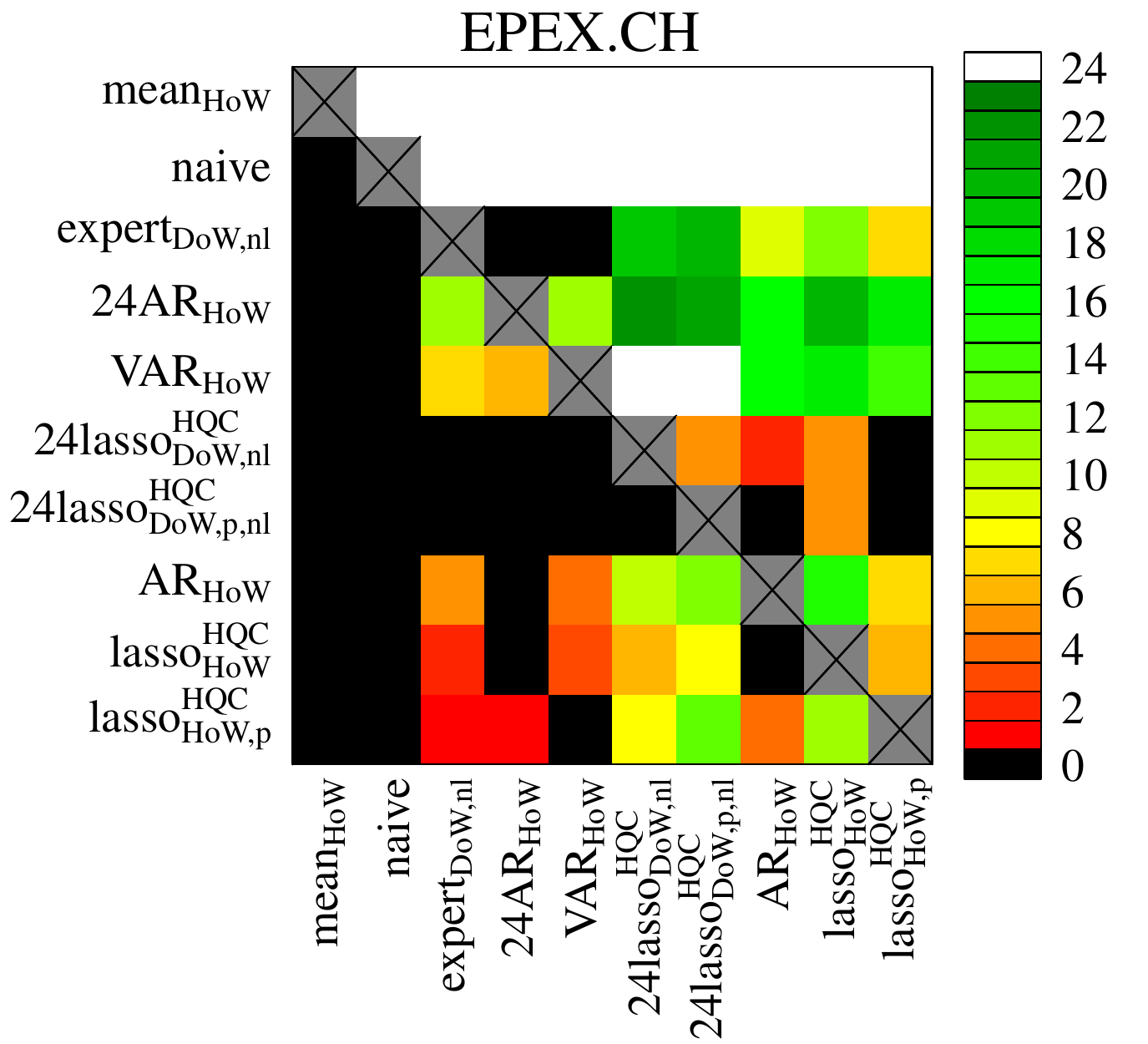} 
  \includegraphics[width=.32\textwidth, height=.22\textheight]{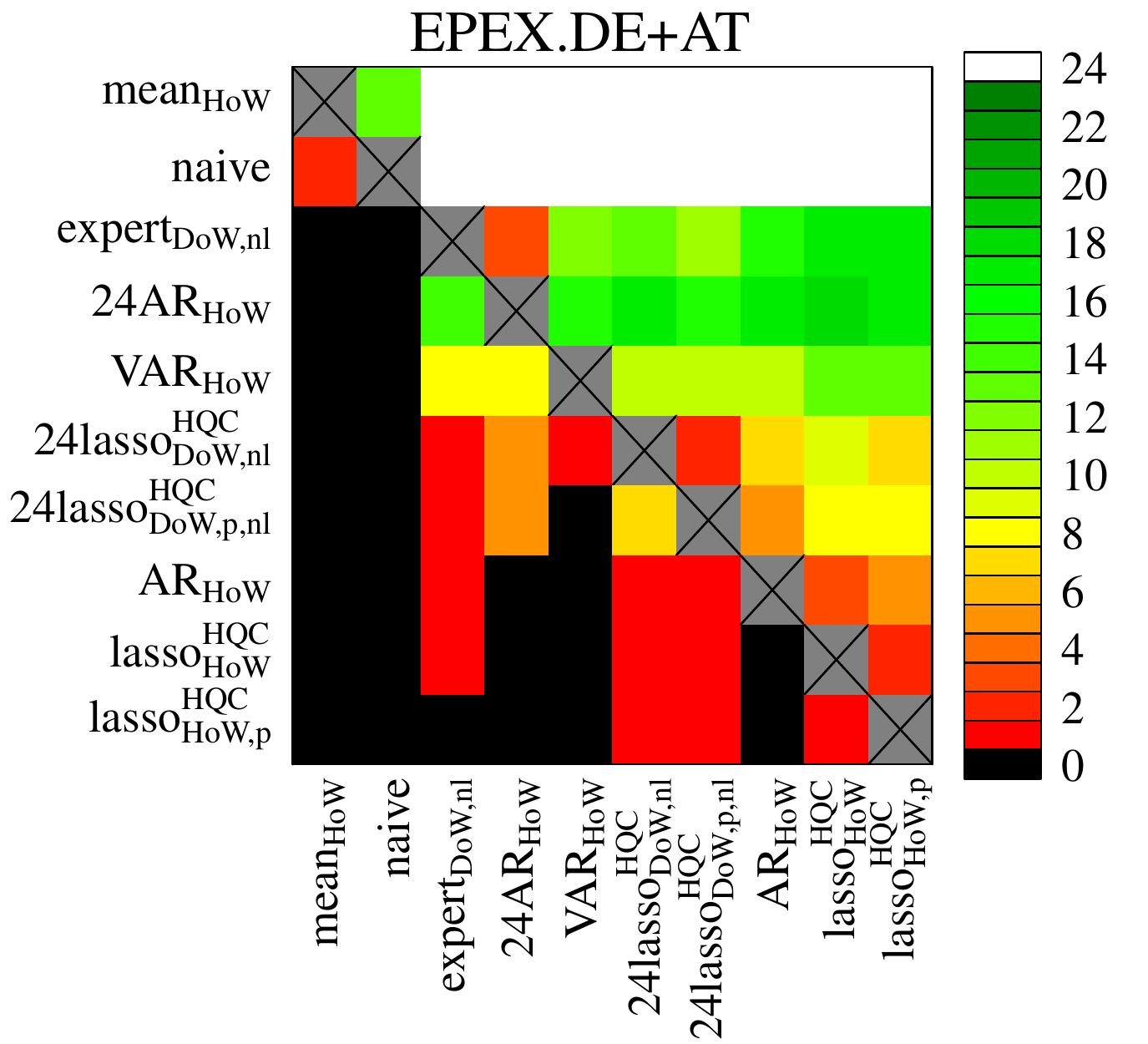} 
  \includegraphics[width=.32\textwidth, height=.22\textheight]{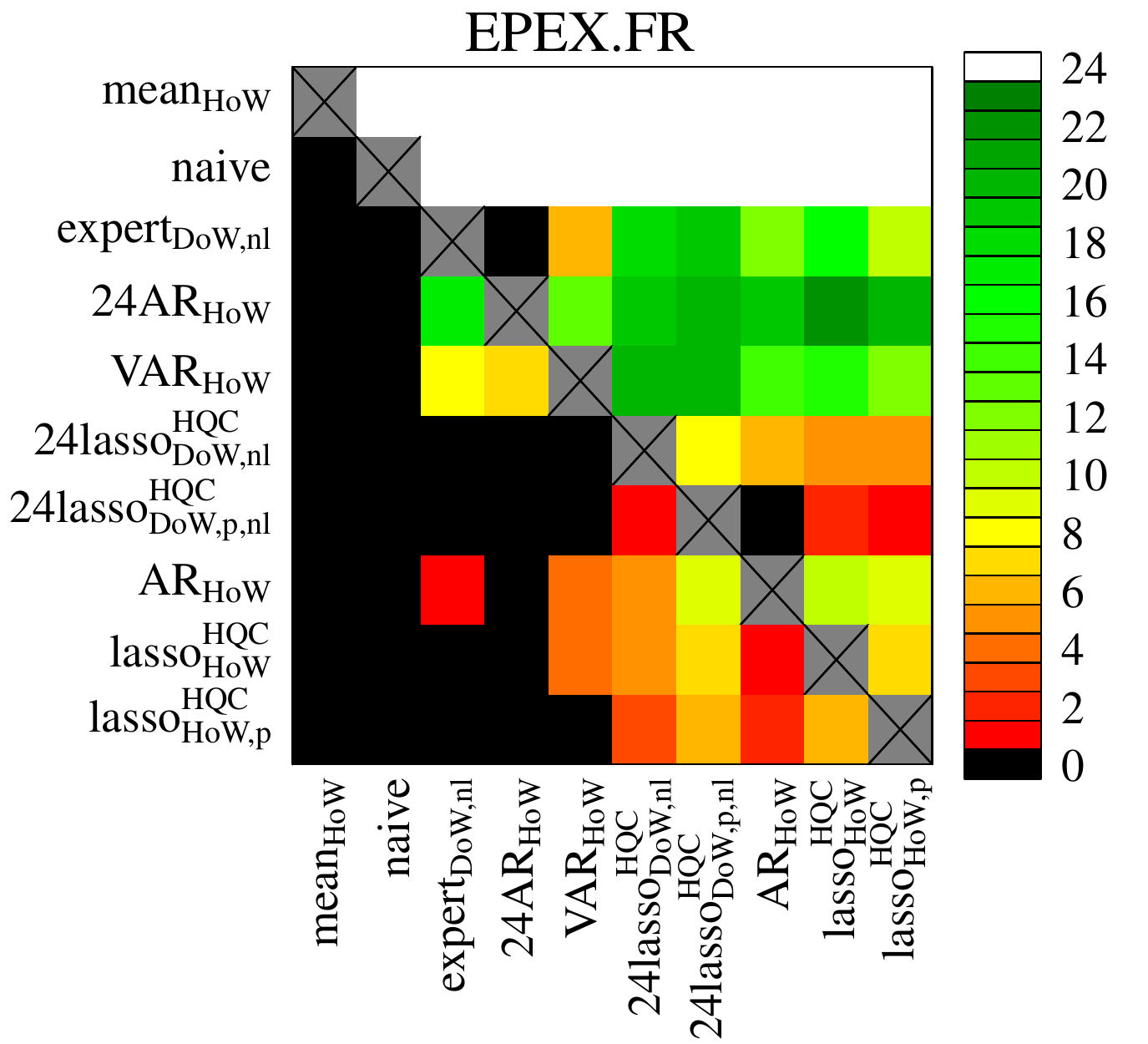} 
  \includegraphics[width=.32\textwidth, height=.22\textheight]{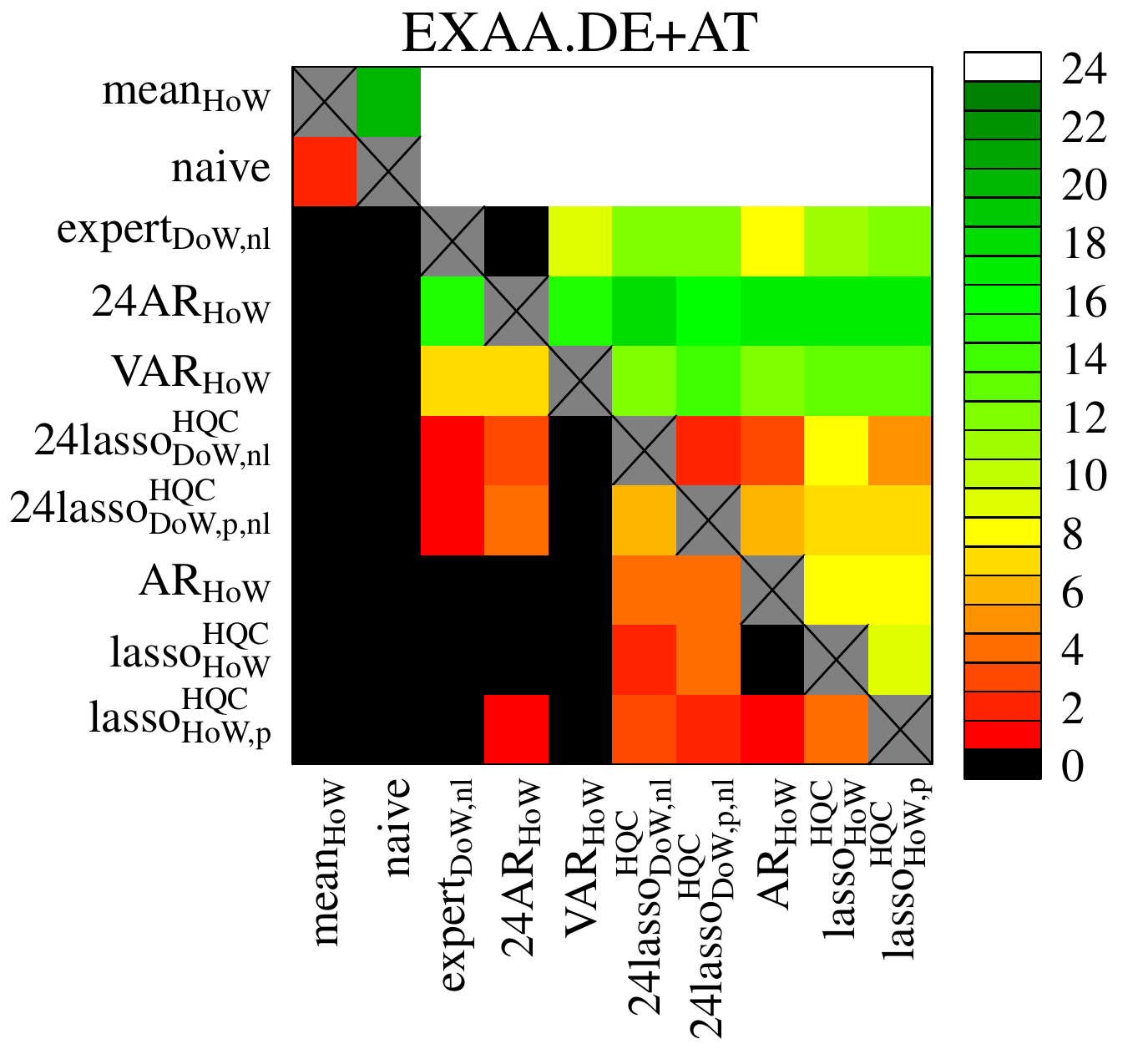} 
  \includegraphics[width=.32\textwidth, height=.22\textheight]{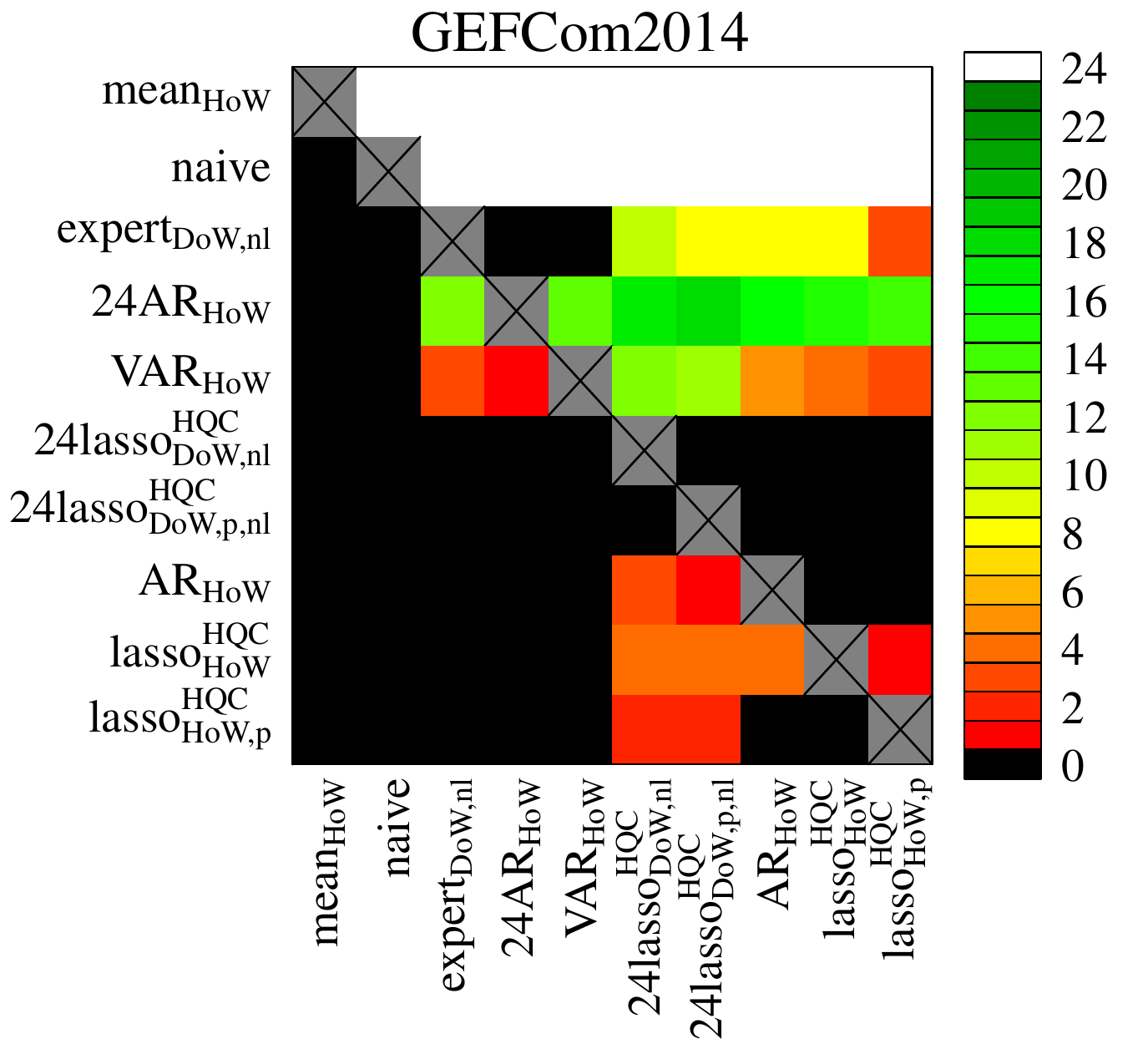} 
  \includegraphics[width=.32\textwidth, height=.22\textheight]{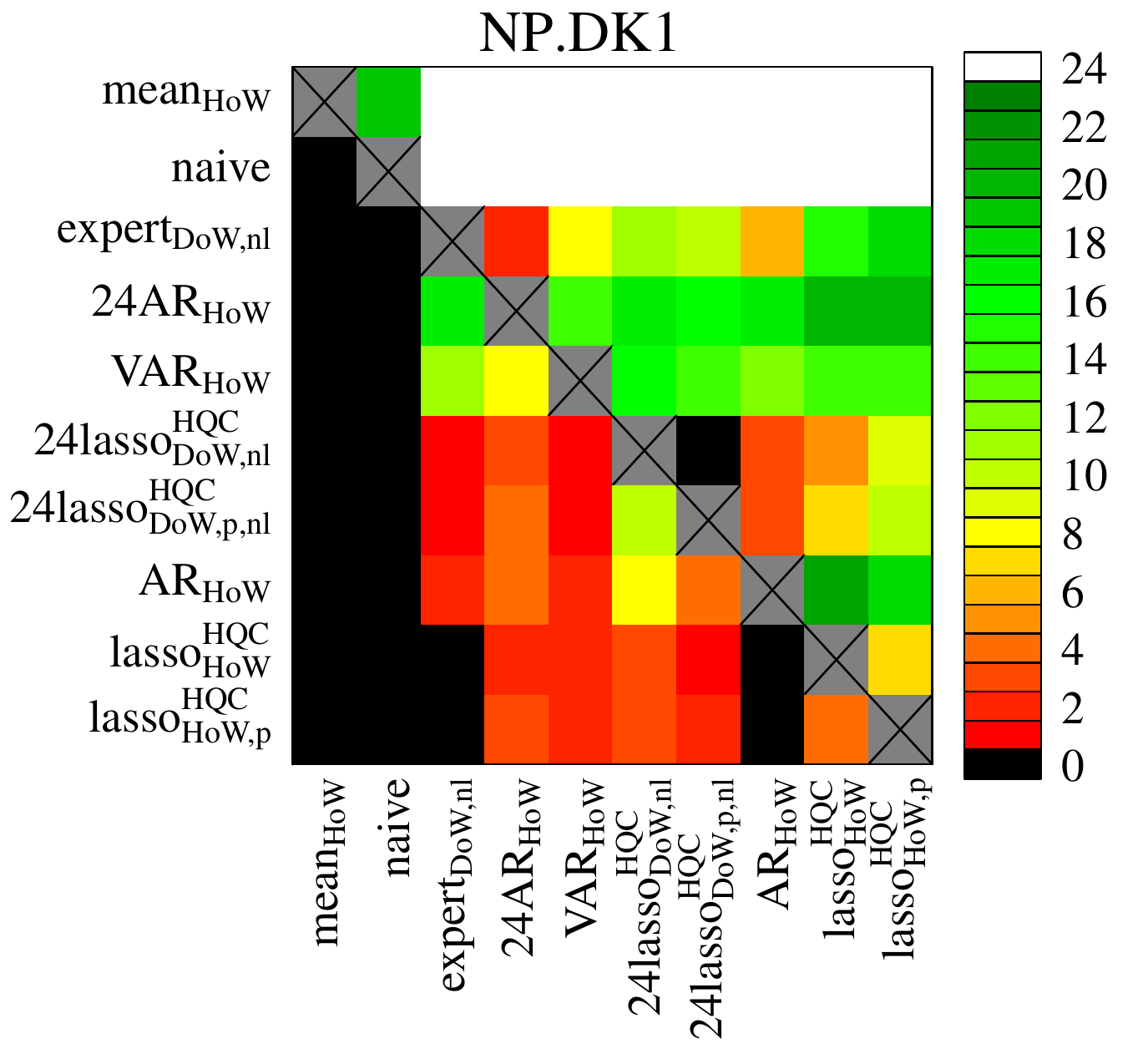} 
  \includegraphics[width=.32\textwidth, height=.22\textheight]{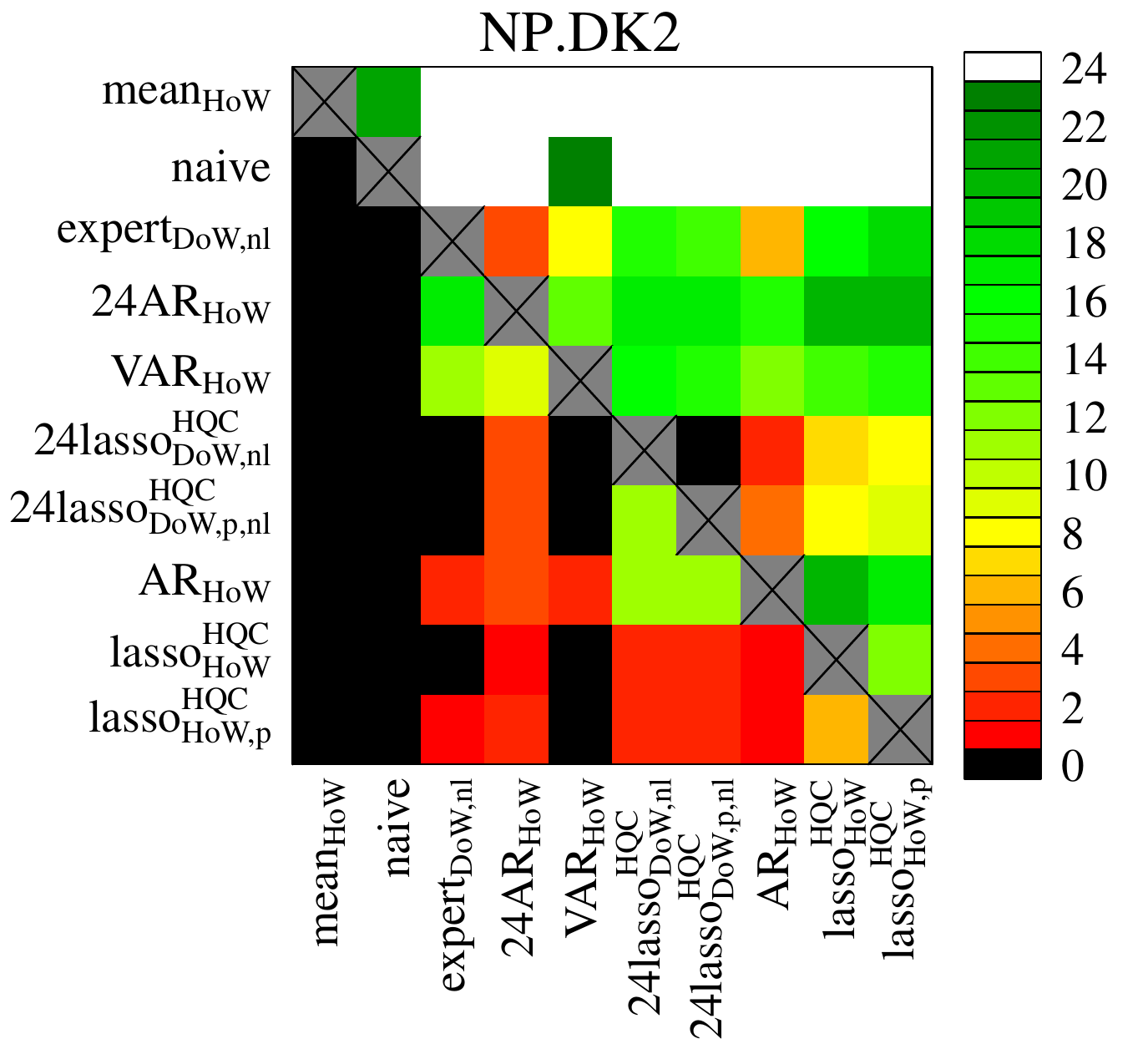} 
  \includegraphics[width=.32\textwidth, height=.22\textheight]{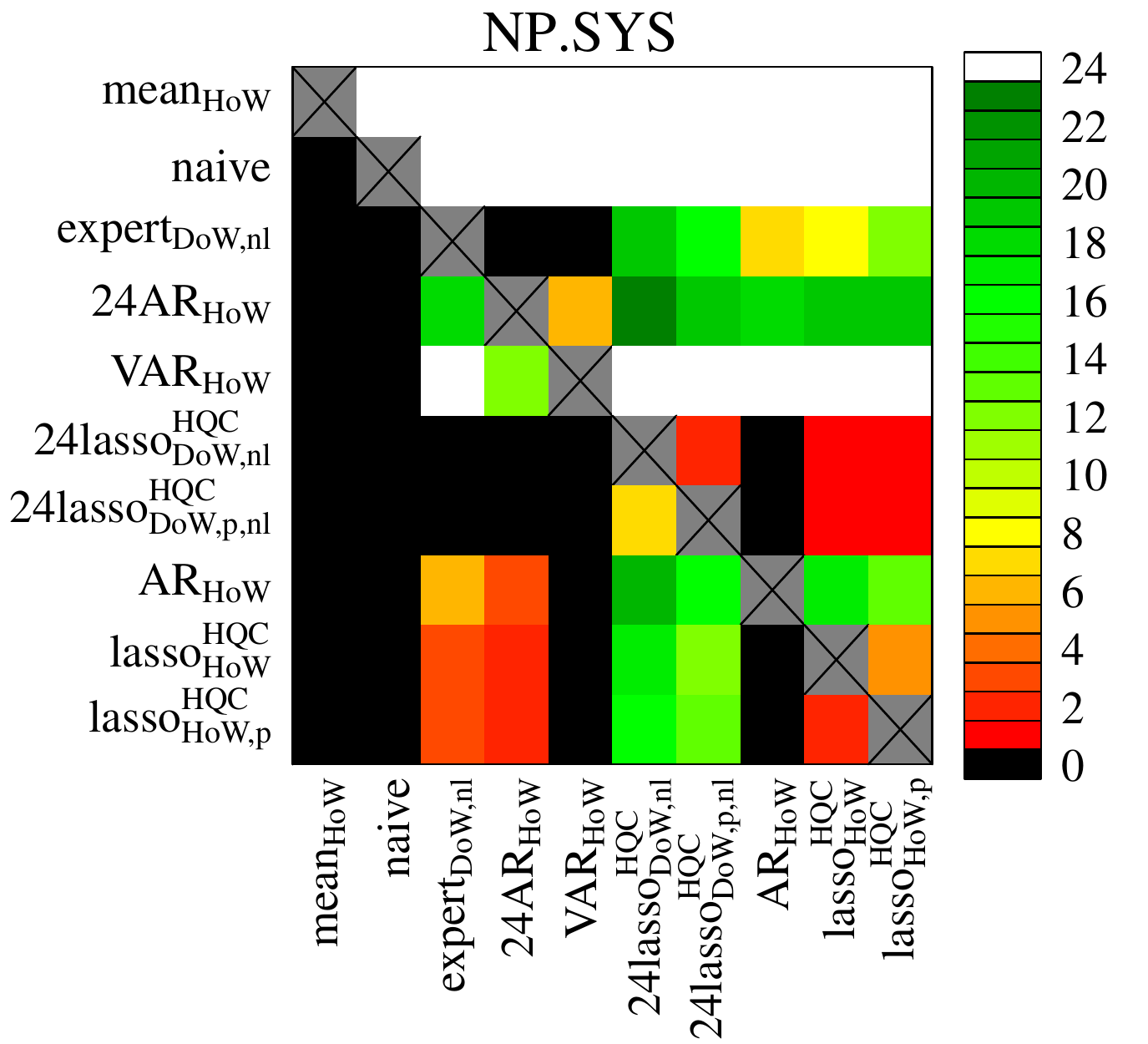} 
  \includegraphics[width=.32\textwidth, height=.22\textheight]{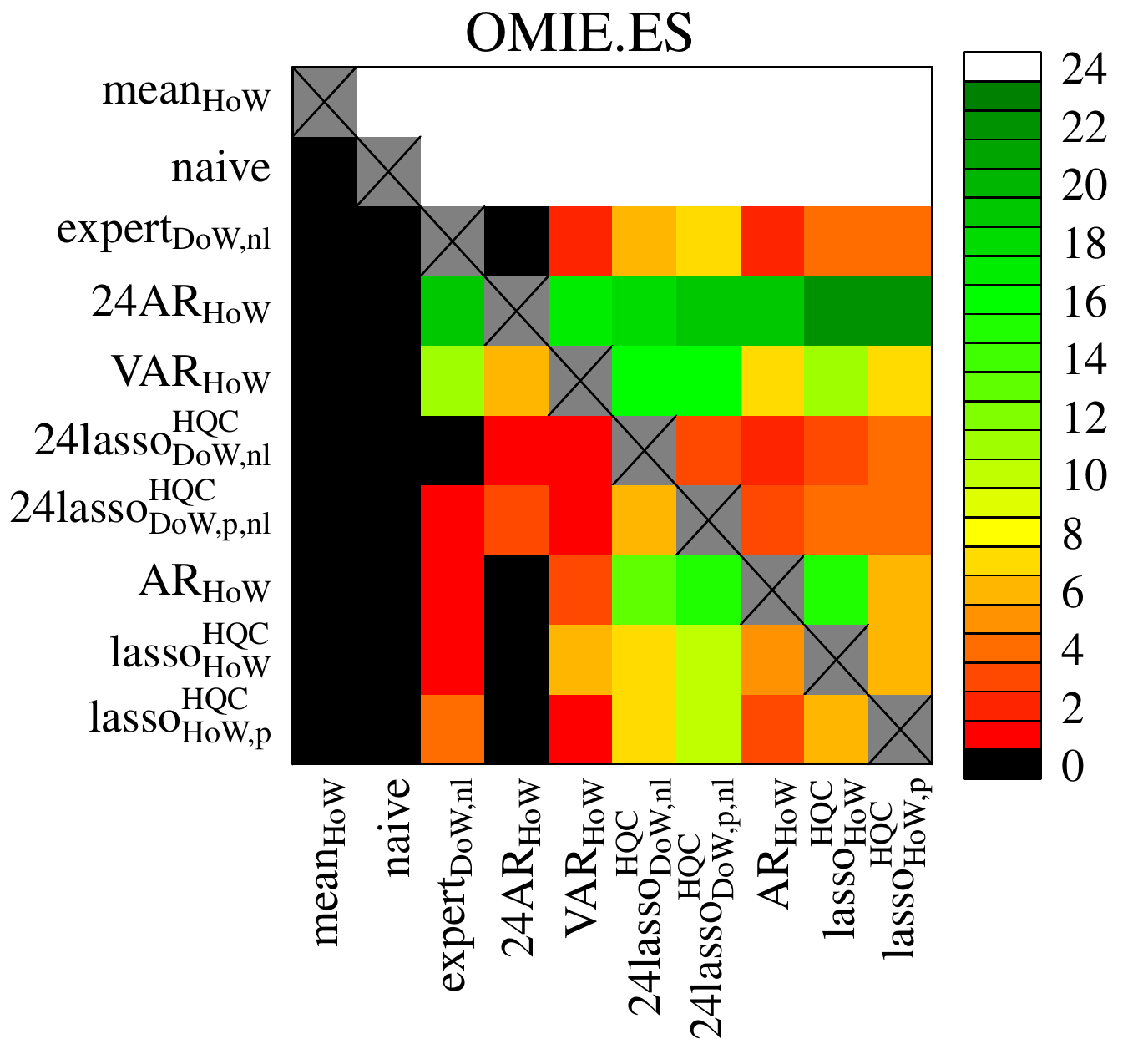} 
  \includegraphics[width=.32\textwidth, height=.22\textheight]{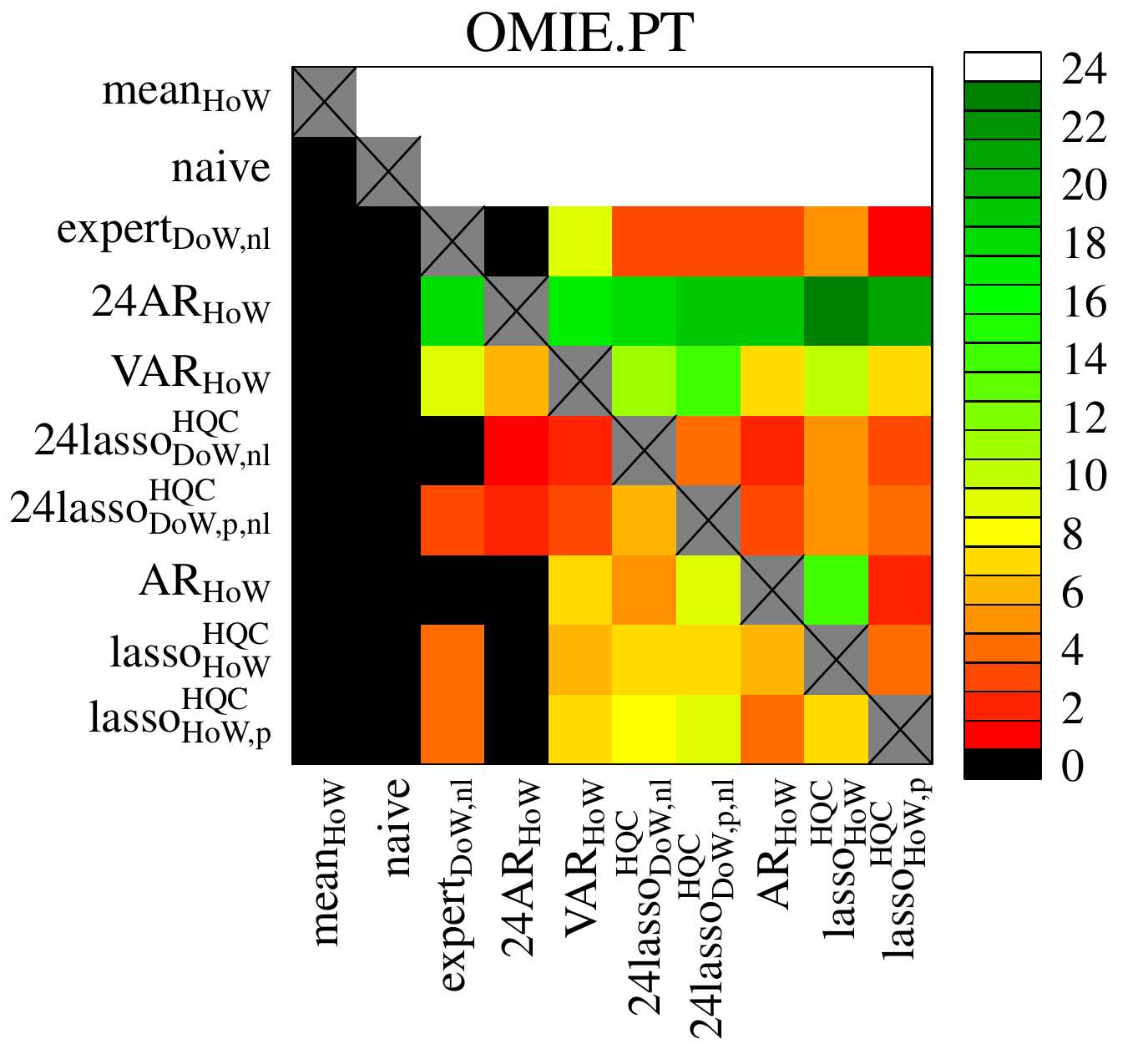} 
  \includegraphics[width=.32\textwidth, height=.22\textheight]{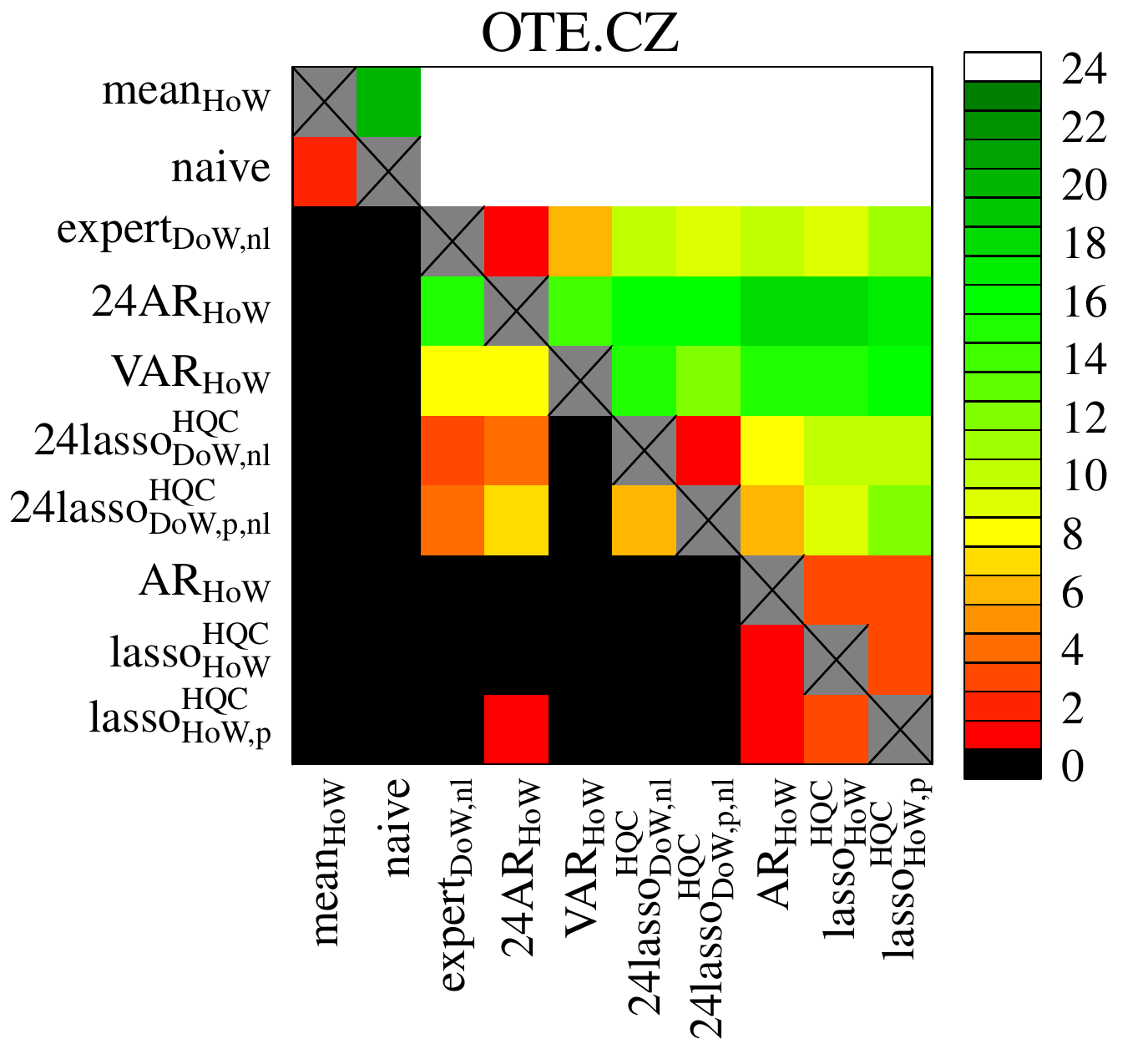} 

\caption{Results of the 24 hourly DM tests at the 5\% level, defined by the loss differential series in Eqn. \eqref{eqn:Delta:DM:std}. We sum the number of significant differences in forecasting performance across the 24 hours and use a heatmap to indicate the number of hours for which the forecasts of a model on the X-axis are significantly better than those of a model on the Y-axis. A white square indicates that forecasts of a model on the X-axis are better for all 24 hours, while a black square that they are not better for a single hour.
}
\label{fig_dmtesth-chess}
\end{figure*}

In Figure \ref{fig_dmtesth-chess} we summarize the DM results for all datasets. Namely, we sum the number of significant differences in forecasting performance across the 24 hours and use a heat map to indicate the number of hours for which the forecasts of a model on the X-axis are significantly better than those of a model on the Y-axis. If the forecasts of a model on the X-axis are significantly better for all 24 hours of the day, we indicate this by a white square. On the other hand, if the forecasts of a model on the X-axis are not significantly better for any hour, we plot a black square. Naturally, the diagonal (gray crosses on black squares) should be ignored as it concerns the same model on both axes. 
Columns with many non-black squares (the more green or white the better) indicate that the forecasts of a model on the X-axis are significantly better than the forecasts of many of its competitors. Conversely, rows with many non-black squares mean that the forecasts of a model on the Y-axis are significantly worse than the forecasts of many of its competitors. For instance, for the EPEX.CH dataset, the white row for the \textbf{mean$_{\text{HoW}}$} benchmark indicates that the forecasts of this simple model are significantly worse than the forecasts of all of its competitors for all 24 hours, while the black column for \textbf{mean$_{\text{HoW}}$} means that not a single competitor produces significantly worse forecasts than this benchmark, even for a single hour of the day.

For the more sophisticated lasso models we cannot draw such clear cut conclusions. 
There is no model that would beat the competitors for all 24 hours of the day.
Moreover, for most markets the forecasts of the best univariate and multivariate models are significantly better than those of other structures only for some hours of the day.
This shows that the statements about the significance of results we were able to draw from the multivariate DM tests presented in Figure \ref{fig_dmtest} are mainly due to a significantly better performance for a few hours of the day. More interestingly, we see that for many markets, there is at least one hour (indicated by a square that is at least red) where both the forecasts of \textbf{24lasso$_{\text{DoW,p,nl}}^{\text{HQC}}$} are significantly better than those of \textbf{lasso$_{\text{HoW,p}}^{\text{HQC}}$} and vice versa. For example, for the BELPEX.BE dataset the univariate lasso model shows significantly better forecasts for seven hours of the day and significantly worse for five hours. 

\begin{figure}[p]
	\centering
	\includegraphics[width=.32\textwidth, height=.22\textheight]{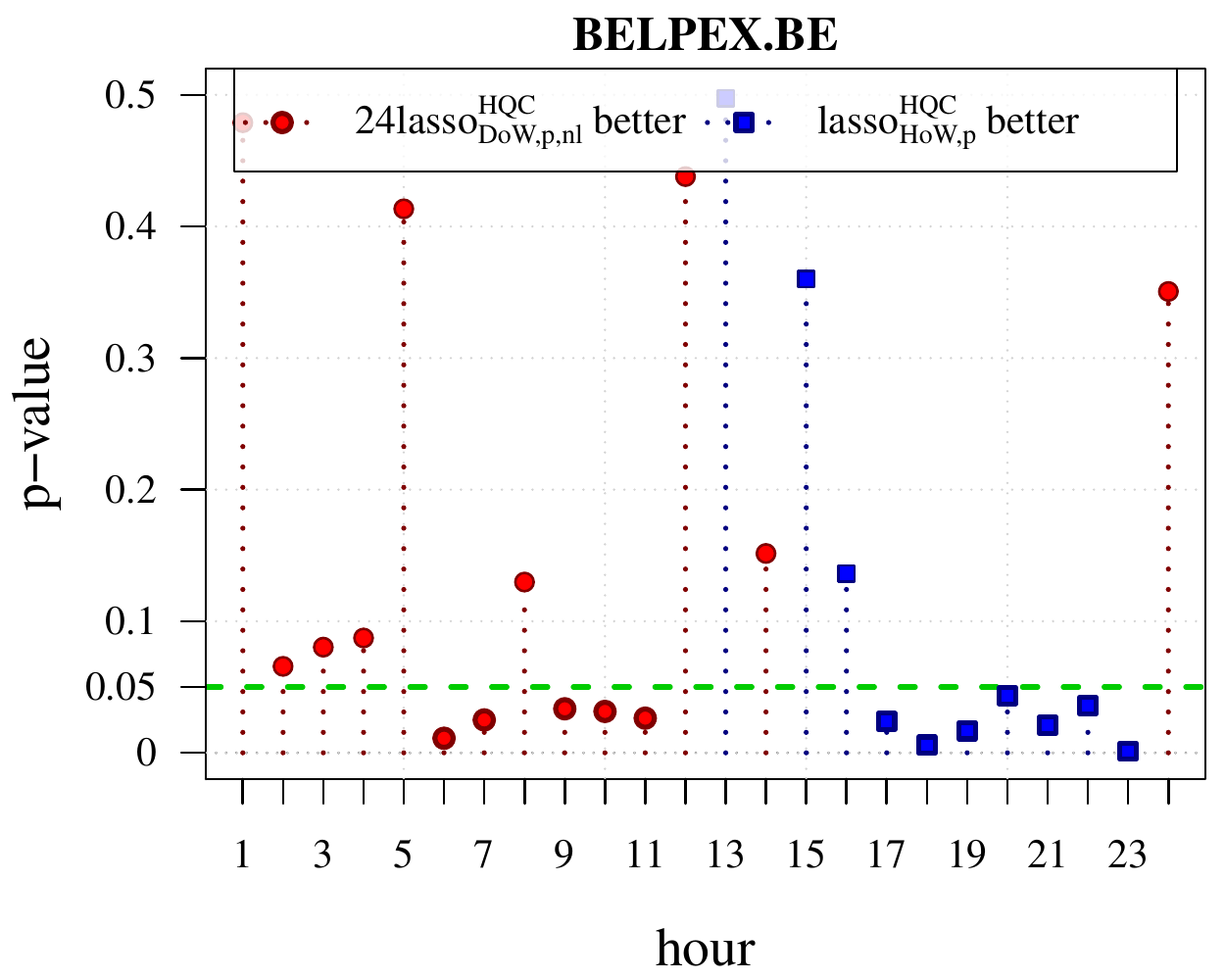} 
	\includegraphics[width=.32\textwidth, height=.22\textheight]{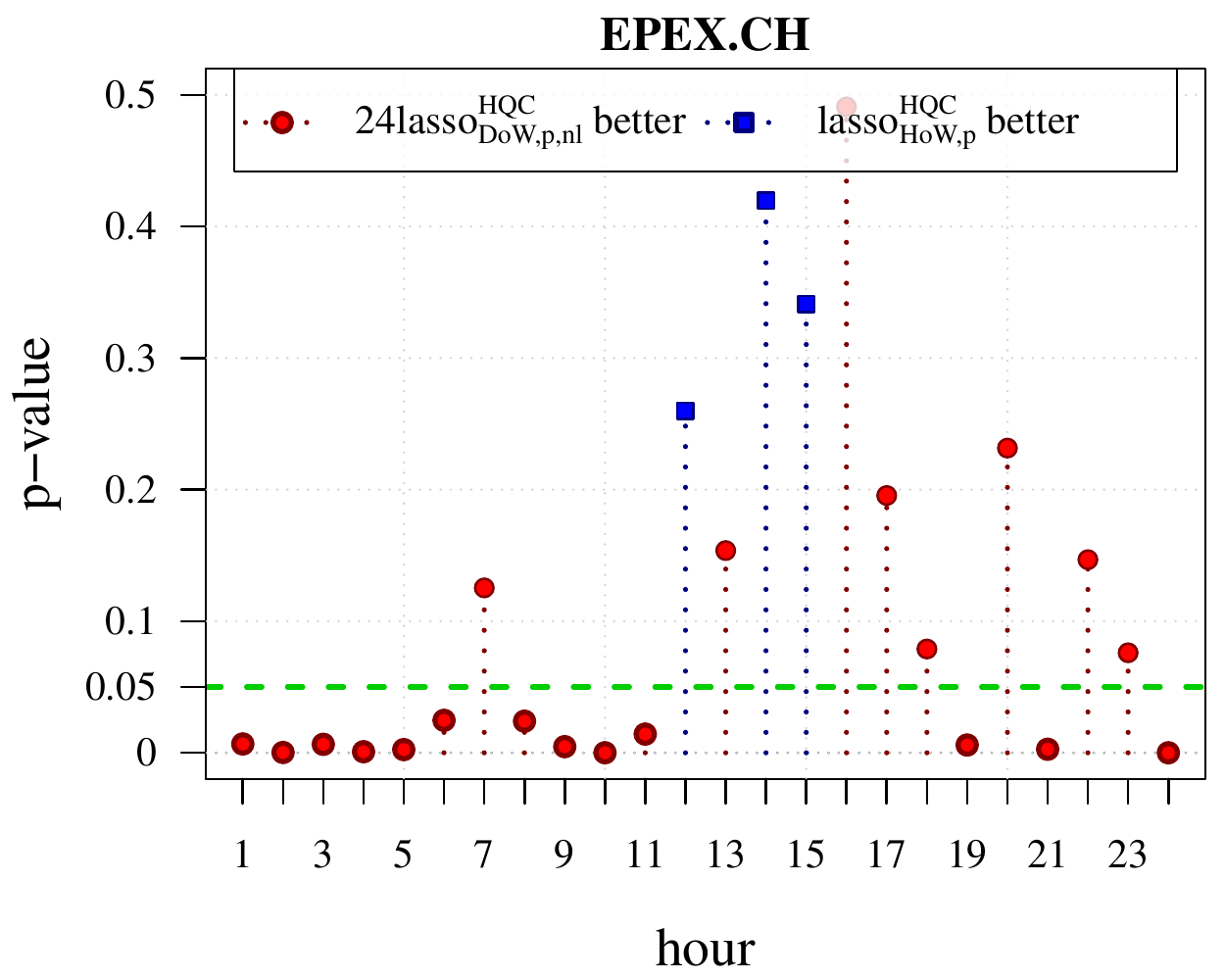} 
	\includegraphics[width=.32\textwidth, height=.22\textheight]{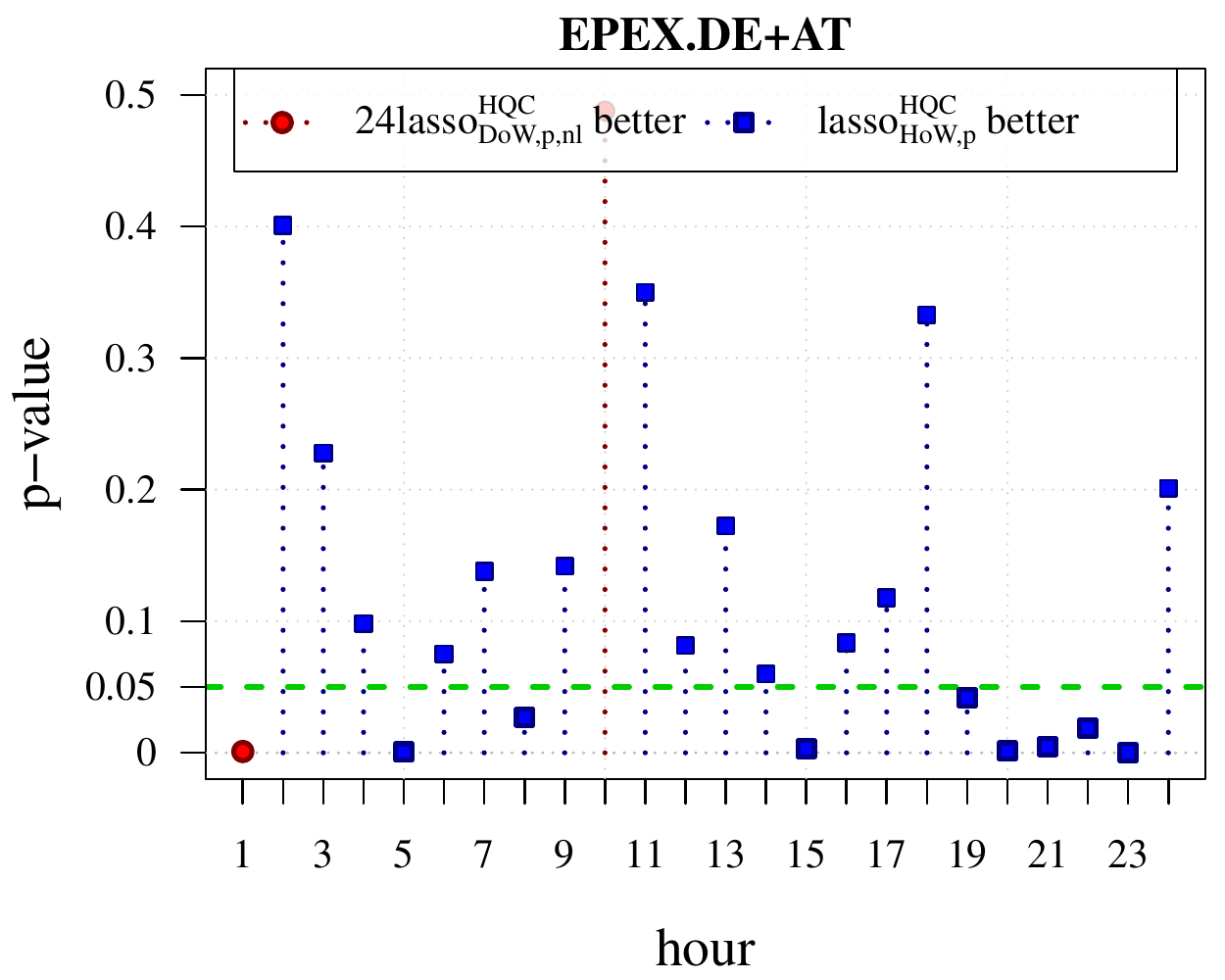} 
	\includegraphics[width=.32\textwidth, height=.22\textheight]{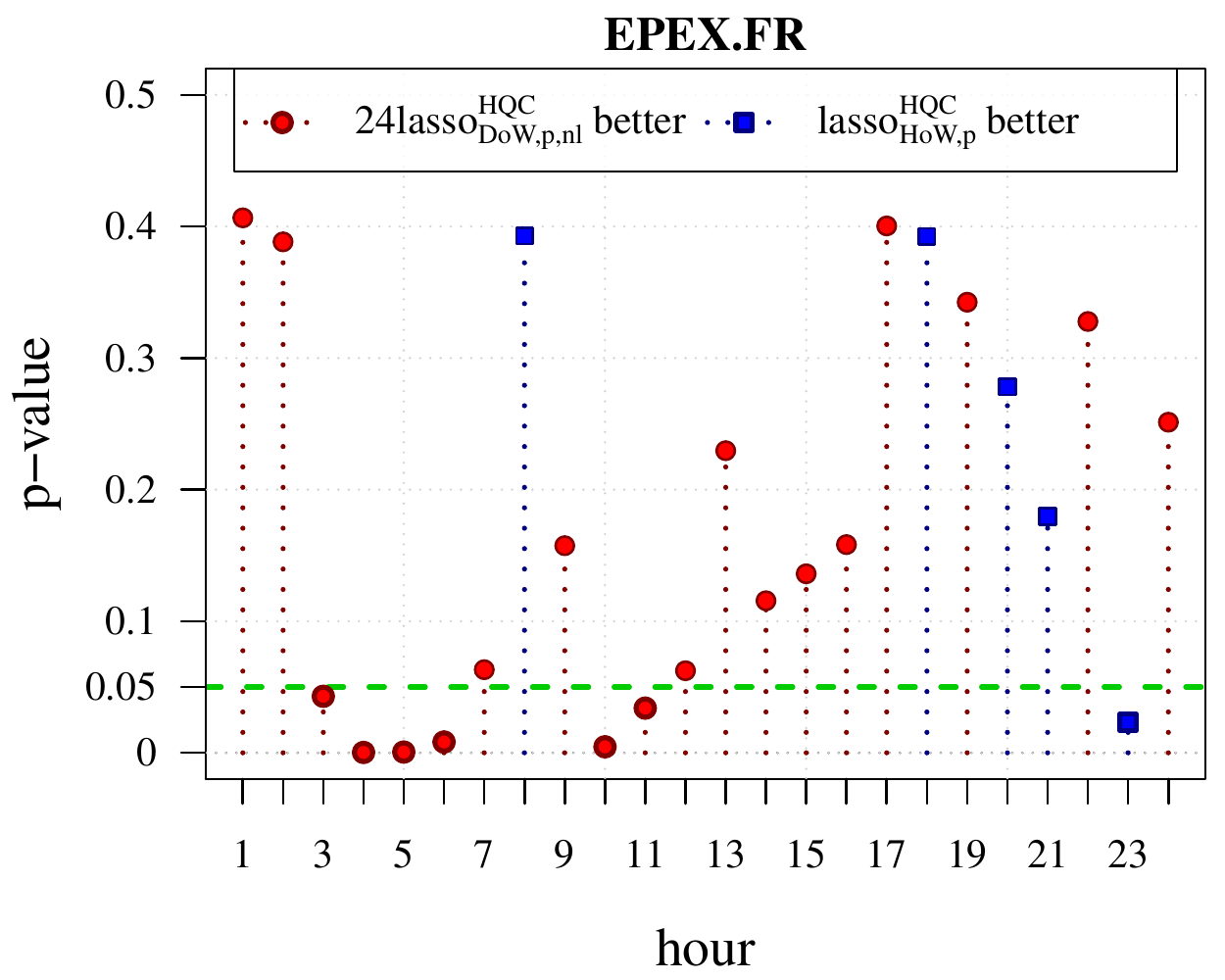} 
	\includegraphics[width=.32\textwidth, height=.22\textheight]{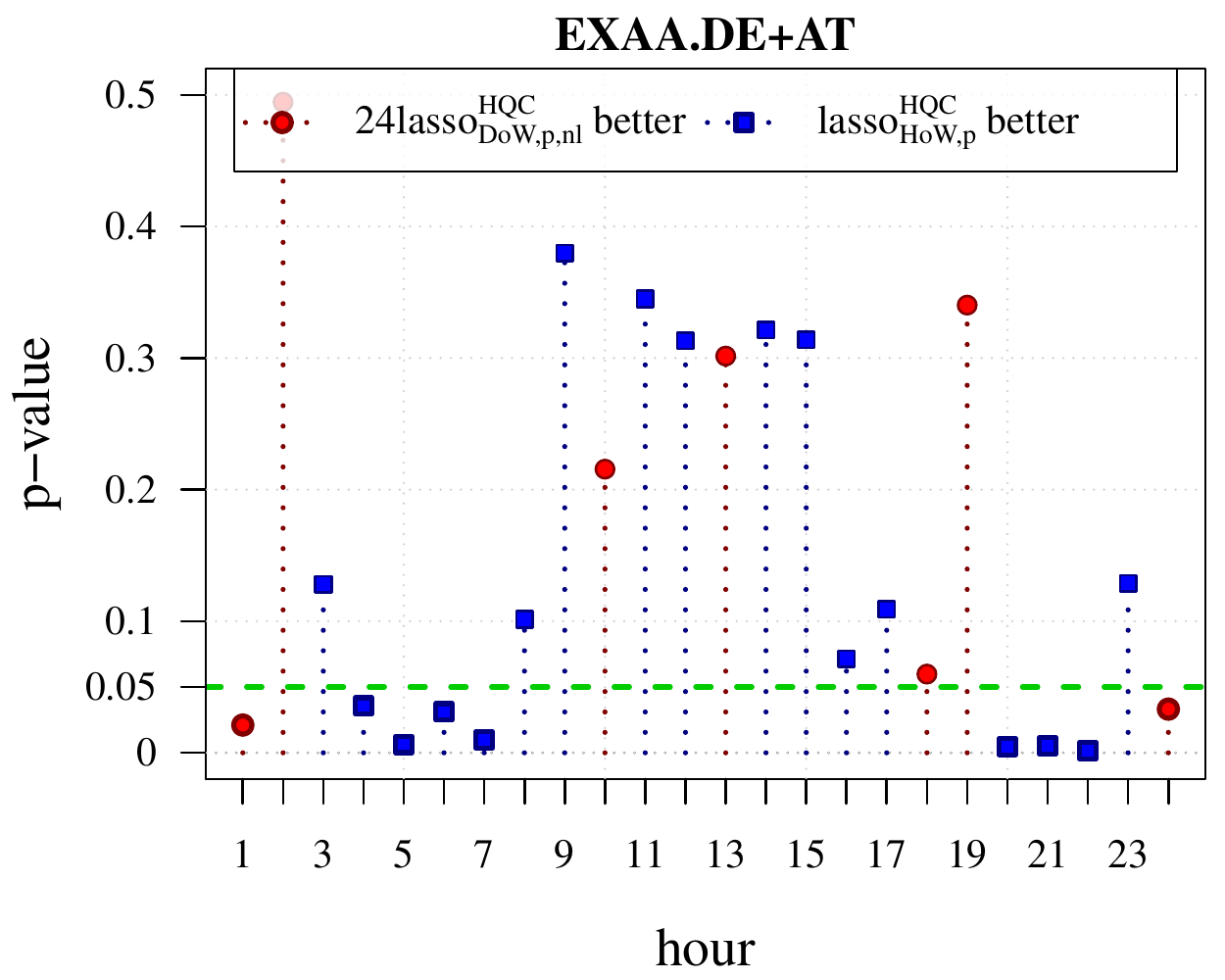} 
	\includegraphics[width=.32\textwidth, height=.22\textheight]{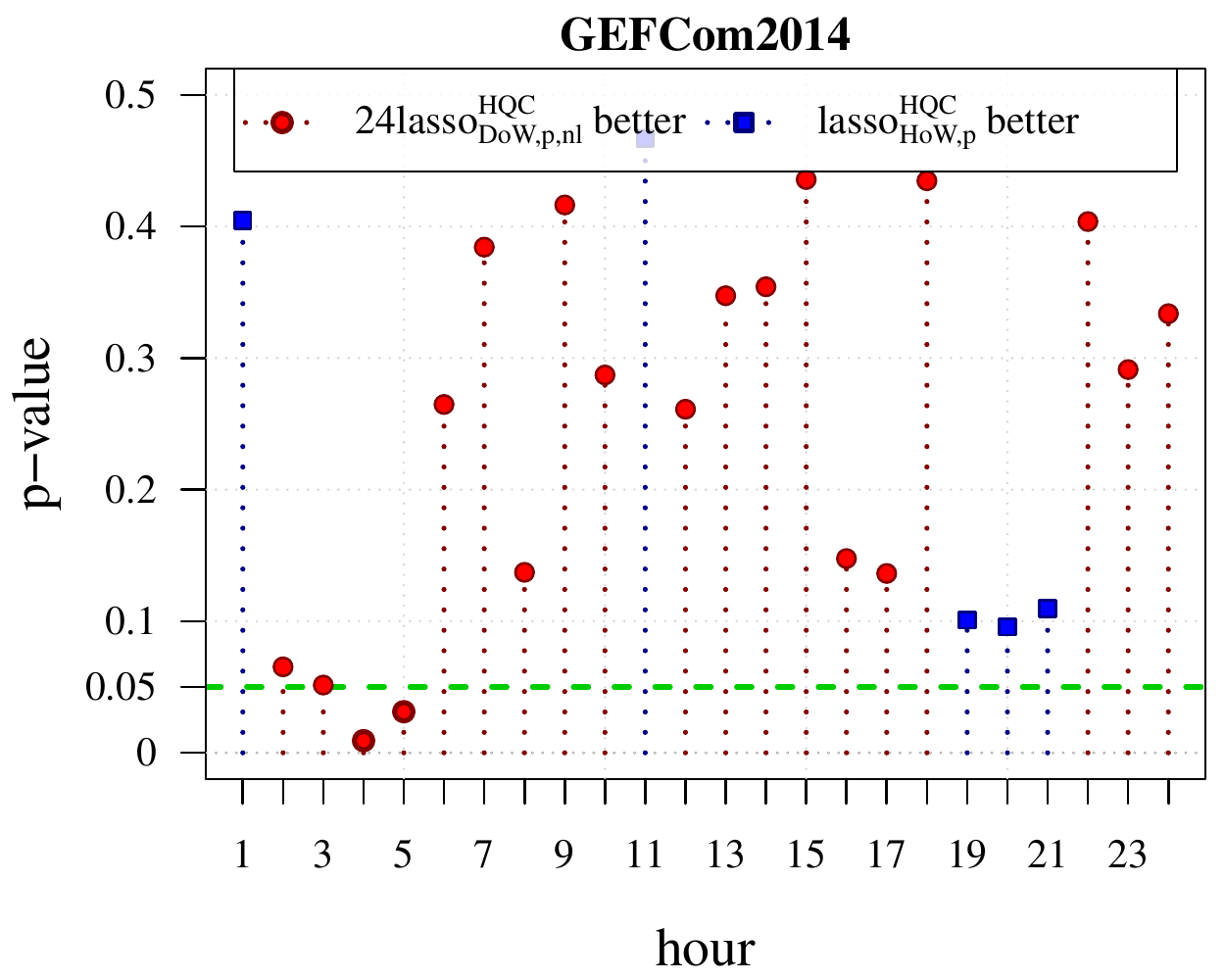} 
	\includegraphics[width=.32\textwidth, height=.22\textheight]{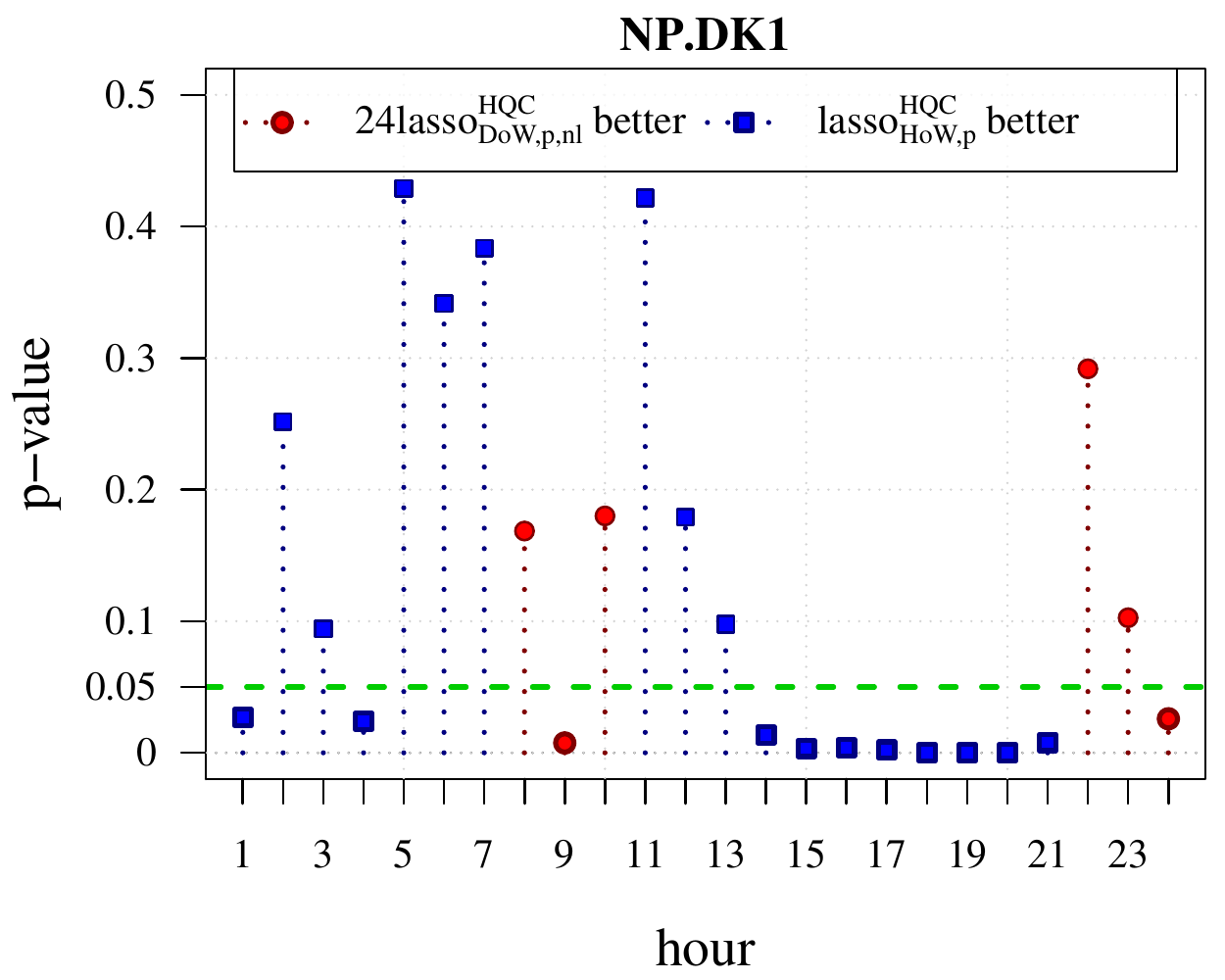} 
	\includegraphics[width=.32\textwidth, height=.22\textheight]{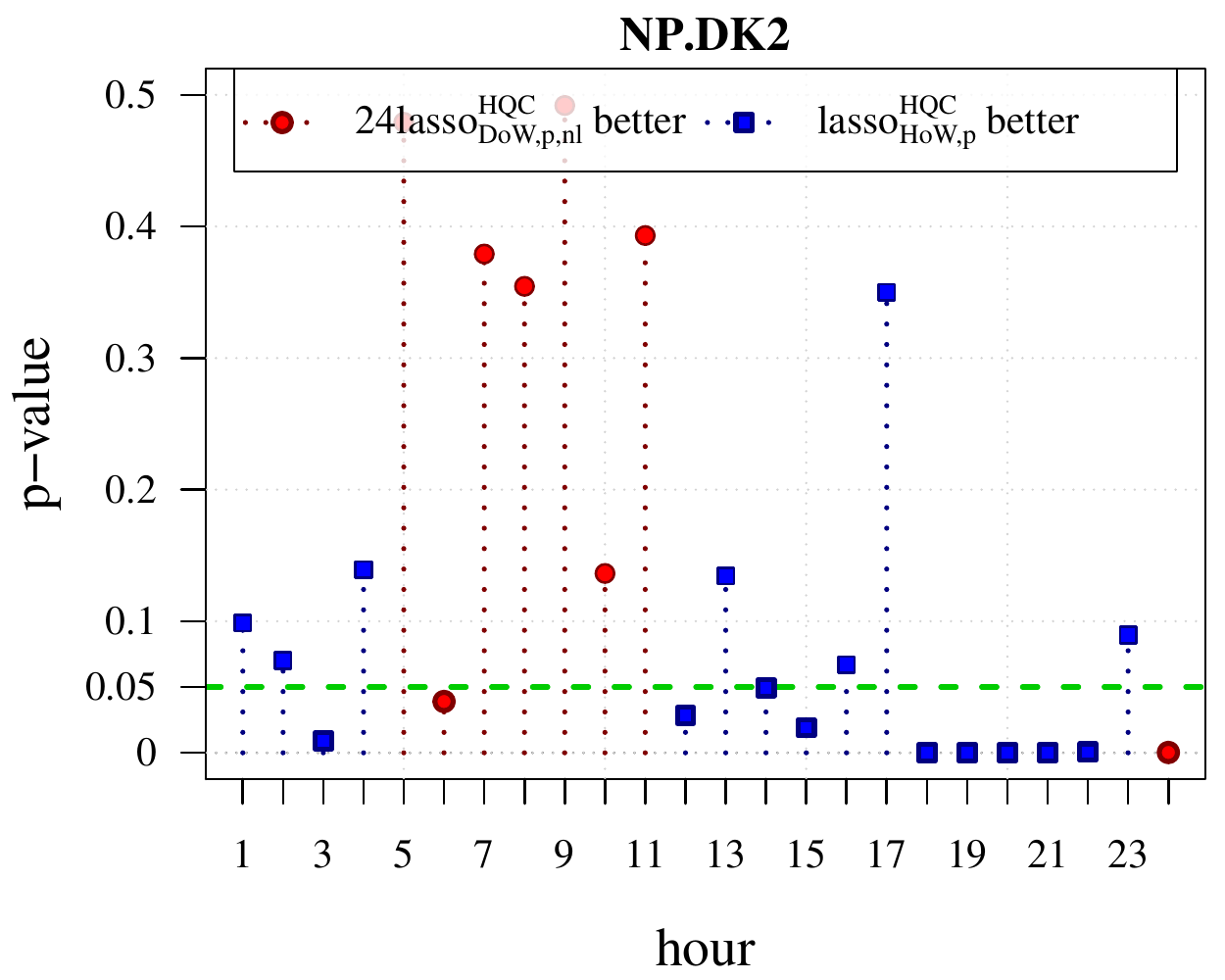} 
	\includegraphics[width=.32\textwidth, height=.22\textheight]{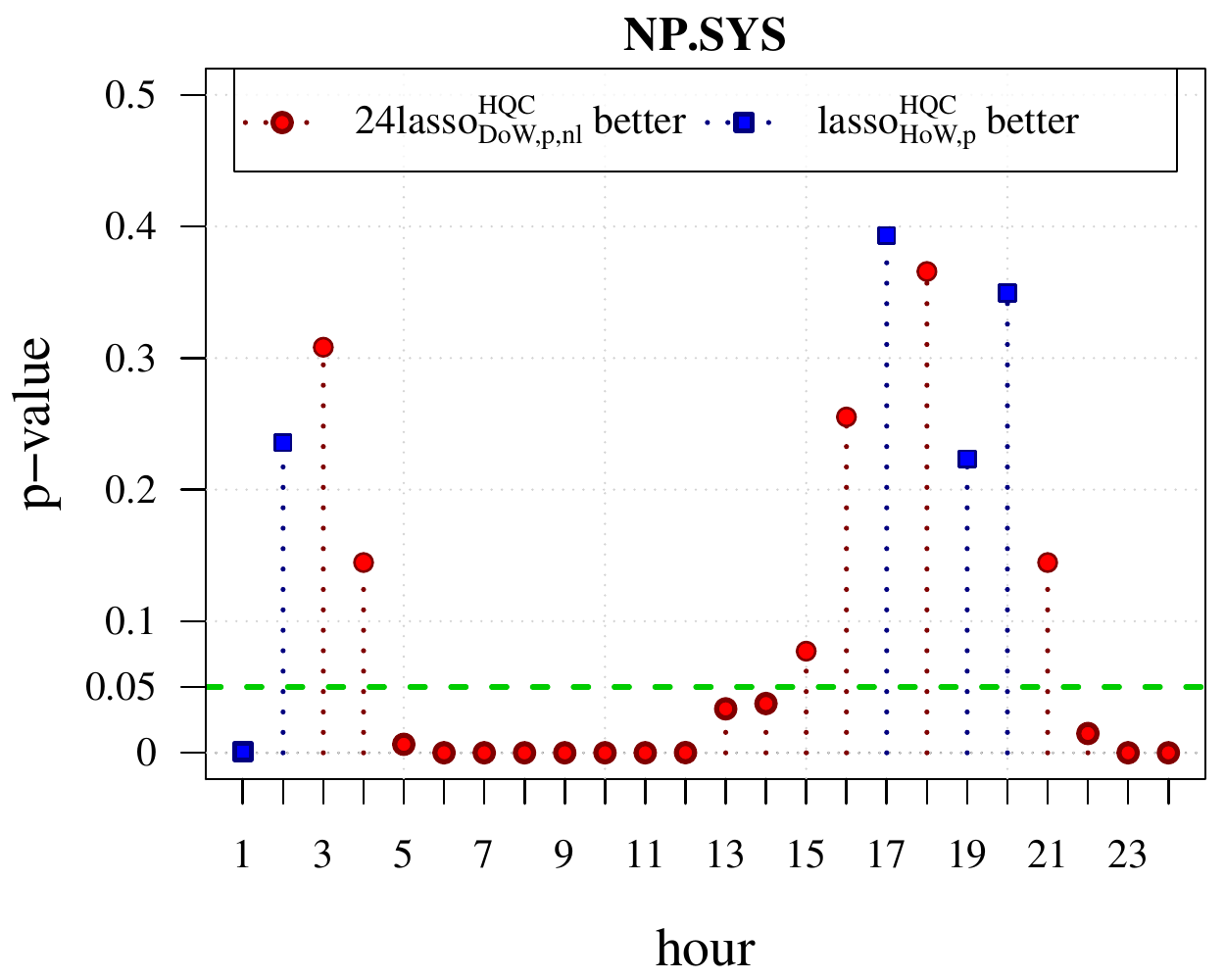} 
	\includegraphics[width=.32\textwidth, height=.22\textheight]{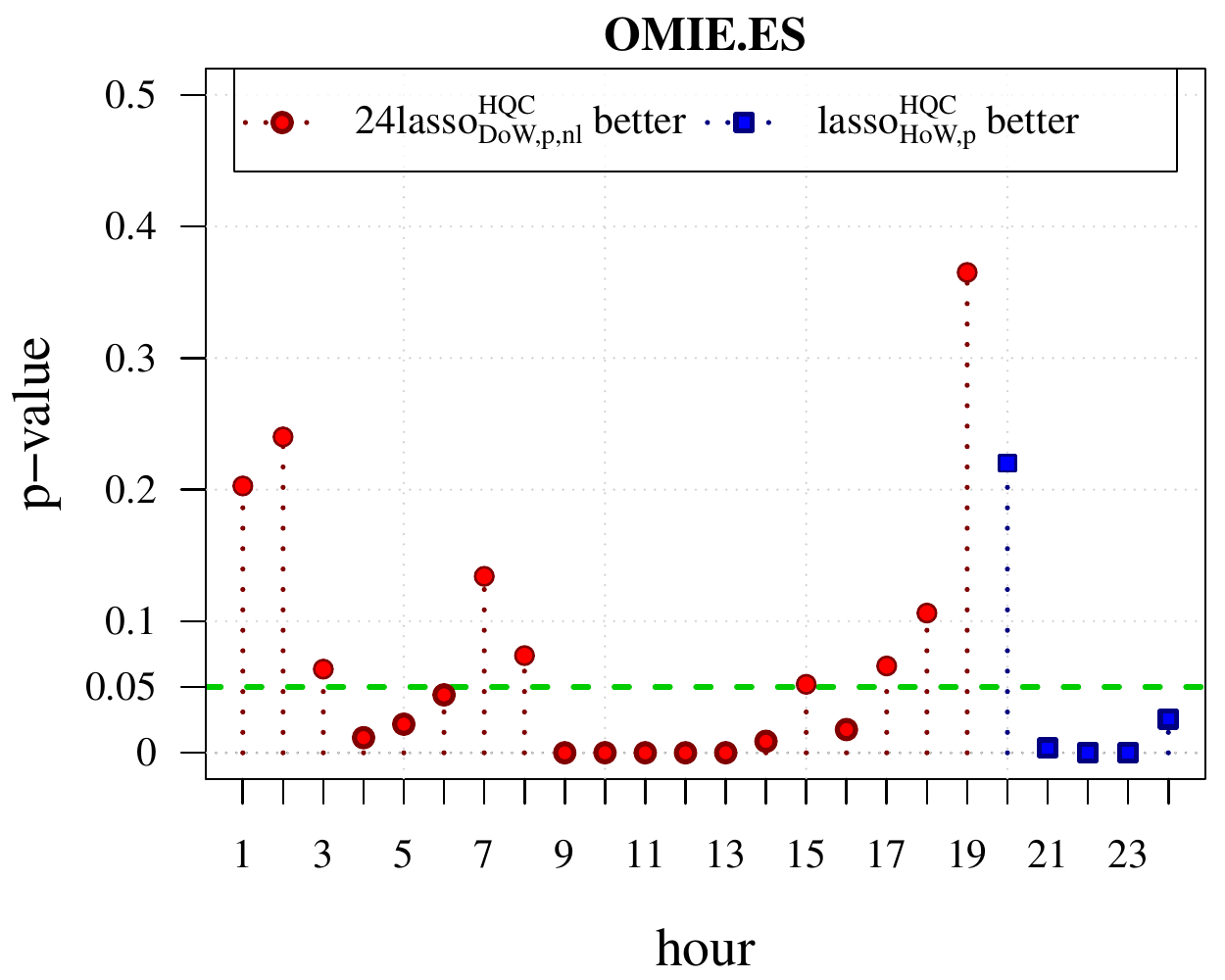} 
	\includegraphics[width=.32\textwidth, height=.22\textheight]{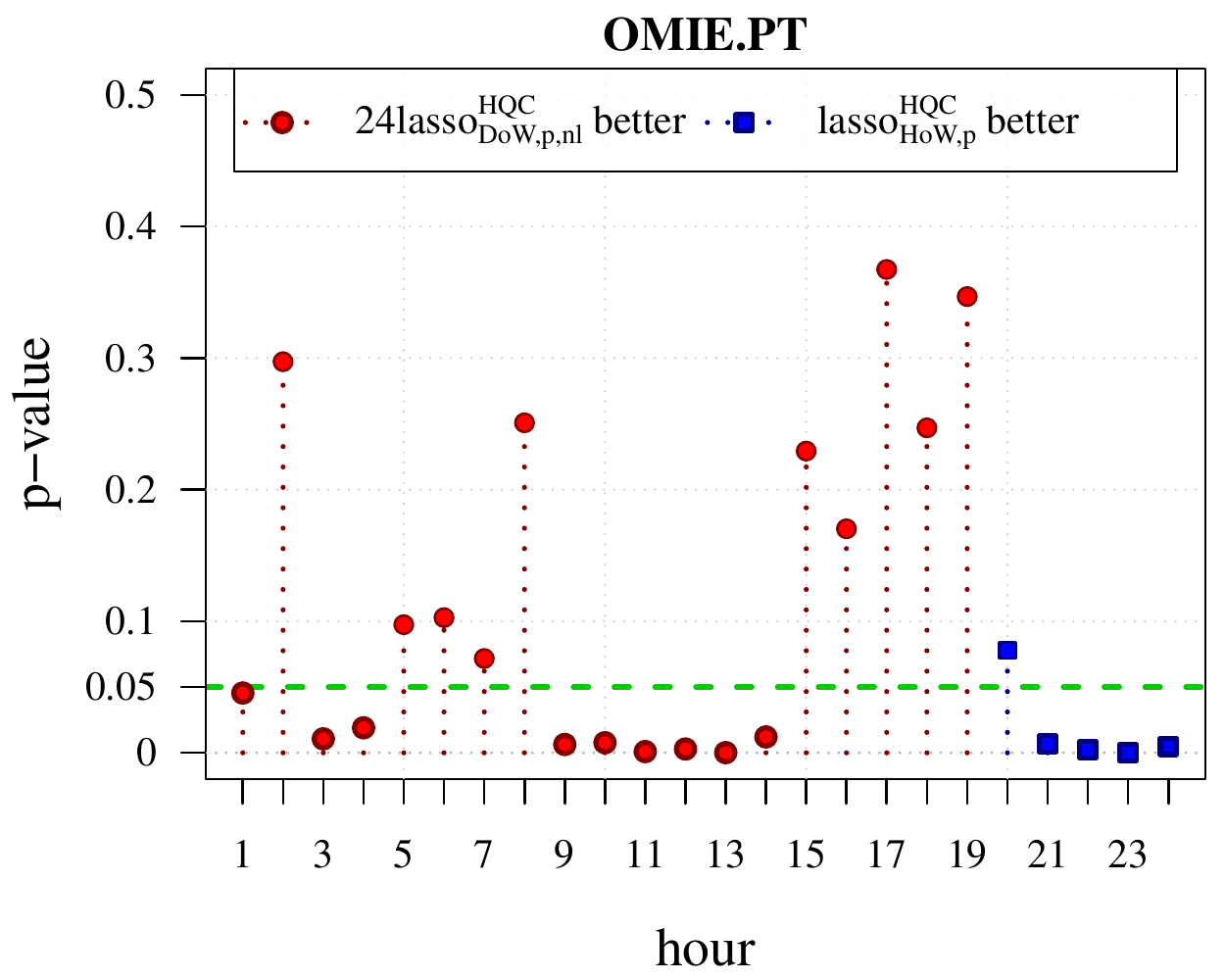} 
	\includegraphics[width=.32\textwidth, height=.22\textheight]{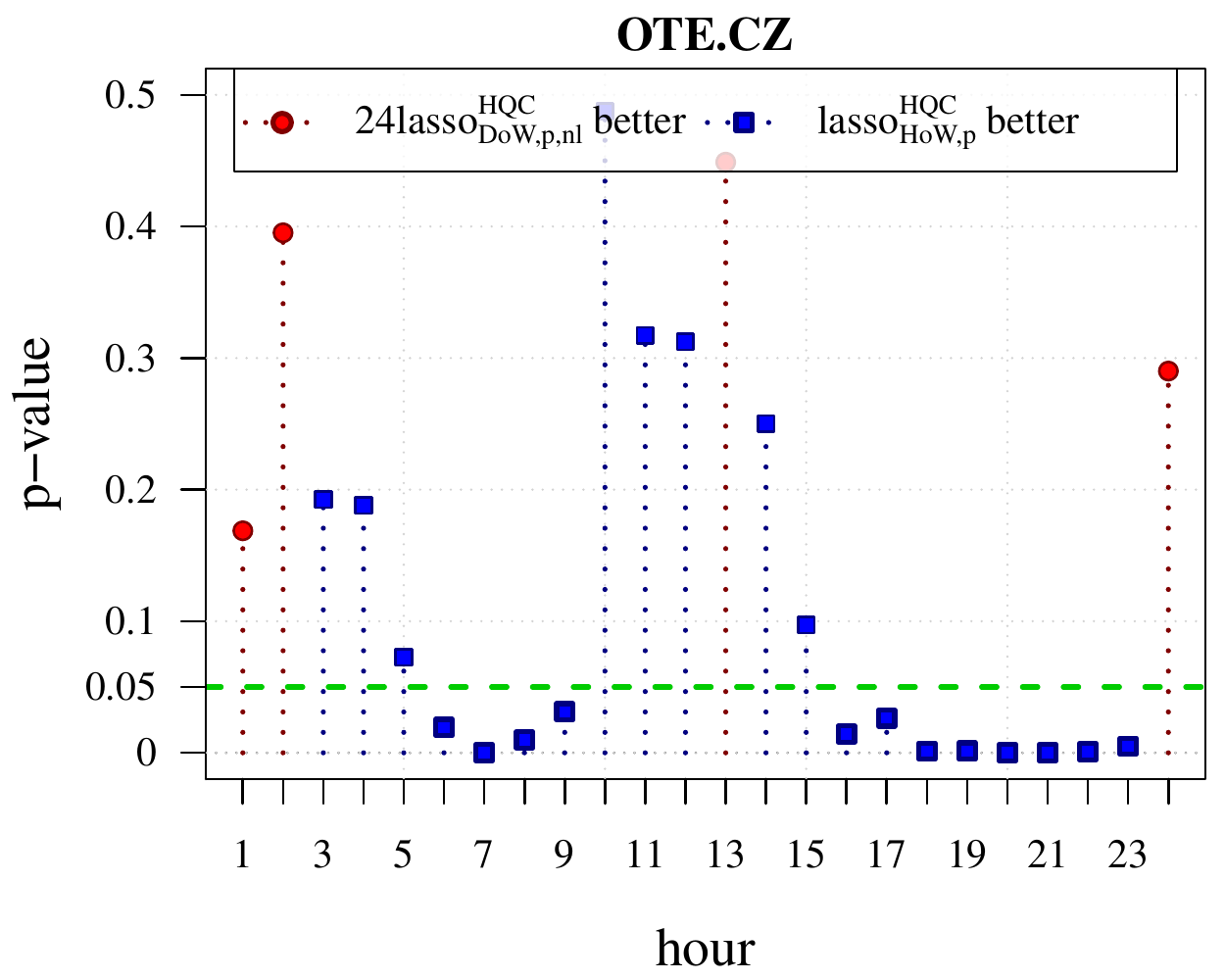} 
	
	\caption{Results of the 24 hourly DM tests, as defined by the loss differential series in Eqn.\ \eqref{eqn:Delta:DM:std}, for the best multivariate vs.\ the best univariate lasso models. We plot the $p$-values for the standard test with null $H_0$ (red circles; $\rightarrow$ \textbf{24lasso$_{\text{DoW,p,nl}}^{\text{HQC}}$} yields better forecasts) or the complementary test with reverse null $H^R_0$ (blue squares; $\rightarrow$  \textbf{lasso$_{\text{HoW,p}}^{\text{HQC}}$} yields better forecasts), whichever is smaller. The dashed green line represents the 5\% significance level.
	}
	\label{fig_dmtesth}
\end{figure}

In an attempt to better understand the performance of the models across the hours of the day, in Figure \ref{fig_dmtesth} we plot the $p$-values of the 24 hourly DM tests for the best multivariate vs.\ the best univariate lasso models. The Y-axis is capped at 0.5, so we either plot a red circle ($\rightarrow$ \textbf{24lasso$_{\text{DoW,p,nl}}^{\text{HQC}}$} yields better forecasts) or a blue square ($\rightarrow$  \textbf{lasso$_{\text{HoW,p}}^{\text{HQC}}$} yields better forecasts). Clearly, no universal daily pattern can be observed. If anything, more often the multivariate specification outperforms the univariate in the morning hours (for BELPEX.BE, EPEX.CH, EPEX.FR, OMIE.ES, OMIE.PT and except for hour 1 also for GEFCom2014), whereas the univariate more often outperforms the multivariate in the late evening/night hours (for BELPEX.BE, EPEX.DE+AT, EXAA.DE+AT, two Danish and two Iberian markets).
This is in contrast to what \cite{zie:16:TPWRS} concludes for relatively simple models from both model classes -- univariate models (like \textbf{AR$_{\text{HoW}}$}) perform better for the first half of the day, whereas similar structures within a multivariate framework (like \textbf{24AR$_{\text{HoW}}$}) are better in the second half of the day. This discrepancy may be due to the complexity (higher number of parameters in this study) or the calibration of the models (different information criteria; introduction of the HQC criterion in this study, which apparently outperforms AIC and BIC, see \ref{sec:App:ModelSelection}).

\begin{table}[tb]
	\caption{Mean Absolute Errors (MAE) for the better (\emph{ex-post}) of the two best lasso models, i.e., \textbf{24lasso$_{\text{DoW,p,nl}}^{\text{HQC}}$} and \textbf{lasso$_{\text{HoW,p}}^{\text{HQC}}$}, 
	and their arithmetic average. Note, that combining yields an improvement for all 12 datasets. {According to the multivariate DM-test, defined by Eqn.\ \eqref{eqn:Delta:DM}, it is significant at the 5\% level for eight markets (denoted by *).} 
	}
	\label{tab_MAE_combination}
	\centering
	\footnotesize
	\setlength{\tabcolsep}{4pt}
	\begin{tabular}{lrrrrrrrrrrrr}
		\toprule
		& \rotatebox{90}{BELPEX.BE}
		& \rotatebox{90}{EPEX.CH}
		& \rotatebox{90}{EPEX.DE+AT}
		& \rotatebox{90}{EPEX.FR}
		& \rotatebox{90}{EXAA.DE+AT}
		& \rotatebox{90}{GEFCom2014}
		& \rotatebox{90}{NP.DK1}
		& \rotatebox{90}{NP.DK2}
		& \rotatebox{90}{NP.SYS}
		& \rotatebox{90}{OMIE.ES}
		& \rotatebox{90}{OMIE.PT}
		& \rotatebox{90}{OTE.CZ} \\ 
		\midrule
Better of the two & 5.948 & 3.926 & 5.048 & 4.818 & 4.120 & 6.724 & 5.079 & 4.751 & 1.692 & 6.016 & 6.237 & 4.452 \\ 
  Combination & 5.844 & 3.891 & 5.007 & 4.757 & 4.070 & 6.637 & 5.045 & 4.713 & 1.687 & 5.985 & 6.193 & 4.414 \\ 
  Difference in MAE & 0.104 & 0.035 & 0.041 & 0.061 & 0.050 & 0.087 & 0.034 & 0.037 & 0.006 & 0.031 & 0.044 & 0.038 \\ 
  Improvement in \% & 1.75* & 0.89* & 0.81* & 1.27* & 1.22* & 1.30 & 0.67* & 0.79* & 0.33 & 0.52 & 0.71 & 0.85* \\ 
%{DM-test p-value in \%} & 0.00 & 0.64 & 0.19 & 0.00 & 0.00 & 25.61 & 0.34 & 0.08 & 58.24 & 8.05 & 5.32 & 0.27 \\
  \bottomrule
		\end{tabular}
\end{table}

Last but not least, the observation that for some hours of the day one model structure dominates another and vice versa gives grounds to believe (or expect) that combining their forecasts will yield further improvements. Although forecast combinations are not the focus of this study, we motivate this conclusion by considering a simple arithmetic average of the best two lasso models: \textbf{24lasso$_{\text{DoW,p,nl}}^{\text{HQC}}$} and \textbf{lasso$_{\text{HoW,p}}^{\text{HQC}}$}; for recent studies involving forecast averaging we refer to \cite{bor:bun:lis:nan:13}, \cite{now:rav:tru:wer:14}, \cite{wer:14}, \cite{rav:bou:dij:15}, \cite{gai:gou:ned:16}, \cite{mar:uni:wer:18} and \cite{now:wer:18}, among others. In Table \ref{tab_MAE_combination} we report the MAE values for the better of the two lasso models (this is in row `Better of the two') and their arithmetic average (row `Combination'). 
Clearly, the simple arithmetic average (the weights, i.e., $\frac12$ for each model, are obviously chosen \emph{ex-ante}) gives an 
improvement in forecasting accuracy (over the \emph{ex-post} selected lasso model) for all 12 datasets. {According to the multivariate DM-test, defined by Eqn.\ \eqref{eqn:Delta:DM}, it is significant at the 5\% level for  the majority of markets. The improvement of 1.30\% for the GEFCom2014 dataset is not significant probably due to the much shorter test period.}\\

\subsection{Variable selection}

In this section we analyze the structures of the best multivariate and univariate lasso models. Like \cite{uni:now:wer:16}, we count the number of times a given explanatory variable was selected (its coefficient is different form zero) for \textbf{24lasso$_{\text{DoW,p,nl}}^{\text{HQC}}$} (see Tables \ref{tab:param:multi:1}-\ref{tab:param:multi:4}) or \textbf{lasso$_{\text{HoW,p}}^{\text{HQC}}$} (see Table \ref{tab:param:uni}).
However, unlike \cite{uni:now:wer:16}, we do not present the numbers themselves, but the percentages (over all days in the out-of-sample test periods and across all 12 datasets). Heat maps are used to indicate more ($\rightarrow$ green) and less ($\rightarrow$ red) commonly-selected variables. Several interesting conclusions can be drawn.

In particular, for the multivariate lasso model, see Eqn.\ \eqref{eq_model1}, we observe that:
\begin{itemize}
	\itemsep0em
	\item For the autoregressive parameters, i.e., $\phi_{h,k,l,0}$, the most important is lag $k=1$, which refers to electricity prices of the previous day, see Table \ref{tab:param:multi:1} and compare with Tables \ref{tab:param:multi:2} and \ref{tab:param:multi:3}. For $k=1$ we observe that the diagonal elements, $\phi_{h,1,h,0}$, and the last hour of the previous day, $\phi_{h,1,24,0}$, are much more `green' than the remaining variables.	However, the diagonal elements seem to be less important during the working hours, whereas $\phi_{h,1,24,0}$ shows high relevance for all $h$. {This impact of hour 24 is likely the reason for the outperformance of the \textbf{24AR$_{\text{HoW}}$} model by \textbf{VAR$_{\text{HoW}}$}, see Table \ref{tab:MAE}. Even though both models have the same structure in terms of variables used, only \textbf{VAR$_{\text{HoW}}$} captures the `last hour of the day' effect across all hours.}

	\item For the less relevant autoregressive parameters at lag $k=1$, i.e., $\phi_{h,1,l,0}$, the lower triangle with $l < h$ carries most of the information. This is interesting because this triangle represents all prices $P_{d-1,l}$ which are closer (in time) to the predicted price, i.e., $P_{d,h}$, than the prices 24 hours ago, i.e., $P_{d-1,h}$, represented the by the `diagonal' in Table \ref{tab:param:multi:1}. This relationship seems to carry the most information for the night hours ($l=1,2,\ldots, 6$) and the evening hours ($l=18,19,\ldots, 24$).
	
	\item For the autoregressive parameters with lag $k>1$, see Tables \ref{tab:param:multi:1}-\ref{tab:param:multi:3}, generally only the diagonal elements $\phi_{h,k,h,0}$ show some importance. Seldom we observe a lag-importance  `island' (e.g., for $\phi_{h,2,5,0}$, $\phi_{h,2,23,0}$ or $\phi_{h,7,23,0}$ in the night hours) that could justify the complex model parametrization.
	
	\item The non-linear minimum and maximum effects, $\phi_{h,k,l,\text{min}}$ and $\phi_{h,k,l,\text{max}}$, are only relevant for $k=1$, i.e., the effect of previous day's minimum/maximum price, see Table \ref{tab:param:multi:4}. In general, it seems that the minimum is more important, especially in the first six hours of the day. In contrast, the maximum seems to be more important for the late morning hours ($h=8,9,10$). To some extent, these temporal differences are also visible in Tables 2 and 3 in \cite{uni:now:wer:16}. Interestingly, they conclude that the maximum `is slightly more influential' than the minimum, however, both in our and their study the differences are rather small.
	
	\item The day-of-the-week dummies, i.e., $\phi_{h,0,0,j}$, are in general very important, especially the Monday, Saturday and Sunday ($j=1,6,7$) dummies, see Table \ref{tab:param:multi:4}. This means that the commonly used design of expert models is appropriate \cite[see e.g.][]{mis:tru:wer:06,wer:mis:08,ser:11,kri:12,now:rav:tru:wer:14,gai:gou:ned:16,mac:now:wer:16,now:wer:18,uni:now:wer:16,zie:16:TPWRS}.
	The Tuesday and Friday ($j=2,5$) dummies are less important, with the latter one only for the evening hours. 
	
	\item The periodic parameters, $\phi_{h,1,h,j}$ and $\phi_{h,1,24,j}$, exhibit relevance, see Table \ref{tab:param:multi:4}. However, in contrast to the day-of-the-week dummies ($\phi_{h,0,0,j}$), $\phi_{h,1,h,j}$'s are more important during the working days, especially Tuesday, Wednesday and Friday ($j=3,4,5$).
\end{itemize}
For the univariate lasso model, see Eqn.\ \eqref{eq_model_uni}, we observe that:
\begin{itemize}
	\itemsep0em
	\item The autoregressive parameters which model the dependency on the previous hour, i.e., $\phi_{1,k}$, have a clear pattern. Lags around multiples of 24 are very important, e.g. 23, 24, 25 and 26 or 47, 48, 49 and 50. However, the remaining lags exhibit moderate importance and only for the first 24 hours, see Table \ref{tab:param:uni}.

	\item For the intercepts and periodic parameters, i.e.,  $\phi_{0,k}$, $\phi_{2,k}$ and $\phi_{3,k}$, the patterns are not that obvious. However, almost every parameter seems to have relevance at least for some markets (as there are almost no red backgrounds). We also see that lags being a multiple of 24 tend to have more importance than the other parameters, which is most clearly visible for the intercepts: $\phi_{0,24}$, $\phi_{0,48}$, etc. 
	
	\item Another interesting observation is that there is another group of lag-importance `islands'. These occur at lags of order $7$ and $8$, e.g. 24+7=31 and 24+8=32 or 48+7=55 and 48+8=56.
\end{itemize}
Finally, we should note that for the multivariate model the non-linear effects exhibit moderate importance and increase the overall model fit. However, for the univariate model, the non-linear effects do not improve the model performance. The reason may be that we only use two parameters to capture the minimum/maximum effects, even though the effect seems to be periodic. An introduction of non-linear periodic effects in the univariate setting may lead to a further improvement of the predictive accuracy.

\begin{table}[p]
	\centering
	\setlength{\tabcolsep}{1pt}
	\caption{Mean occurrence (in \%) of the multivariate lasso model parameters across all 12 datasets and the full out-of-sample test period. Columns represent the hours and rows the parameters of the  \textbf{24lasso$_{\text{DoW,p,nl}}^{\text{HQC}}$} model, see Eqn.\ \eqref{eq_model1} for details.	A heat map is used to indicate more ($\rightarrow$ green) and less ($\rightarrow$ red) commonly-selected variables. Continued in Table \ref{tab:param:multi:2}.
	}
	\label{tab:param:multi:1}
	\tiny
\begin{tabular}{|rrrrrrrrrrrrrrrrrrrrrrrrrr|}
		\hline
	&	$h$ & 1 & 2 & 3 & 4 & 5 & 6 & 7 & 8 & 9 & 10 & 11 & 12 & 13 & 14 & 15 & 16 & 17 & 18 & 19 & 20 & 21 & 22 & 23 & 24 \\
		\hline
 \parbox[t]{2mm}{\multirow{24}{*}{\rotatebox[origin=c]{90}{\textbf{Day $(d-1)$}}}}&		$\phi_{h,1,1,0}$ & \cellcolor[rgb]{1,0.543,0.3} 21.42 & \cellcolor[rgb]{1,0.37,0.3} 10.61 & \cellcolor[rgb]{1,0.345,0.3} 9.04 & \cellcolor[rgb]{1,0.549,0.3} 21.81 & \cellcolor[rgb]{1,0.545,0.3} 21.58 & \cellcolor[rgb]{1,0.383,0.3} 11.41 & \cellcolor[rgb]{1,0.402,0.3} 12.62 & \cellcolor[rgb]{1,0.409,0.3} 13.04 & \cellcolor[rgb]{1,0.393,0.3} 12.07 & \cellcolor[rgb]{1,0.342,0.3} 8.86 & \cellcolor[rgb]{1,0.342,0.3} 8.89 & \cellcolor[rgb]{1,0.338,0.3} 8.62 & \cellcolor[rgb]{1,0.314,0.3} 7.10 & \cellcolor[rgb]{1,0.335,0.3} 8.43 & \cellcolor[rgb]{1,0.298,0.3} 6.13 & \cellcolor[rgb]{1,0.295,0.3} 5.93 & \cellcolor[rgb]{1,0.251,0.3} 3.17 & \cellcolor[rgb]{1,0.215,0.3} 0.92 & \cellcolor[rgb]{1,0.226,0.3} 1.63 & \cellcolor[rgb]{1,0.252,0.3} 3.26 & \cellcolor[rgb]{1,0.278,0.3} 4.87 & \cellcolor[rgb]{1,0.347,0.3} 9.18 & \cellcolor[rgb]{1,0.413,0.3} 13.31 & \cellcolor[rgb]{1,0.518,0.3} 19.85 \\
\hline
	\end{tabular}
\end{table}

\begin{table}[p]
	\centering
	\setlength{\tabcolsep}{1pt}
	\caption{Mean occurrence (in \%) of the multivariate lasso model parameters across all 12 datasets and the full out-of-sample test period. Columns represent the hours and rows the parameters of the  \textbf{24lasso$_{\text{DoW,p,nl}}^{\text{HQC}}$} model, see Eqn.\ \eqref{eq_model1} for details.	A heat map is used to indicate more ($\rightarrow$ green) and less ($\rightarrow$ red) commonly-selected variables. Continued in Table \ref{tab:param:multi:3}.
	}

	\label{tab:param:multi:2}
	\tiny
	\begin{tabular}{|rrrrrrrrrrrrrrrrrrrrrrrrrr|}
		\hline
	&	$h$ & 1 & 2 & 3 & 4 & 5 & 6 & 7 & 8 & 9 & 10 & 11 & 12 & 13 & 14 & 15 & 16 & 17 & 18 & 19 & 20 & 21 & 22 & 23 & 24 \\
		\hline
\parbox[t]{2mm}{\multirow{24}{*}{\rotatebox[origin=c]{90}{\textbf{Day $(d-4)$}}}}&		$\phi_{h,4,1,0}$ & \cellcolor[rgb]{1,0.313,0.3} 7.04 & \cellcolor[rgb]{1,0.255,0.3} 3.46 & \cellcolor[rgb]{1,0.265,0.3} 4.05 & \cellcolor[rgb]{1,0.262,0.3} 3.88 & \cellcolor[rgb]{1,0.282,0.3} 5.11 & \cellcolor[rgb]{1,0.298,0.3} 6.11 & \cellcolor[rgb]{1,0.284,0.3} 5.25 & \cellcolor[rgb]{1,0.257,0.3} 3.59 & \cellcolor[rgb]{1,0.286,0.3} 5.40 & \cellcolor[rgb]{1,0.291,0.3} 5.71 & \cellcolor[rgb]{1,0.301,0.3} 6.34 & \cellcolor[rgb]{1,0.291,0.3} 5.70 & \cellcolor[rgb]{1,0.323,0.3} 7.68 & \cellcolor[rgb]{1,0.288,0.3} 5.49 & \cellcolor[rgb]{1,0.288,0.3} 5.49 & \cellcolor[rgb]{1,0.248,0.3} 3.00 & \cellcolor[rgb]{1,0.254,0.3} 3.37 & \cellcolor[rgb]{1,0.252,0.3} 3.27 & \cellcolor[rgb]{1,0.261,0.3} 3.79 & \cellcolor[rgb]{1,0.206,0.3} 0.36 & \cellcolor[rgb]{1,0.251,0.3} 3.20 & \cellcolor[rgb]{1,0.302,0.3} 6.37 & \cellcolor[rgb]{1,0.344,0.3} 9.02 & \cellcolor[rgb]{1,0.546,0.3} 21.62 \\
\hline
\end{tabular}
\end{table}

\begin{table}[p]
	\centering
	\setlength{\tabcolsep}{1pt}
	\caption{Mean occurrence (in \%) of the multivariate lasso model parameters across all 12 datasets and the full out-of-sample test period. Columns represent the hours and rows the parameters of the  \textbf{24lasso$_{\text{DoW,p,nl}}^{\text{HQC}}$} model, see Eqn.\ \eqref{eq_model1} for details.	A heat map is used to indicate more ($\rightarrow$ green) and less ($\rightarrow$ red) commonly-selected variables. Continued in Table \ref{tab:param:multi:4}.
	}
	\label{tab:param:multi:3}
	\tiny
	\begin{tabular}{|rrrrrrrrrrrrrrrrrrrrrrrrrr|}
		\hline
	&	$h$ & 1 & 2 & 3 & 4 & 5 & 6 & 7 & 8 & 9 & 10 & 11 & 12 & 13 & 14 & 15 & 16 & 17 & 18 & 19 & 20 & 21 & 22 & 23 & 24 \\
		\hline
\parbox[t]{2mm}{\multirow{24}{*}{\rotatebox[origin=c]{90}{\textbf{Day $(d-7)$}}}}&		$\phi_{h,7,1,0}$ & \cellcolor[rgb]{1,0.322,0.3} 7.63 & \cellcolor[rgb]{1,0.281,0.3} 5.08 & \cellcolor[rgb]{1,0.253,0.3} 3.28 & \cellcolor[rgb]{1,0.387,0.3} 11.69 & \cellcolor[rgb]{1,0.329,0.3} 8.09 & \cellcolor[rgb]{1,0.333,0.3} 8.32 & \cellcolor[rgb]{1,0.503,0.3} 18.94 & \cellcolor[rgb]{1,0.524,0.3} 20.24 & \cellcolor[rgb]{1,0.401,0.3} 12.53 & \cellcolor[rgb]{1,0.298,0.3} 6.10 & \cellcolor[rgb]{1,0.295,0.3} 5.94 & \cellcolor[rgb]{1,0.329,0.3} 8.06 & \cellcolor[rgb]{1,0.311,0.3} 6.96 & \cellcolor[rgb]{1,0.315,0.3} 7.17 & \cellcolor[rgb]{1,0.404,0.3} 12.78 & \cellcolor[rgb]{1,0.376,0.3} 10.98 & \cellcolor[rgb]{1,0.378,0.3} 11.13 & \cellcolor[rgb]{1,0.326,0.3} 7.86 & \cellcolor[rgb]{1,0.281,0.3} 5.03 & \cellcolor[rgb]{1,0.386,0.3} 11.65 & \cellcolor[rgb]{1,0.414,0.3} 13.40 & \cellcolor[rgb]{1,0.306,0.3} 6.63 & \cellcolor[rgb]{1,0.366,0.3} 10.35 & \cellcolor[rgb]{1,0.369,0.3} 10.54 \\
\hline
\end{tabular}
\end{table}

\begin{table}[p]
	\centering
	\setlength{\tabcolsep}{1pt}
	\caption{Mean occurrence (in \%) of the multivariate lasso model parameters across all 12 datasets and the full out-of-sample test period. Columns represent the hours and rows the parameters of the  \textbf{24lasso$_{\text{DoW,p,nl}}^{\text{HQC}}$} model, see Eqn.\ \eqref{eq_model1} for details.	A heat map is used to indicate more ($\rightarrow$ green) and less ($\rightarrow$ red) commonly-selected variables.
	}
	\label{tab:param:multi:4}
	\tiny
	\begin{tabular}{|rrrrrrrrrrrrrrrrrrrrrrrrrr|}
		\hline
	&	$h$ & 1 & 2 & 3 & 4 & 5 & 6 & 7 & 8 & 9 & 10 & 11 & 12 & 13 & 14 & 15 & 16 & 17 & 18 & 19 & 20 & 21 & 22 & 23 & 24 \\
		\hline
\parbox[t]{2mm}{\multirow{8}{*}{\rotatebox[origin=c]{90}{\textbf{Daily minimums}}}}&		$\phi_{h,1,\text{min},0}$ & \cellcolor[rgb]{0.888,1,0.3} 57.99 & \cellcolor[rgb]{0.774,1,0.3} 66.16 & \cellcolor[rgb]{0.568,1,0.3} 80.84 & \cellcolor[rgb]{0.474,1,0.3} 87.56 & \cellcolor[rgb]{0.561,1,0.3} 81.33 & \cellcolor[rgb]{0.941,1,0.3} 54.23 & \cellcolor[rgb]{1,0.616,0.3} 26.01 & \cellcolor[rgb]{1,0.611,0.3} 25.68 & \cellcolor[rgb]{1,0.528,0.3} 20.48 & \cellcolor[rgb]{1,0.404,0.3} 12.72 & \cellcolor[rgb]{1,0.311,0.3} 6.94 & \cellcolor[rgb]{1,0.293,0.3} 5.83 & \cellcolor[rgb]{1,0.287,0.3} 5.42 & \cellcolor[rgb]{1,0.271,0.3} 4.45 & \cellcolor[rgb]{1,0.312,0.3} 7.03 & \cellcolor[rgb]{1,0.289,0.3} 5.54 & \cellcolor[rgb]{1,0.32,0.3} 7.51 & \cellcolor[rgb]{1,0.292,0.3} 5.76 & \cellcolor[rgb]{1,0.253,0.3} 3.28 & \cellcolor[rgb]{1,0.227,0.3} 1.68 & \cellcolor[rgb]{1,0.288,0.3} 5.52 & \cellcolor[rgb]{1,0.247,0.3} 2.91 & \cellcolor[rgb]{1,0.248,0.3} 3.02 & \cellcolor[rgb]{1,0.312,0.3} 7.03 \\
\hline 
\end{tabular}
\end{table}

\begin{table}[p]
	\centering
	\caption{Mean occurrence (in \%) of the univariate lasso model parameters across all 12 datasets and the full out-of-sample test period. The columns represent the intercept ($\phi_{0,*}$), the autoregressive terms ($\phi_{1,*}$), the periodic effects of lag $1$ ($\phi_{2,*}$) and the periodic effects of lag $24$ ($\phi_{3,*}$) of the \textbf{lasso$_{\text{HoW,p}}^{\text{HQC}}$} model, see Eqn.\ \eqref{eq_model_uni} for details. Note, that this model does not include the non-linear effects, i.e., $\phi_{4,1}=\phi_{4,2}=0$. A heat map is used to indicate more ($\rightarrow$ green) and less ($\rightarrow$ red) commonly-selected variables.
	}
	\label{tab:param:uni}
	\tiny
\begin{tabular}{|rrrrrr|rrrrrr|rrrrrr|}
		\hline
		& $k$ & $\phi_{0,k}$ &  $\phi_{1,k}$ &$\phi_{2,k}$ & $\phi_{3,k}$ &&
		$k$ & $\phi_{0,k}$ &  $\phi_{1,k}$ &$\phi_{2,k}$ & $\phi_{3,k}$ &&
		$k$ & $\phi_{0,k}$ &  $\phi_{1,k}$ &$\phi_{2,k}$ & $\phi_{3,k}$ \\
		\hline
	\parbox[t]{0mm}{\multirow{24}{*}{\rotatebox[origin=c]{90}{\textbf{Day $(d-1)$}}}} & 1 & \cellcolor[rgb]{0.484,1,0.3} 86.83 & \cellcolor[rgb]{0.3,0.8,0.3} 100 & \cellcolor[rgb]{0.761,1,0.3} 67.10 & \cellcolor[rgb]{0.747,1,0.3} 68.08 & \parbox[t]{0mm}{\multirow{24}{*}{\rotatebox[origin=c]{90}{\textbf{Day $(d-4)$}}}} & 73 & \cellcolor[rgb]{0.604,1,0.3} 78.28 & \cellcolor[rgb]{0.372,0.914,0.3} 94.83 & \cellcolor[rgb]{0.642,1,0.3} 75.55 & \cellcolor[rgb]{0.859,1,0.3} 60.06 & \parbox[t]{0mm}{\multirow{24}{*}{\rotatebox[origin=c]{90}{\textbf{Day $(d-7)$}}}} & 145 & \cellcolor[rgb]{0.66,1,0.3} 74.26 & \cellcolor[rgb]{0.397,0.953,0.3} 93.06 & \cellcolor[rgb]{0.635,1,0.3} 76.05 & \cellcolor[rgb]{0.861,1,0.3} 59.95 \\
		&2 & \cellcolor[rgb]{0.551,1,0.3} 82.06 & \cellcolor[rgb]{0.609,1,0.3} 77.90 & \cellcolor[rgb]{1,0.993,0.3} 49.57 & \cellcolor[rgb]{0.95,1,0.3} 53.58 && 74 & \cellcolor[rgb]{0.98,1,0.3} 51.43 & \cellcolor[rgb]{0.733,1,0.3} 69.06 & \cellcolor[rgb]{0.963,1,0.3} 52.67 & \cellcolor[rgb]{1,0.86,0.3} 41.26 && 146 & \cellcolor[rgb]{0.924,1,0.3} 55.42 & \cellcolor[rgb]{0.55,1,0.3} 82.13 & \cellcolor[rgb]{1,0.93,0.3} 45.64 & \cellcolor[rgb]{1,0.843,0.3} 40.19 \\
		&3 & \cellcolor[rgb]{1,0.936,0.3} 45.99 & \cellcolor[rgb]{1,0.969,0.3} 48.08 & \cellcolor[rgb]{0.826,1,0.3} 62.44 & \cellcolor[rgb]{1,0.867,0.3} 41.66 && 75 & \cellcolor[rgb]{1,0.763,0.3} 35.19 & \cellcolor[rgb]{1,0.377,0.3} 11.07 & \cellcolor[rgb]{0.763,1,0.3} 66.94 & \cellcolor[rgb]{1,0.785,0.3} 36.55 && 147 & \cellcolor[rgb]{0.842,1,0.3} 61.26 & \cellcolor[rgb]{1,0.431,0.3} 14.44 & \cellcolor[rgb]{0.972,1,0.3} 51.99 & \cellcolor[rgb]{1,0.605,0.3} 25.31 \\
\hline
	\end{tabular}
\end{table}

\section{Conclusions and guidelines for energy forecasters}
\label{sec:Conclusions}

%Given that there is no consensus in the existing literature as to the optimal model structure for short-term electricity price forecasting, we have conducted an extensive empirical study, involving datasets from 12 power markets, state-of-the-art parsimonious expert models, univariate and multivariate autoregressive benchmarks,  multi-parameter regression models estimated via the lasso and formal statistical testing for significance using two variants of the \cite{die:mar:95} test. The aim was to address three pertinent questions put forward in the Introduction, which concern the optimality of the multivariate vs. univariate modeling framework.

{We have conducted an extensive empirical study on short-term electricity price forecasting (EPF) to address the long-standing question if the optimal model structure for EPF is univariate or multivariate.} We provide evidence that despite a minor edge in predictive performance overall, as measured by the linear (MAE) and quadratic (RMSE) error measures and the mean percentage deviation from the best performing model (m.p.d.f.b.), the multivariate modeling approach does not uniformly outperform the univariate one across all datasets, seasons of the year or hours of the day, and at times is outperformed by the latter. These fluctuations in forecasting performance across the hours of the day can be utilized, however, via model averaging or combining forecasts. As illustrated in the paper, a simple arithmetic average of the forecasts of the best multivariate and the best univariate lasso model beats the better (\emph{ex-post}) of the two for all 12 considered datasets. 

When analyzing model performance in the four seasons of the year, we find that the annual ranking of the models is preserved in the Spring, Summer and Winter. However, in the Fall the univariate models have an edge over the multivariate lasso models. A plausible explanation may be that in the majority of analyzed markets the electricity prices and their volatility tend to increase towards the end of the calendar year. The univariate models, by taking into account all hourly prices in the past week, seem to be able to adapt quicker to the increasing prices. Interestingly, team TOLOLO used a similar approach to win the Price Track of the GEFCom2014 competition \cite[see][]{gai:gou:ned:16}. For three tasks corresponding to days in December (i.e., late Fall/early Winter) they used a specifically designed model, different from the models for the remaining tasks (corresponding to Summer months). These results suggest that the predictive efficiency may be further increased by designing different models for 
different seasons of the year.

Also regarding performance across the markets, there is some variability in forecasting accuracy, which may be a result of differences in the generation mix or market regulations. The latter may be accounted for when constructing fundamental models \cite[like the X-model of][that focuses on modeling the bidding behavior]{zie:ste:16}, however, we do not see a straightforward way of incorporating such information into statistical models. Yet, it is always one of the multivariate lasso models estimated using the HQC criterion or the univariate lasso model with periodic effects, also estimated using the HQC criterion, that yields the best performance. Given that the three multivariate lasso models never perform badly, they are recommended for EPF in general.

Concerning variable (or feature) selection, analyzing the structures of the best multivariate and univariate lasso models, we have identified the most important variables and thus provided guidelines to structuring better performing expert models. In particular, we have confirmed the high explanatory power of last day's prices for the same or neighboring hours, of last day's prices for midnight and of the price for the same hour a week earlier. However, more importantly, we have found the periodic effects (daily dummies multiplied by the last day's prices for the same hour or for midnight) to play a very important role. Hence, like \cite{uni:now:wer:16}, we strongly suggest to incorporate periodic structures not only in expert models, but also in general model designs. 

Finally, we should comment on converting univariate models into multivariate and vice versa. This possibility can give ideas for modeling approaches in the `other modeling world'. For instance, if we observe a clear effect for a particular parameter in the multivariate setting, then the corresponding effect should be important {for} univariate modeling as well. However, when rewriting multivariate models into univariate form and vice versa, the error structures change. So implicitly, in a multivariate specification the residual variance is assumed to be different for the 24 models, whereas for the univariate models the considered estimation methods assume that the variance is the same across all hours. Here, iterative reweighting schemes can help to incorporate variance changing effects, especially for univariate approaches \cite[see e.g.][]{zie:ste:hus:15,zie:16:CSDA}.

\appendix
\section{The set of models}
\label{sec:App:AllModels}

In this Appendix we define the remaining models from classes C3-C8 that were considered in the empirical study, but are not discussed and compared in Sections \ref{sec:Models} and \ref{sec:Empirical}. They are briefly evaluated in terms of WMAE and m.p.d.f.b.\ in \ref{sec:App:ModelSelection} below.

\subsection{Expert models (class C3)}
\label{ssec:App:ExpertModels}

For the asinh-transformed price on day $d$ and hour $h$, the generic class of expert models used in this paper is defined by:
\begin{align}
Y_{d,h} = & ~~\beta_{h,1} + \underbrace{\beta_{h,2} Y_{d-1,h} + \beta_{h,3} Y_{d-2,h} + \beta_{h,4} Y_{d-7,h}}_{\text{autoregressive effects}} + \underbrace{\beta_{h,5} Y_{d-1,\min} + \beta_{h,6} Y_{d-1,\max}}_{\text{non-linear effects}} +  \beta_{h,7} Y_{d-1,24} \nonumber \\ 
&+ \underbrace{\sum_{j=1}^7 \beta_{h,7+j} \text{DoW}^{j}_{d,h}}_{\text{weekday dummies}}
+ \underbrace{\sum_{j=1}^7 \beta_{h,14+j} \text{DoW}^{j}_{d,h} Y_{d-1,h} 
	+ \sum_{j=1}^7 \beta_{h,21+j} \text{DoW}^{j}_{d,h} Y_{d-1,24}}_{\text{periodic effects}} + \eps_{d,h}.
\label{eq_expert-wd}  
\end{align}
Note, that due to collinearity $\beta_{h,14}$, $\beta_{h,21}$ and $\beta_{h,28}$ can be dropped, so the model has effectively 25 parameters (for $h=24$ only 18). We estimate the parameters using OLS.

Such a model is denoted by \textbf{expert$_{\text{DoW,p,nl}}$}. Compared to the \textbf{expert$_{\text{DoW,nl}}$} model considered in Sections \ref{sec:Models} and \ref{sec:Empirical}, the above formula additionally includes \emph{periodic effects} (hence subscript \textbf{p} in the model name). The second sum in Eqn.\ \eqref{eq_expert-wd} is a consequence of the experience gained by TEAM POLAND during the GEFCom2014 competition; one of the conclusions of \cite{mac:now:16} was that it could be beneficial to use different model structures for different days of the week, not only different parameter sets. The third sum in Eqn.\ \eqref{eq_expert-wd} is an innovation introduced in this study to allow for different model structures including the previous day's price at midnight, i.e., $Y_{d-1,24}$, for different days of the week.
We consider three special cases of the \textbf{expert$_{\text{DoW,p,nl}}$} model:
\begin{enumerate}
	\setcounter{enumi}{1}
	\item \textbf{expert$_{\text{DoW,nl}}$} without periodic effects (used in Sections \ref{sec:Models} and \ref{sec:Empirical}),
	
	\item \textbf{expert$_{\text{DoW,p}}$} without non-linear effects, 
	
	\item \textbf{expert$_{\text{DoW}}$} without periodic and non-linear effects.
\end{enumerate}
Furthermore, if we restrict the three sums in the second row of Eqn.\ \eqref{eq_expert-wd} to sum only over Monday, Saturday and Sunday, i.e., $j=1,6,7$, then we receive the next four experts:
\begin{enumerate}
	\setcounter{enumi}{4}
	\item \textbf{expert$_{\text{p,nl}}$} -- the full model,
	
	\item \textbf{expert$_{\text{nl}}$} without periodic effects,
	
	\item \textbf{expert$_{\text{p}}$} without non-linear effects, 
	
	\item \textbf{expert} without periodic and non-linear effects.
\end{enumerate}
There is an important difference between models defined in Eqn.\ \eqref{eq_expert-wd} and the expert models considered by \cite{uni:now:wer:16}. In the latter article no intercept is included in the formulas, i.e.,  $\beta_{h,1}=0$. Instead $Y_{d,h}$ is de-meaned by the daily mean of hourly frequency, i.e., $\ov{Y}_{\text{HoD},d,h}$ in our notation of Section \ref{ssec:Dummies}.
Replacing $Y_{d,h}$ by $(Y_{d,h} - \ov{Y}_{\text{HoD},d,h})$ and removing the intercept $\beta_{h,1}$ in Eqn.\ \eqref{eq_expert-wd} yields the corresponding models. We denote them by superscript $*$:
\begin{enumerate}
	\setcounter{enumi}{8}
	\item[9-10.] \textbf{expert$_{\text{DoW,p,nl}}^*$} and \textbf{expert$_{\text{p,nl}}^*$} -- the full models,
	
	\item[11-12.] \textbf{expert$_{\text{DoW,nl}}^*$} and \textbf{expert$_{\text{nl}}^*$} without periodic effects,
	
	\item[13-14.] \textbf{expert$_{\text{DoW,p}}^*$} and \textbf{expert$_{\text{p}}^*$} without non-linear effects, 
	
	\item[15-16.] \textbf{expert$_{\text{DoW}}^*$} and \textbf{expert$^*$} without periodic and non-linear effects.
\end{enumerate}

\subsection{The second 24AR-type model (class C4)}
\label{ssec:App:24AR}

As an alternative to the \textbf{24AR$_{\text{HoW}}$} model defined in Section \ref{sssec:24AR}, we consider a specification in which the asinh-transformed price is demeaned with respect to $\ov{Y}_{\text{HoD},d,h}$, not $\ov{Y}_{\text{HoW},d,h}$, and modeled as an AR($p_h$) process independently for each hour $h$. The resulting \textbf{24AR$_{\text{HoD}}$} model is given by:
\begin{equation}
Y_{d,h} = \ov{Y}_{\text{HoD},d,h} + \phi_{0,h} +  \sum_{k=1}^{p_h} \phi_{k,h}( Y_{d-k,h} - \ov{Y}_{\text{HoD},d,h}) + \eps_{d,h},
\label{eq_24AR}
\end{equation}
and estimated analogously to \textbf{24AR$_{\text{HoW}}$}.

\subsection{The second VAR-type model (class C5)}
\label{ssec:App:VAR}

As an alternative to the \textbf{VAR$_{\text{HoW}}$} model defined in Section \ref{sssec:VAR}, we consider the \textbf{VAR$_{\text{HoD}}$} model:
\begin{equation}
\bsY_d = \ov{\bsY}_{\text{HoD},d} + \bsphi_0 + \sum_{k=1}^{p} \bsPhi_k (\bsY_{d-k} - \ov{\bsY}_{\text{HoD},d}) + \bseps_d,
\label{eq_VAR_HoD}
\end{equation}
where $\bsY_d = [Y_{d,1}, \ldots, Y_{d,24}]'$ with its mean vector $\ov{\bsY}_{\text{HoD},d}$ across all available days in the calibration sample (which corresponds to the 24 possible values of $\ov{Y}_{\text{HoD},d,h}$). Analogously to \textbf{VAR$_{\text{HoW}}$}, we calibrate the model by solving the multivariate Yule-Walker equations with $p_{\max}=8$.

\subsection{Multivariate lasso models (class C6)}
\label{ssec:App:24lasso}

Based on Eqn.\ \eqref{eq_model1} and using certain restrictions, we can define the full set of 16 models:
\begin{enumerate}
	\itemsep 0mm
	\item[1-4.] \textbf{24lasso$_{\text{DoW,p,nl}}^{\text{IC}}$} -- the full model,
	
	\item[5-8.] \textbf{24lasso$_{\text{DoW,nl}}^{\text{IC}}$} without the periodic parameter terms (i.e., with $\phi_{h,1,h, j}=\phi_{h,1,24,j}=0$), 
	
	\item[9-12.] \textbf{24lasso$_{\text{DoW,p}}^{\text{IC}}$} without the non-linear effects (i.e., with $\phi_{h,k,\min, 0}=\phi_{h,k,\max, 0}=0$), 
	
	\item[13-16.] and \textbf{24lasso$_{\text{DoW}}^{\text{IC}}$} without the non-linear and periodic effects,
\end{enumerate}
where the superscript \textbf{IC} (= \textbf{AIC}, \textbf{HQC}, \textbf{BIC}, \textbf{OLS}) denotes the information criterion used.

Because of the nature of the shrinkage factor in Eqn.\ \eqref{eqn:lasso}, making $\lambda$ sufficiently large will cause some of the coefficients to be exactly zero. Obviously, selecting a good value of $\lambda$ for the lasso is critical. We choose it such that an in-sample information criterion is maximized. To this end, we consider the \emph{generalized information criterion}  \cite[GIC; see][]{sto:sel:04}:
\begin{equation}
	GIC_{\kappa}(\bsbeta) =   \text{RSS}(\bsbeta) + \kappa K_{\bsbeta} \sigma^2,
	\label{eqn:GIC}
\end{equation}
where $\kappa$ is an information criterion type parameter, $\text{RSS}(\bsbeta)$ is the residual sum of squares associated with $\bsbeta$, $K_{\bsbeta}$ are the non-zero parameters in $\bsbeta$ and $\sigma^2$ is the variance of the residuals that we estimate by the sample variance. We consider the following special cases:
\begin{itemize}
	\itemsep0em
	\item the Akaike Information Criterion (\textbf{AIC}) with $\kappa = 2$, as used in the EPF context by \cite{zie:ste:hus:15},
	\item the Hannan-Quinn Information Criterion (\textbf{HQC}) with $\kappa = 2 \log( \log( n ) )$,
	\item and the Bayesian Information Criterion (\textbf{BIC}; also known as the Schwarz criterion) with $\kappa =  \log( n )$, as used in the EPF context by \cite{zie:16:TPWRS},
\end{itemize}
where $n$ is the sample size and $p$ is the number of possible parameters in the model (in fact, both are the dimensions of the model parameters of the regression matrix). 
Next to the lasso estimation approach, we also consider the OLS solution, with \textbf{IC}=\textbf{OLS}. It can be interpreted as an information criterion based solution with $\kappa = 0$. Note, that two HQC-based models, i.e., \textbf{24lasso$_{\text{DoW,p,nl}}^{\text{HQC}}$} and \textbf{24lasso$_{\text{DoW,nl}}^{\text{HQC}}$}, have been analyzed in the empirical study in Section \ref{sec:Empirical}.

\subsection{Three other univariate AR models (class C7)}
\label{ssec:App:AR}

The \textbf{AR$_{\text{HoW}}$} model is complemented by three univariate AR structures:
\begin{align}
Y_t &= \ov{Y} + \phi_0 +  \sum_{k=1}^p \phi_k( Y_{t-k} - \ov{Y}) + \eps_t,
\label{eq_AR} \\
Y_t &= \ov{Y}_{\text{DoW},t} + \phi_0 + \sum_{k=1}^p \phi_k( Y_{t-k} - \ov{Y}_{\text{DoW},t}) + \eps_t,  \label{eq_ARDoW} \\
Y_t &= \ov{Y}_{\text{HoD},t} + \phi_0 + \sum_{k=1}^p \phi_k( Y_{t-k} - \ov{Y}_{\text{HoD},t}) + \eps_t,  \label{eq_ARHoD} 
\end{align}
denoted by \textbf{AR}, \textbf{AR$_{\text{DoW}}$} and \textbf{AR$_{\text{HoD}}$}, respectively. 
Like the \textbf{AR$_{\text{HoW}}$} model, we estimate these three variants by solving the Yule-Walker equations and minimizing the AIC with a maximum order of $p_{\max}=196$.

\subsection{Univariate lasso models (class C8)}
\label{ssec:App:lasso}

Given the following general formula:
\begin{align}
Y_{t} =
& \sum_{k =1 }^{168} \phi_{0,k} \text{HoW}^k_t  
+ \sum_{k =1 }^{196} \phi_{1,k} Y_{t-k} 
+ \underbrace{ \sum_{k =1 }^{168} \phi_{2,k} \text{HoW}^k_t Y_{t-1} 
	+ \sum_{k =1 }^{168} \phi_{3,k} \text{HoW}^k_t Y_{t-24}  }_{\text{periodic effects}} \nonumber \\
&+ \underbrace{\phi_{4,1} Y_{t-24, \min} + \phi_{4,2} Y_{t-24, \max}}_{\text{non-linear effects}}  + \eps_{t},
\label{eq_model_uni_full}
\end{align}
we can define the full set of 16 models: 
\begin{enumerate}
	\itemsep 0mm
	\item[1-4.] \textbf{lasso$_{\text{HoW,p,nl}}^{\text{IC}}$} -- the full model,
	
	\item[5-8.] \textbf{lasso$_{\text{HoW,p}}^{\text{IC}}$} without non-linear effects (i.e., with $\phi_{4,1}=\phi_{4,2}=0$), 
	
	\item[9-12.] \textbf{lasso$_{\text{HoW, nl}}^{\text{IC}}$} without periodic effects  (i.e., with $\phi_{2,k}=\phi_{3,k}=0$), 
	
	\item[13-16.] and \textbf{lasso$_{\text{HoW}}^{\text{IC}}$} without non-linear and periodic effects,
\end{enumerate}
where the superscript \textbf{IC} (= \textbf{AIC}, \textbf{HQC}, \textbf{BIC}, \textbf{OLS}) denotes the information criterion used; for estimation details see Section \ref{sssec:24lasso} and \ref{ssec:App:24lasso}).
Note, that two HQC-based models, i.e., \textbf{24lasso$_{\text{DoW,p}}^{\text{HQC}}$} and \textbf{24lasso$_{\text{DoW}}^{\text{HQC}}$}, have been analyzed in the empirical study in Section \ref{sec:Empirical}.

	\section{Alternative representations}
	\label{sec:App:AlternativeRepresentations}
	
	In this Appendix
we provide alternative representations for some of the models considered in the study. This may lead to a better understanding of the autoregressive structures and the relationships between similar the multivariate and univariate modeling frameworks.
		
		For instance, the expert models defined in Section \ref{sssec:ExpertModels} and \ref{ssec:App:ExpertModels} can be written in a univariate way as well. However, the representation is quite complex, therefore we show the representation only for the \textbf{expert$_{\text{DoW,p,nl}}$} model defined by Eqn.\ \eqref{eq_expert-wd}. The model is a sparse 168-periodic AR model with non-linear impact:
		\begin{align}
		Y_t &= \phi_{0,t} + \sum_{k}^{168} \phi_{k,t} Y_{t-k} 
		+ \underbrace{ \phi_{\min,t} Y_{\min,t} 
			+ \phi_{\max,t} Y_{\max,t}}_{\text{non-linear effects}} + \eps_{t}, \\
		\phi_{k,t} &= 
		\begin{cases}
		\beta_{h,1} + \sum_{j=1}^{7} \beta_{h,7+j} \text{DoW}^j_t, & \mbox{for } k=0\\
		\beta_{h,7} \text{HoD}_t^k  + \sum_{j=1}^7 \beta_{h, 21+j}  \text{HoD}_t^k \text{DoW}^{j}_t ,  & \mbox{for } k = 1,2,\ldots, 23  \\
		\beta_{h,2} +  \beta_{h,7} \text{HoD}_t^k +\sum_{j=1}^7 ( \beta_{h,14+j} + \beta_{h, 21+j}  \text{HoD}_t^k ) \text{DoW}^{j}_t,  & \mbox{for } k = 24 \\
		\beta_{h,3} ,  & \mbox{for } k = 48 \\
		\beta_{h,4} ,  & \mbox{for } k = 168 \\
		%\phi_{k, k} \text{HoD}^{\text{mod}(k-1,24)+1}_t, & \mbox{for } k \in \{24+1,\ldots, 48, 6\cdot 24+1,\ldots, 7\cdot 24\}  \nonumber \\
		0, & \ow,
		\end{cases} \nonumber
		\end{align}
		$\phi_{\min,t} = \beta_{h,5}$, $\phi_{\max,t} = \beta_{h,6}$, $Y_{\min,t}=\min(\bsY_{d-1})$, $Y_{\max,t}=\max(\bsY_{d-1})$ with
		$h= \modulo(t-1,24) + 1$, $d=(t-h)/24$,
		% denote the univariate versions of the processes 
		% $Y_{d, \min}$ and $Y_{d, \max}$, 
		and $\modulo$ as modulo operator with respect to the second argument (here $24$), e.g., for $t=26$ we receive $\text{mod}(26-1,24)+1 = 2$. 
	
	On the other hand, the \textbf{24AR$_{\text{HoD}}$} model defined by Eqn.\ \eqref{eq_24AR} and the \textbf{VAR$_{\text{HoD}}$} model defined by Eqn.\ \eqref{eq_VAR_HoD} can be easily rewritten as univariate models. The former as a sparse 24-periodic AR model:
		\begin{equation}
		Y_t = \phi_{0,t} + \sum_{k=1}^{p} \phi_{24k,t} Y_{t-24k} + \eps_{t} \ \ \text{ with } \ \ \phi_{24k,t} = \phi_{24k,h} ,
		\end{equation}
		where $p = \max_h(p_h)$ and $h= \modulo(t-1,24) + 1$, 
		and the latter as a 24-periodic AR model:
		\begin{align}
		Y_t = \phi_{0,t} + \sum_{k=1}^{K} \phi_{k,t} Y_{t-k} + \eps_{t} \ \ \text{ with } \ \ \phi_{0,t} = \phi_{0,h} ,  \\
		\phi_{k,t} = \phi_{ \text{div}(k+h-1,24),h, \modulo(k+h-1,24)+1 },
		\nonumber
		\end{align} 
		$h= \modulo(t-1,24) + 1$, $K = 24(p+1)-1$, $\phi_{k,l,m}$ as elements of $\bsPhi_k$ 
		where we set $\phi_{0,l,m} = \phi_{p+i,l,m} = 0$ for all $i>0$
		and $\text{div}$ as quotient of the Euclidean division with respect to the second argument.
	Alternatively, we can represent the \textbf{VAR$_{\text{HoD}}$} model as a set of 24 single equations, where for each $h$ we have:
	\begin{equation}
	Y_{d,h} = 
	\ov{Y}_{\text{HoD},d,h} + \phi_{h,0} +
	\sum_{k=1}^{p} \sum_{j=1}^{24} \phi_{h,j,k} ( Y_{d-k,j} - \ov{Y}_{\text{HoD},d-k,h} ) + \eps_{d,h}.
	\label{eq_VAR_HoD:24AR}
	\end{equation}
	
	Finally, the univariate \textbf{AR} model, defined by Eqn.\ \eqref{eq_AR}, can be rewritten as a special case of a 24-dimensional VAR process. However, the representation is:
	\begin{equation} 
	\bsY_{d} = \bsPhi_0^{-1} \bsphi_0 + \sum_{k=1}^{r} \bsPhi_0^{-1} \bsPhi_k ( \bsY_{d-1} - \ov{Y} \bsone)+ \bseps_d,
	\end{equation}
	with
	$\bsphi_0 = (\ov{Y} + \phi_0)\bsone$,
	$$\bsPhi_0 = \begin{bmatrix}
	1 & 0 & 0 & \cdots & 0 & 0  \\
	-\phi_1 & 1 & 0 & \cdots & 0 & 0  \\
	-\phi_2 & -\phi_1 & 1 & \cdots & 0 & 0  \\
	\vdots & \vdots & \vdots & \ddots & \vdots & \vdots  \\
	-\phi_{22} & -\phi_{21} & -\phi_{20} & \cdots & 1 & 0  \\
	-\phi_{23} & -\phi_{22} & -\phi_{21} & \cdots & -\phi_{1} & 1  \\
	\end{bmatrix}
	\ \ \text{ and } \ \ 
	\bsPhi_k = \begin{bmatrix}
	\phi_{24(k-1)} & \phi_{24(k-1) -1} & \cdots & \phi_{24(k-1) -23}  \\
	\phi_{24(k-1)+1} & \phi_{24(k-1)} & \cdots & \phi_{24(k-1) -22}  \\
	\vdots & \vdots & \ddots & \vdots  \\
	\phi_{24(k-1)+23} & \phi_{24(k-1) +22} & \cdots & \phi_{24(k-1) }  \\
	\end{bmatrix},
	$$
	and we assume that $\phi_k = 0$ for $k>p$.

\section{Model selection}
\label{sec:App:ModelSelection}

\begin{figure}[p]
	\centering
	\includegraphics[width=1\textwidth]{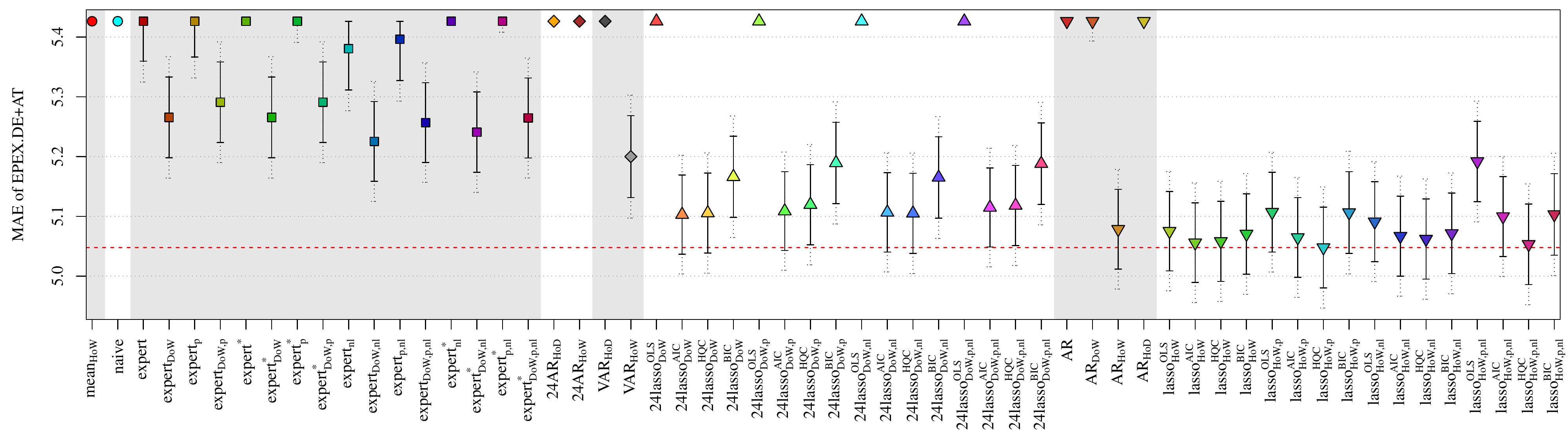} 
	\includegraphics[width=1\textwidth]{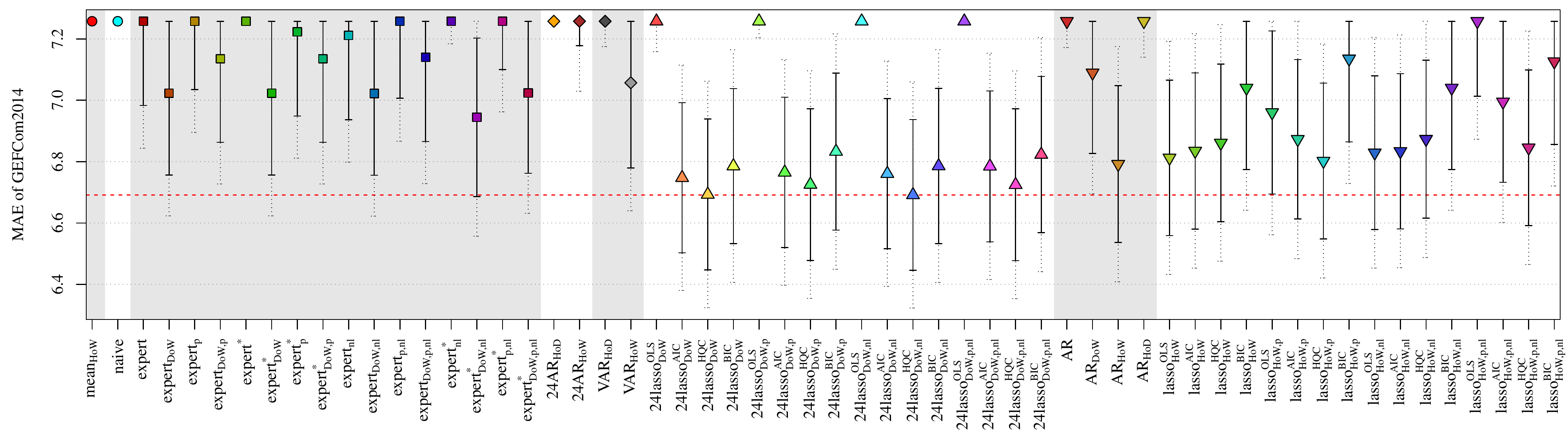} 
	\includegraphics[width=1\textwidth]{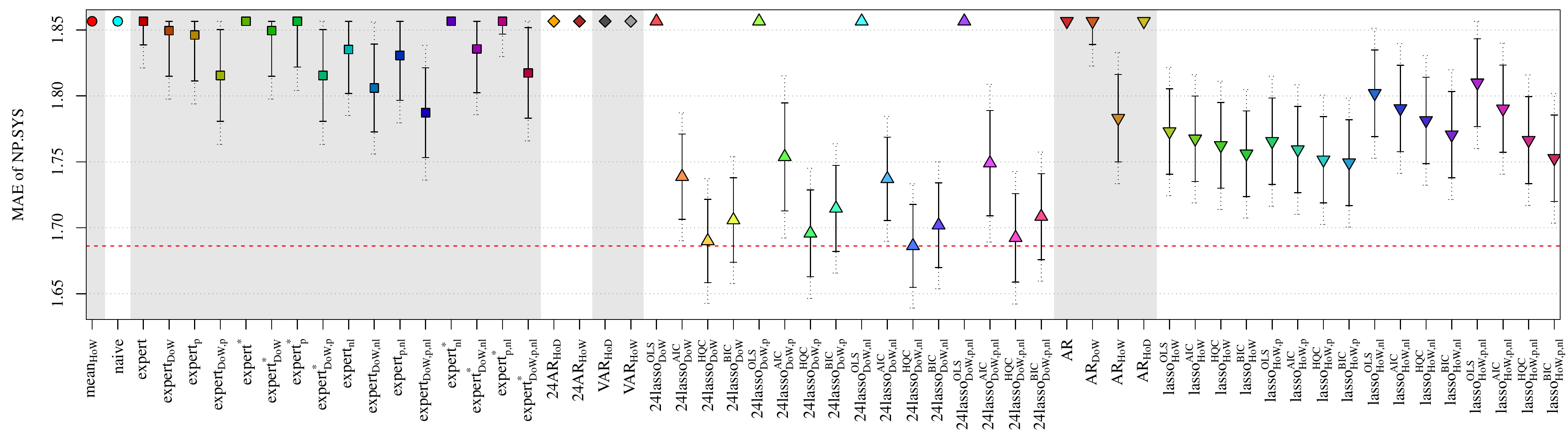} 
	\includegraphics[width=1\textwidth]{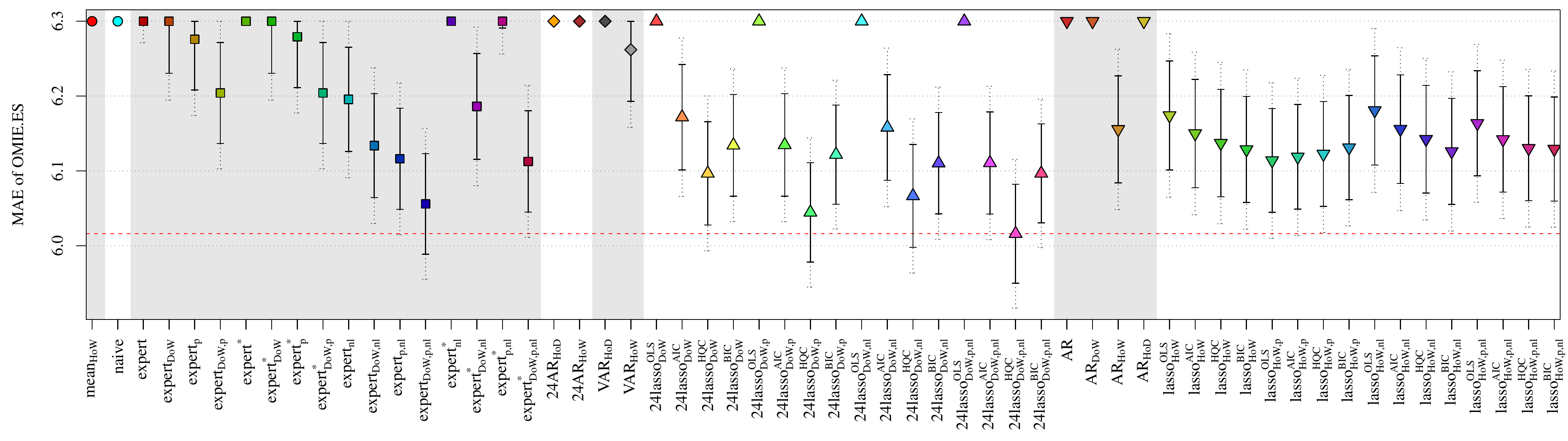}
	\caption{Mean Absolute Errors (MAE) for the full out-of-sample period, as defined by Eqn.\ \eqref{eqn:MAE}, for all 58 models and four major markets (\emph{from top to bottom}): EPEX.DE+AT, GEFCom2014, NP.SYS and OMIE.ES. The solid and dotted whiskers represent the $2\sigma$ and $3\sigma$ ranges, respectively. The dashed red line indicates the MAE of the best performing model. Gray/white background is used to group models from the same class. For clarity of presentation, in all four panels we use an upper cap, e.g., for the EPEX.DE+AT market all MAE values exceeding 5.5 (EUR/MWh) are set equal to 5.5.}
	\label{fig:MAE:4}
\end{figure}

\begin{figure}[tb]
	\centering
	\includegraphics[width=1\textwidth]{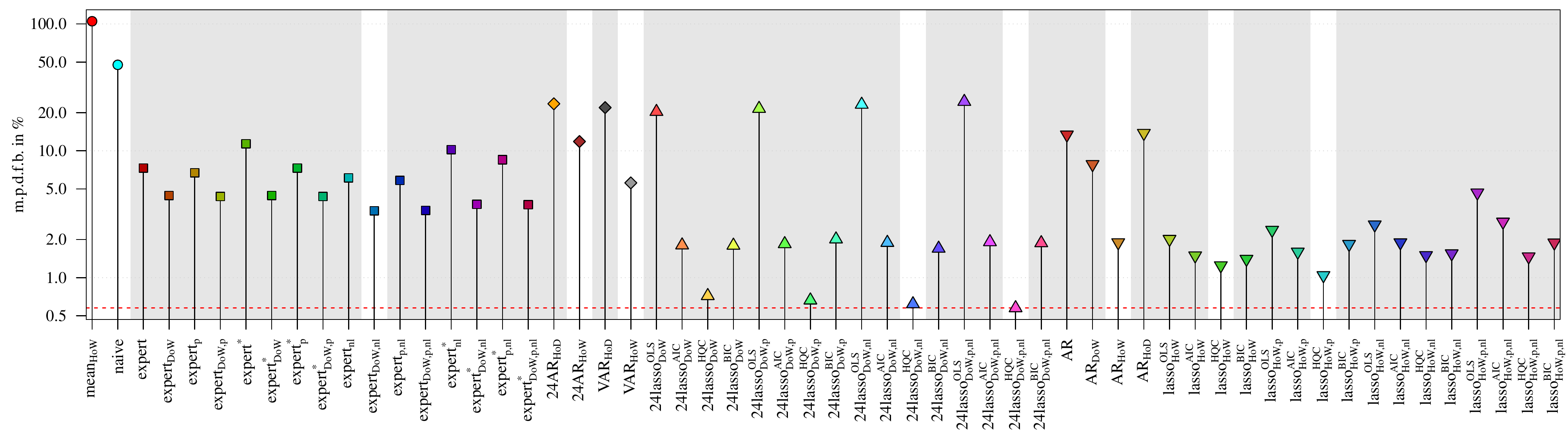}
	\caption{Mean percentage deviation from the best (m.p.d.f.b.) model in terms of MAE, for all 58 models and across all 12 datasets, as defined in Eqn.\ \eqref{eqn:mpdfb}. Note the logarithmic scale on the Y-axis and compare with Table \ref{tab:MAE}, where the m.p.d.f.b.\ values for 10 selected models are also provided. White background is used to indicate the 10 selected models, gray -- the worse performing models in each class. 
	}
	\label{fig:mpdfb}
\end{figure}

In Figures \ref{fig:MAE:4} and \ref{fig:mpdfb} we summarize the results for all 58 models.  In particular, in Figure \ref{fig:MAE:4} we plot Mean Absolute Errors (MAE) for the full out-of-sample period, as defined by Eqn.\ \eqref{eqn:MAE}, for all 58 models and four major markets: EPEX.DE+AT, GEFCom2014, NP.SYS and OMIE.ES.

From Figure \ref{fig:mpdfb} we can clearly see that the HQC criterion, see Section \ref{sssec:24lasso}, leads to the best performing on average lasso models, both multivariate and univariate. Interestingly, only the AIC and BIC criteria have been tried in this context before \cite[see][]{zie:ste:hus:15,zie:16:TPWRS}. This behavior is, however, not uniform across the markets. For instance, for the NP.DK1 and NP.DK2 datasets \textbf{lasso$_{\text{DoW}}^{\text{BIC}}$} is as good as  \textbf{lasso$_{\text{DoW,p}}^{\text{HQC}}$}. From Figure \ref{fig:mpdfb} we can also see that the HQC-estimated multivariate lasso models on average outperform all univariate lasso models. But the differences are small. On the other hand, when the full model is estimated (i.e., the OLS `criterion'), the performance deteriorates but the  loss of forecasting accuracy is extreme only for the multivariate lasso models. Probably the 24 times shorter calibration sample of 730 observations is too small for the multi-parameter structure. 

The importance of including dummies for all days of the week is clearly visible in Fig.\ \ref{fig:mpdfb}. Every second expert model is better than the preceding model without the \textbf{DoW} component. These results support the observation of \cite{uni:now:wer:16} that the weekly seasonality requires better modeling than offered by typically-used expert models. This effect is also visible for the multi-parameter structures -- \textbf{24AR}$_{\text{HoW}}$ is better than \textbf{24AR}$_{\text{HoD}}$, \textbf{VAR}$_{\text{HoW}}$ is better than \textbf{VAR}$_{\text{HoD}}$, \textbf{AR}$_{\text{DoW}}$ is better than \textbf{AR} and \textbf{AR}$_{\text{HoW}}$ is better than \textbf{AR}$_{\text{HoD}}$.

\section*{Acknowledgments}

This work was partially supported by the National Science Center (NCN, Poland) through Grant 2015/17/B/HS4/00334 (to RW).

\section*{References}
\bibliographystyle{elsarticle-harv}
\bibliography{bibliography}

\end{document}